\RequirePackage[l2tabu,orthodox]{nag} %
\documentclass[doctorate,g5paper,twoside, nocover]{chalmers-thesis}

\usepackage[utf8]{inputenc} %
\usepackage{microtype} %
\usepackage{subfiles} %
\usepackage{emptypage} %
\usepackage{hyperref} %
\usepackage[english]{babel}

\usepackage{thesis_commands}
\newacronym{ldpc}{LDPC}{low-density parity-check}
\newacronym{qldpc}{qLDPC}{quantum \glsentrylong{ldpc}}
\newacronym{qec}{QEC}{quantum error-correcting}
\newacronym{bp}{BP}{belief propagation}
\newacronym{osd}{OSD}{ordered statistics decoding}
\newacronym{css}{CSS}{Calderbank-Shor-Steane}
\newacronym{mbqc}{MBQC}{measurement-based quantum computing}
\newacronym{cv}{CV}{continuous-variable}
\newacronym{cqed}{cQED}{circuit quantum electrodynamics}

\makeatletter
\let\blx@rerun@biber\relax
\makeatother

\DeclareBibliographyCategory{fullcited}
\newcommand{\mybibexclude}[1]{\addtocategory{fullcited}{#1}}

\title{Bosonic quantum computing with near-term devices and beyond}
\author{Timo Hillmann}
\department{Department of Microtechnology and Nanoscience (MC2)}
\division{Applied Quantum Physics Laboratory}
\reportno{5632}
\ISBN{978-91-8103-174-4} %
\copyrightyear{2025}

\coverfigure{
\includegraphics[width=0.635\textwidth]{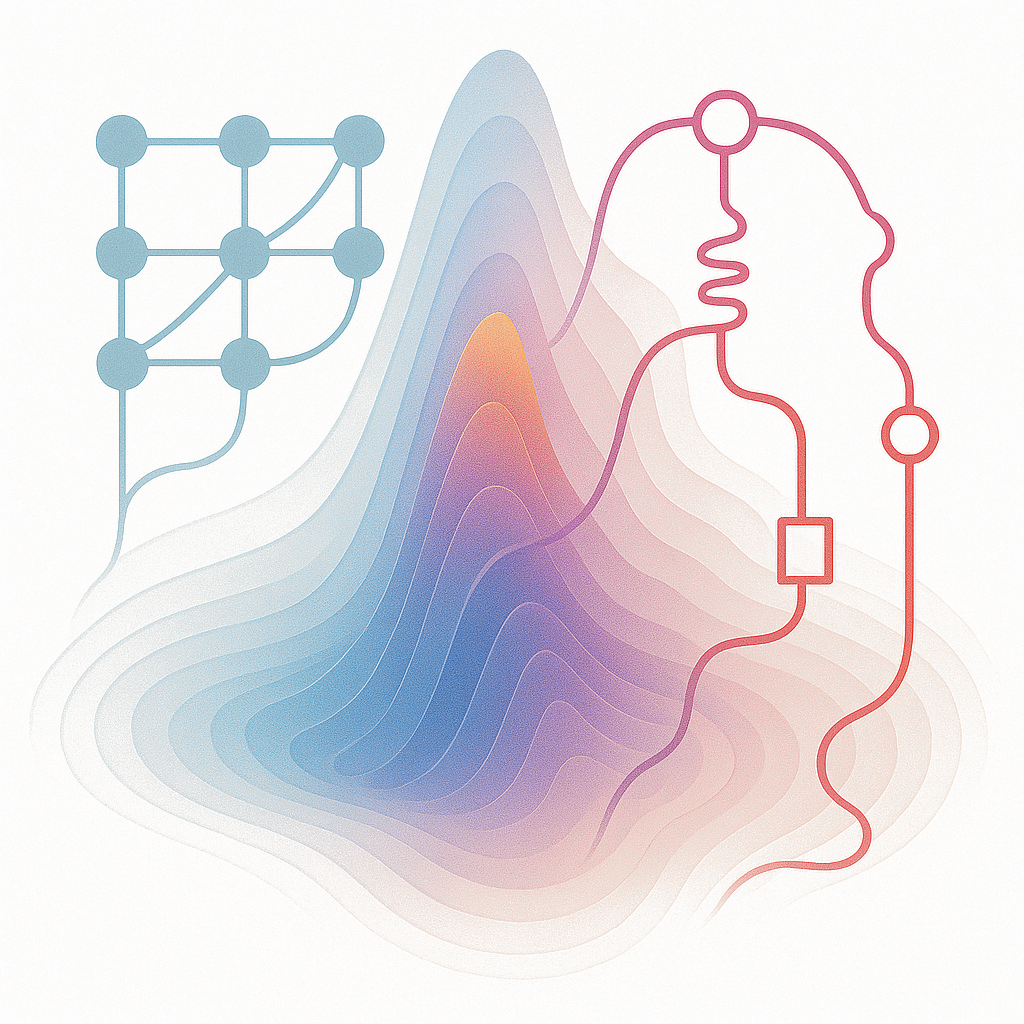}
}
\covercaption{This illustration captures the thematic intersection explored in this thesis: quantum error correction, bosonic codes, and superconducting circuits.
The left side features an artistic rendition of a Tanner graph, representing the aspects of quantum error correction, while the right side shows a stylized superconducting quantum circuit.
These two domains are unified by the central height profile of a Wigner function, evoking the continuous-variable nature at the heart of bosonic quantum error correction.
This illustration was generated with the assistance of ChatGPT.}

\firstabstract{
This thesis investigates scalable strategies for fault-tolerant quantum computation by developing and analyzing bosonic quantum codes, quantum low-density parity-check (LDPC) codes, and decoding protocols that aim to unify bosonic and discrete-variable quantum error correction.

In the continuous-variable regime, we explore the use of native nonlinearities in superconducting microwave circuits to realize a universal gate set for continuous-variable quantum computing, including the deterministic generation of a cubic phase state. 
Separately, we propose and analyze the dissipatively squeezed cat qubit, a noise-biased bosonic encoding that offers improved error suppression and faster gate implementations compared to standard cat qubits. 
To evaluate the broader viability of bosonic encodings, we study the performance of rotation-symmetric and Gottesman-Kitaev-Preskill (GKP) codes under realistic noise and measurement models, revealing important trade-offs in measurement-based approaches.

Recognizing the need to integrate bosonic codes into larger fault-tolerant architectures, we develop decoding techniques that explicitly leverage analog syndrome information from the readout of bosonic modes. These methods reduce the need for repeated measurements and enable quasi-single-shot decoding in concatenated schemes, forming a bridge between continuous-variable encodings and discrete-variable stabilizer codes.

To advance scalable discrete-variable fault tolerance, we introduce localized statistics decoding, a flexible and highly parallelizable decoding framework for general quantum LDPC codes with state-of-the-art accuracy.
Based on a novel on-the-fly matrix elimination strategy, this decoder efficiently identifies and resolves local error configurations, enabling low-latency and hardware-friendly implementations. 
Additionally, we present quantum radial codes, a new family of single-shot quantum LDPC codes constructed from lifted products of classical quasi-cyclic codes. 
These codes offer low overhead, tunable parameters, and competitive performance under circuit-level noise, making them promising candidates for near-term implementation.

Finally, we propose the concept of fault complexes, a homological framework for representing and analyzing faults in dynamic quantum error correction protocols. Extending the role of homology in static CSS codes, fault complexes provide a general language for the design and analysis of fault-tolerant schemes.

}

\keywords{
Quantum Error Correction,
Bosonic Codes,
Superconducting Circuits,
Quantum low-density parity-check codes,
Squeezed cat qubit,
Gottesman-Kitaev-Preskill code,
Belief propagation,
Ordered statistics decoding,
Localized statistics decoding,
CSS codes,
Quantum Radial Codes,
Schrieffer-Wolff transformation,
}

\dedication{
\vfill
\setlength{\epigraphwidth}{0.45\textwidth}
\epigraph{Being a physicist is a great privilege. Be worthy of it. Most of humanity spends its life doing boring repetitive tasks.}{David P. Stern, \emph{All I Really Need to Know...}, Physics Today \textbf{46} (5), 63 (1993)}} %
\acknowledgements{
First and foremost, I would like to thank Fernando Quijandría for engaging with my ideas --- however vague they may have been --- throughout the past five years. 
Though you moved to Japan before formally becoming my thesis supervisor, your encouragement and guidance at key moments gave me the confidence to pursue this work independently. Thank you for your support, even from afar.
I would also like to thank Giulia Ferrini for taking on the role of supervisor and for providing the framework that allowed this thesis to come together.

The time throughout my PhD has been shaped by various opportunities and people.
At the beginning of it stands the fellowship from the Excellence Initiative Nano that funded this research, granting me the freedom and independence that I am now so used to.
Furthermore, an important aspect of my PhD experience has been working with the team at Xanadu Quantum Computing.
I would like to especially thank Rafael Alexander, Ben Baragiola, Eli Bourassa, Ilan Tzitrin, and Mike Vasmer for acting as great collaborators, teachers, friends, and mentors throughout this time.
Possibly more than anything else, my participation in the IBM Quantum Error Correction Summer School 2022 had a profound influence on the shape this thesis ultimately took.  I am grateful to the organizers and lecturers for creating such a formative experience.

I am also indebted to the organizers of the many conferences, group visits, and workshops I had the privilege to attend.
In particular, QIP Ghent 2023, YITP QEC 2024, FTQT Benasque 2024, and Coogee 2025 stand out as events that not only contributed to my research but also led to lasting memories, friendships, and future collaborations.
What made attending these events even more enriching was knowing that, no matter where I went, I could always find someone I knew --- whether to discuss new research ideas or to enjoy a coffee together.
Thank you, Asmae Benhemou, Lucas Berent, Matt McEwen, Julio Carlos Magdalena de la Fuente, Teague, Tom Scruby, and many more.

I was also glad to have made many friends at AQP, QTL, and around Gothenburg generally.  
Thank you, Alberto Del Ángel Medina, Oliver Hahn, Théo Sépulcre, David Fitzek, Therese Karmstrand, Henning Kirchberg, Claudia Castillo-Moreno, Amr Osman, Chris Warren, Werner Dobrautz, and others
for the enjoyable coffee and fika breaks, lunch conversations, and climbing sessions.
Thanks to Vitaly Shumeiko for being someone with whom I could always have meaningful physics discussions.
To my first Master's student, Thomas Jaeken, thank you for meeting the challenge with trust, energy, and curiosity.

Many thanks to A43 Coffee and Beanaut Coffee Roasters, and the specialty coffee community of Sweden, for introducing me to the world of specialty coffee.
I've made sure to carry that passion into the quantum error correction community.

To all my other friends, thank you for the visits, messages, and distractions --- for keeping me connected to a life beyond this thesis.
Finally, I want to thank my family for their continuous support throughout my 10 years of university, despite the distance that kept us apart for much of that time.
\vfill

\begin{flushright}
Timo Hillmann, Göteborg, May 2025
\end{flushright}

}

\makeatletter

\newrobustcmd*{\nobibliography}{%
  \@ifnextchar[%
    {\blx@nobibliography}
    {\blx@nobibliography[]}}

\def\blx@nobibliography[#1]{}

\appto{\skip@preamble}{\let\printbibliography\nobibliography}

\makeatother

\fancypagestyle{plain}{%
\fancyhf{}
\renewcommand{\headrulewidth}{0pt}
\fancyfoot[RO]{\thepage}
\fancyfoot[LE]{\thepage}
}

\fancypagestyle{fancy-a}{%
\fancyhf{}
\renewcommand{\headrulewidth}{0.5pt}
\fancyhead[RO]{\normalfont \rightmark}
\fancyhead[LE]{\normalfont \leftmark}
\fancyfoot[RO]{\thepage}
\fancyfoot[LE]{\thepage}
}

\begin{document}
\maketitle

\cleardoublepage
\pagestyle{fancy-a}

\chapter{Introduction \label{chap:intro}}
Classical computers have become indispensable tools for solving a wide range of problems.
However, some problems are out of reach for these devices.
Quantum computing offers an alternative paradigm of computing that utilizes the features of quantum mechanics, which promises to solve some of those problems more efficiently than classical computers.
Most prominently, quantum computers promise an exponential speed up for integer factorization~\cite{shor_algorithms_1994, shor_polynomial-time_1997} and the simulation of quantum systems~\cite{lloyd_universal_1996}.
The latter is particularly challenging for conventional computers because they require incorporating the laws of quantum mechanics into the simulation --- a task that another quantum system, a quantum computer, is potentially much better suited for~\cite{benioff_computer_1980, feynman_simulating_1982}.
Being able to more efficiently and accurately simulate quantum systems could have a significant impact on material engineering~\cite{bauer_quantum_2020}, aiding the design of improved batteries and photovoltaics, quantum chemistry~\cite{aspuru-guzik_simulated_2005, reiher_elucidating_2017, cao_quantum_2019, motta_emerging_2022, mcardle_quantum_2020}, and the development of pharmaceuticals~\cite{santagati_drug_2024}.
However, realizing these advantages requires overcoming profound physical and theoretical challenges, foremost among them the fragility of quantum information.  
This fragility necessitates the development of fault-tolerant quantum computers, where encoded quantum information can be protected and manipulated reliably over time.

It has become evident that quantum error correction will be necessary for almost any practical application of quantum computers, with one of the few exceptions possibly being small-scale physics simulations.
In general, the need for quantum error correction introduces a substantial overhead~\cite{beverland_cost_2021, gidney_how_2021} that reduces the advantage of quantum algorithms over conventional classical algorithms.
Quantum algorithms are further challenged by the classical algorithm design based on heuristics, which are efficient methods guided by intuition and insight that lack formal guarantees yet typically yield accurate results.
Indeed, in classical computing, many algorithms began as heuristics, with their strengths understood only after broad application, see e.g., Ref.~\cite{spielman_smoothed_2001}.
Approaching the development of quantum algorithms in a similar spirit is possibly hindered in two ways.
First, quantum mechanics often defies direct intuition, and as a result, important discoveries, such as gate teleportation~\cite{nielsen_programmable_1997}, were made numerically by coincidence~\cite{zierler_michael_2022}.
Second, the absence of large-scale fault-tolerant quantum hardware limits the discovery of new protocols to those tractable by classical simulation.

The choice of the quantum error-correcting code carries many practical trade-offs.
The current gold standard, the surface code~\cite{kitaev_fault-tolerant_2003, bravyi_quantum_1998, freedman_projective_2001}, requires only nearest-neighbor connectivity in a planar architecture and has a high resilience against realistic noise, but will require thousands of physical qubits for each logical qubit it protects against noise.
On the other hand, \gls{qldpc} codes~\cite{breuckmann_quantum_2021} can reduce the number of physical qubits needed but have increased connectivity requirements and are typically less resilient against realistic noise processes compared to the surface code.
Theoretical progress in this domain is rapid, not only on code construction and design~\cite{panteleev_asymptotically_2022, dinur_good_2022, leverrier_quantum_2022, bravyi_high-threshold_2024, yoshida_concatenate_2024, shaw_lowering_2025}, but also decoding~\cite{wolanski_introducing_2024, ott_decision-tree_2025, beni_tesseract_2025, yin_symbreak_2024, iolius_almost-linear_2024, gong_toward_2024}, and logical operations~\cite{cowtan_ssip_2024, cowtan_parallel_2025, he_extractors_2025, ide_fault-tolerant_2024, swaroop_universal_2024, cross_linear-size_2024, williamson_low-overhead_2024}.

Interestingly, we are approaching a tipping point in the design of quantum error-correcting codes where theoretical progress is beginning to translate into small-scale experimental implementations, for example, in superconducting circuits~\cite{acharya_quantum_2025, putterman_hardware-efficient_2025, sivak_real-time_2023, krinner_realizing_2021, besedin_realizing_2025, gupta_encoding_2024, sundaresan_demonstrating_2023}, neutral atoms~\cite{bluvstein_logical_2024, rodriguez_experimental_2024, reichardt_demonstration_2024}, or trapped ions~\cite{berthusen_experiments_2024, matsos_robust_2024, ryan-anderson_high-fidelity_2024, da_silva_demonstration_2024, moses_race-track_2023, ryan-anderson_realization_2021}.
Importantly, this allows us not only to refine our understanding of fault tolerance in practical settings but also to design more efficient quantum error-correcting codes tailored to the physical architecture.

One of the subfields within quantum computing that has made exceptional experimental and theoretical progress in the last decade is the field of bosonic quantum computing and bosonic quantum error correction~\cite{brady_advances_2024, cai_bosonic_2021, joshi_quantum_2021}.
Bosonic quantum computing possibly resembles the quantum analog of the almost forgotten field of classical analog computing more closely than the notion of discrete-variable (quantum) information processing discussed so far.
It is therefore also often referred to as continuous-variable quantum information processing or quantum computing.
On the other hand, bosonic quantum error correction, or simply bosonic codes, represent discrete quantum information encoded into a system of quantum continuous variables.
Some of the most prominent bosonic codes, such as the Gottesman-Kitaev-Preskill code~\cite{gottesman_encoding_2001} or the cat qubit~\cite{leghtas_hardware-efficient_2013, leghtas_confining_2015}, can be viewed as utilizing ideas present in classical transistors to protect quantum information.
For instance, in flash memory, a logical bit is defined by the charge on a floating gate --- for practical purposes, a continuous quantity --- and a threshold is used to distinguish between zero and one.
The physical encoding is analog, but the (logical) computation is digital, similar to bosonic codes.
The premise of bosonic codes is to exploit hardware-level protection to achieve fault tolerance with fewer physical resources than their (unprotected) discrete-variable counterparts.
However, their improved protection comes with other trade-offs such as increased physical complexity, susceptibility to other noise channels, and more challenging gate implementations.
Nonetheless, much progress has been made theoretically~\cite{puri_engineering_2017, guillaud_repetition_2019, royer_stabilization_2020, putterman_stabilizing_2022, dubovitskii_bit-flip_2025, ruiz_ldpc-cat_2025, xu_letting_2024, hann_hybrid_2024, matsuura_continuous-variable_2024, shaw_logical_2024, gautier_designing_2023, hastrup_all-optical_2021, hastrup_deterministic_2020, hastrup_improved_2020, weigand_generating_2018, weigand_realizing_2019, duivenvoorden_single-mode_2017} as well as experimentally~\cite{putterman_hardware-efficient_2025, sivak_real-time_2023, rousseau_enhancing_2025, putterman_hardware-efficient_2025, reglade_quantum_2024, de_neeve_error_2020, berdou_one_2023, marquet_autoparametric_2024, vanselow_dissipating_2025, vlastakis_deterministically_2013, ofek_extending_2016, pan_protecting_2022, milul_superconducting_2023}.

\section{Towards Fault-Tolerant Quantum Computing}
The pursuit of practical quantum computing is inextricably linked to the development of robust fault-tolerant architectures. 
While small quantum processors have demonstrated increasingly sophisticated capabilities~\cite{arute_quantum_2019, morvan_phase_2024, gao_establishing_2025}, the realization of large-scale, useful quantum computations continues to be limited by the fragility of quantum information and the substantial resource overheads required for its protection. 

This thesis explores bosonic codes and \gls{qldpc} codes as candidates for building a fault-tolerant quantum computer.
In particular, at the core, this thesis investigates how bosonic codes and \gls{qldpc} codes can contribute to reducing the hardware and time-overhead of quantum error correction.
Instead of focusing on these approaches individually, it aims to address these questions in a unified approach.
On a broader scale, it contributes to bridging the gap between theoretical abstraction and practical implementations in fault-tolerant quantum computing.
To this end, contributions range from physical-level encoding strategies and decoding protocols to the introduction of a framework for the representation of quantum error-correcting codes in space-time.

The following section provides a detailed, chapter-by-chapter outline of the thesis structure and the specific contributions of each paper.

\section{Outline}

\Cref{chap:ftqt} introduces foundational ideas in fault-tolerant quantum computing that frame the contributions of this thesis within the broader context of quantum error correction. It begins by outlining models of quantum computation, including both gate-based and measurement-based paradigms, and proceeds to cover stabilizer codes as a key class of quantum error-correcting codes. The chapter further introduces the language of chain complexes to describe code structure and delves into various approaches to the decoding problem. These topics form the theoretical foundation for the contributions presented in \refpaper{VII}, \refpaper{VIII}, and \refpaper{IX}, and quantum computing more generally.

\Cref{chap:quantum_continuous_variables} focuses on the formalism of quantum continuous-variable systems and their role in quantum information processing. 
It introduces relevant physical and mathematical concepts for describing infinite-dimensional quantum systems, followed by an overview of bosonic quantum error-correcting codes designed to protect information encoded in such systems. Particular emphasis is placed on cat codes and the Gottesman-Kitaev-Preskill encoding, which play a central role in the work presented in \refpaper{IV}, \refpaper{V}, and \refpaper{VI}. The chapter serves as a conceptual bridge between theoretical developments and their application in hardware-efficient fault-tolerant schemes.

\Cref{chap:quantum_computing_architectures} explores how the continuous-variable systems introduced in the previous chapter can be implemented using superconducting quantum circuits. It focuses on circuit quantum electrodynamics as a versatile platform for building quantum hardware, emphasizing techniques for engineering nonlinear interactions that are essential for quantum control. These capabilities enable the realization of bosonic codes and operations, connecting directly to the contributions of \refpaper{I}, \refpaper{II}, and \refpaper{III}.

\cleardoublepage

\chapter{Fault-tolerant Quantum Computing \label{chap:ftqt}}

\glsresetall
Quantum computers process information using quantum bits, \emph{qubits}, instead of conventional (classical) bits. 
A classical bit stores a single unit of binary information, typically represented as \(0\) or \(1\). 
Similarly, a qubit can be represented by two distinguishable quantum states, \(\ket{0}\) and \(\ket{1}\). 
However, unlike classical bits, a qubit can exist in a superposition of these states.  
The general quantum state \(\ket{\psi}\) of a single qubit is given by  
\begin{align}  
    \ket{\psi} = c_0 \ket{0} + c_1\ket{1},  
\end{align}  
where \(c_0\) and \(c_1\) are complex coefficients satisfying the normalization condition $\lvert c_0 \rvert^2 + \lvert c_1 \rvert^2 = 1$.  
We can also use a qubit to represent a classical bit by requiring that at most one of the coefficients $c_i$ is nonzero.
However, even though the coefficients $c_i$ can take on infinitely many different values and thus allow us to store, in principle, an infinite amount of information, we can at most retrieve a single bit of information.
When measured, a qubit collapses to the classical state $0$ with probability \(\lvert c_0 \rvert^2\) and $1$ with probability \(\lvert c_1 \rvert^2\). 
This measurement process erases any phase information and prevents direct access to the values of \(c_0\) and \(c_1\) from a single measurement.
Instead, only their magnitudes can be inferred statistically by performing multiple measurements on identical copies of the same quantum state.

The difference in computational power of a qubit over a bit is hence much more subtle and is partially due to the possibility that the complex-valued coefficients can interfere, something that is not possible for classical bits.
An operation that exemplifies quantum interference is the Hadamard gate \( \Ha \).  
The Hadamard operation transforms qubit states as follows 
\begin{align}  
    \Ha \ket{0} &= \frac{1}{\sqrt{2}} \ket{0} + \frac{1}{\sqrt{2}} \ket{1} = \ket{+}, \\  
    \Ha \ket{1} &= \frac{1}{\sqrt{2}} \ket{0} - \frac{1}{\sqrt{2}} \ket{1} = \ket{-}.  
\end{align}  
If we begin initially in the equal superposition state $\ket{\psi} = (\ket{0} + \ket{1}) / \sqrt{2}$ and then apply a Hadamard operation to it, we obtain
\begin{align}
    \Ha \ket{\psi} = \left(\frac12 + \frac12\right)\ket{0} + \left(\frac12 - \frac12\right) \ket{1} = \ket{0},
\end{align}
due to the cancellation of the \(\ket{1}\) term, an example of destructive interference.

Another key ingredient that distinguishes quantum computation from classical computation is \emph{entanglement}, a form of quantum correlation that arises in multi-qubit superposition states without a classical counterpart.
To illustrate this, let us consider the so-called Bell state $\ket{\Phi^+}$ defined as
\begin{align} 
    \label{eq:def_bell_plus}
    \ket{\Phi^+} = \frac{1}{\sqrt{2}} (\ket{0}\ket{0} + \ket{1}\ket{1}).  
\end{align}
Measuring the first qubit, the outcome is $0$ or $1$ with equal probability.
However, the result of the measurement determines the state of the second qubit,  and the measurement outcome of it will be perfectly correlated with the measurement result of the first qubit.
On the level of the measurement statistics, one obtains the classical bit string $00$ or $11$ with equal probability.
Yet, these correlations alone do not fully capture the essence of entanglement.  
A classical system could mimic the same statistics by randomly selecting \(0\) or \(1\) and preparing either \(\ket{0}\ket{0}\) or \(\ket{1}\ket{1}\) accordingly. 
What distinguishes entanglement from mere correlation is its robustness against local operations such as the Hadamard operation.
To see this, consider applying the Hadamard operation to both qubits before measuring them. 
A quick calculation shows that the Bell state $\ket{\Phi^{+}}$ remains unchanged under the application of the Hadamard operations as the anti-correlation terms $\ket{0}\ket{1}$ and $\ket{1}\ket{0}$ are canceled by interference effects.
This cancellation, however, is not possible if the Hadarmard operation is applied to one of the non-superposition states $\ket{0}\ket{0}$ or $\ket{1}\ket{1}$ such that the measurement outcome of the first and second qubit become perfectly uncorrelated~\cite{bell_einstein_1964}.

The quantum mechanical concepts of superposition and entanglement thus can be viewed as the origin of the separation of computational power of qubits and classical bits, as they are impossible to reproduce efficiently with classical bits alone.
Unfortunately, harnessing this separation in computational power into useful quantum algorithms appears to be a highly nontrivial task.
This might come as a surprise, especially considering the common misconception that superposition enables quantum computers to perform an exponentially large number of operations in parallel. 
In fact, while superposition does allow a quantum system to represent all possible solutions to a problem simultaneously, valid and invalid ones, this does not mean that a quantum computer can simply evaluate all solutions at once to output the desired answer.
This oversimplification completely ignores the role of quantum interference in quantum information processing.
Instead, designing quantum algorithms should be viewed as the intricate task of steering constructive and destructive interference by quantum operations in such a way that the measurement of the quantum state at the end of the algorithm deterministically yields the correct solution.
A weaker statement of the above could be that the role of the algorithm is to amplify the probability of obtaining the correct solution while suppressing the likelihood of wrong solutions.

In the following, we provide a brief overview of quantum operations.

\section{Fundamentals of Quantum Computing \label{sec:qec_fundamentals}}
Running a quantum algorithm, and thus performing quantum information processing, requires a set of operations executed in a particular order that will determine the output of the algorithm.
This set of operations must be general enough to be able to reach any state with arbitrary precision and is called \emph{universal} if that is the case.
Surprisingly, even though the coefficients $c_i$ of a quantum state can take arbitrary complex values, a finite set of operations suffices.
This celebrated result is known as the Solovay-Kitaev theorem~\cite{kitaev_quantum_1997-1, dawson_solovay-kitaev_2006} that states that given a universal set of operations $\mathcal{G}$, there is a finite sequence of operations $S$ of length $O(\log^{c}(1/\epsilon))$ that can be found efficiently such that $S$ approximates an arbitrary unitary $U \in \mathrm{SU}(d)$ with accuracy $\epsilon > 0$ in the operator norm with constant $c \approx  3.97$.
There are several universal gate sets, and we will focus on a particular one commonly referred to as ``Clifford + T".  
This set consists of the controlled-NOT (\CNOT~or $\CX$) gate, the phase gate $S$, the previously discussed Hadamard gate $\Ha$, and the $\pi/8$-phase gate $T$.
We will define the notion of Clifford gates more precisely below.

\subsection{Pauli group and Heisenberg representation}
Consider a quantum computer, that is, a quantum system, in the state $N \ket{\psi}$,
to which a unitary operation $U$ is applied, where $N$ is another operation we have previously applied.
Then the state of the system is described by
\begin{align}
    U N \ket{\psi} = U N U^{\dagger} U \ket{\psi},
\end{align}
that is, the operator $U N U^{\dagger}$ now acts on the (transformed) system as $N$ did before applying $U$.
As a result, instead of considering how the state vector $\ket{\psi}$ evolves through the applications of arbitrary operations $U$, one might equivalently track the evolution
\[
N \to U N U^{\dagger},
\]
for a sufficiently large set of operators $\{ N \}$.
Since the evolution of quantum systems is linear, it is sufficient to follow a set of operators that spans the space of linear operators for the Hilbert space $\mathcal{H}$ in which $\ket{\psi}$ would evolve, i.e., $\ket{\psi} \in \mathcal{H}$.

To this end, let us introduce the set of Hermitian and unitary operations that can be represented as $2 \times 2$ matrices:
\begin{align}
    \label{eq:pauli_matrices}
    \I &= \mqty[1 & 0 \\ 0 & 1], \quad \X = \mqty[0 & 1 \\ 1 & 0], \quad \Y = \mqty[0 & -i \\ i & 0], \quad \Z = \mqty[1 & 0 \\ 0 & -1],
\end{align}
where $\X$, $\Y$, and $\Z$ are collectively known as the Pauli matrices.  
Importantly, the matrices in Eq.~\eqref{eq:pauli_matrices} form a basis for the space $\mathbb{C}^{2 \times 2}$, and thus can be used to describe any operation on a single qubit.  
Let us also define an $n$-qubit Pauli operator $P = \alpha P_n$, where $P_n$ is the $n$-fold tensor product $P_n \in \{I, X, Y, Z \}^{\otimes n}$, and $\alpha \in \{\pm 1, \pm i \}$ is a coefficient.  
We will often refer to the $n$-qubit Pauli operator $P_n$ as a Pauli string and write it by omitting the tensor product, e.g., $X \otimes Z \otimes Z \otimes X$ is represented as $XZZX$.  
The set of all $n$-qubit Pauli operators forms the $n$-qubit Pauli group $\mathcal{P}_n$, and, as in the single-qubit case, the elements of $\mathcal{P}_n$ form a basis for $\mathbb{C}^{2^n \times 2^n}$.
A convenient basis is $\{X_1, \dots, X_n, Z_1, \dots, Z_n \}$ where $X_i$ ($Z_i$) denote the Pauli $X$ ($Z$) acting on the $i^{\mathrm{th}}$ qubit and trivially elsewhere.
Therefore, to completely specify a general operator on $\mathbb{C}^{2^n \times 2^n}$ we only need to track the evolution of $2n$ single-qubit operators.
Since the single-qubit Pauli matrices anti-commute with each other, two $n$-qubit Pauli operators anti-commute only if an odd number of their tensor factors anti-commute.
We will refer to the set of operators $\{X_1, \dots, X_n, Z_1, \dots, Z_n \}$ as \emph{logical} operators, a term that will be justified in a later section. 
We denote by $\overline{X}_i$ and $\overline{Z}_i$ the operators obtained from transformations of $X_i$ and $Z_i$, respectively.

Within this framework, unitary operators $U$ that map Pauli operators to Pauli operators are elements of the Clifford group, that is, $U$ is Clifford if $U P U^{\dagger} \in \mathcal{P}_n, \; \forall P \in \mathcal{P}_n$.
The $\CX$ conjugates the Pauli $X$ and $Z$ operators as follows
\begin{align}
    \CX: XI \to XX,  \hspace{0.5em} IX \to IX,  \hspace{0.5em} ZI \to ZI,  \hspace{0.5em} IZ \to ZZ.
\end{align}
Furthermore, the phase gate $S$ maps $X \to Y$ and $Z \to Z$ under conjugation, while the $H$ gate maps $X \to Z$ and $Z \to X$ under conjugation.
As expected, the $T$ gate is not a Clifford gate.
It is commonly represented in the computational basis as 
\begin{align}
    \label{eq:t_gate_computational_basis}
    T = \sqrt{S} = \mqty[1 & 0 \\ 0 & e^{i\pi/4}],
\end{align}
and, for example, conjugates the Pauli $X$ operator as $T X T^{\dagger} = \cos(\pi/4)X + \sin(\pi/4)Y \notin \mathcal{P}_1$.
In the following, we will also make use of the controlled-$Z$ ($\CZ$) gate that acts similarly to the $\CX$ gate up to a local Hadamard gate, that is, the $\CZ$ gate maps
\begin{align}
    XI \to XZ,  \hspace{0.5em} IX \to ZX,  \hspace{0.5em} ZI \to ZI,  \hspace{0.5em} IZ \to IZ,
\end{align}
under conjugation.

\subsection{Stabilizer circuits and measurements \label{ssec:stabilizer_circuits_measurements}}
For some quantum circuits, we do not need to follow the evolution of all $2n$ Pauli operators.  
This is the case if some of the circuit inputs are fixed to be $+1$ eigenstates of elements of the Pauli group.  
The set of such operators constrains the state of some qubits and is closed under multiplication.  
This set of operators defines the stabilizer group $\mathcal{S}$ as  
\begin{align}
    \mathcal{S} = \left\lbrace 
    S \in \mathcal{P}_n \mid S \ket{\psi} = \ket{\psi}
    \right\rbrace,
\end{align}  
which must hold for all valid input states $\ket{\psi}$.  
We will return in more detail to the stabilizer group when discussing quantum error-correcting codes in \cref{sec:quantum_error_correction}.  
For now, we will use it to outline an approach that incorporates measurements into the Heisenberg representation of quantum computing.

\begin{figure}
    \centering
    \includegraphics{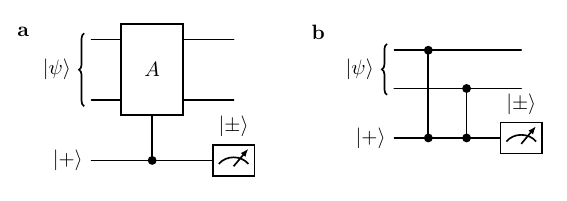}
    \caption{\textbf{a} Quantum circuit for the measurement of an observable $A$ with eigenvalues $\pm 1 $ by an ancilla.
    The circuit performs a controlled-$A$ gate between the $n$ qubit system and an ancilla initially prepared in the state $\ket{+}$ and measures the ancilla in the $X$ basis.
    \textbf{b} Circuit for measuring $A = Z Z$ using $\CZ$ gates.
    }
    \label{fig:measurement_of_observable}
\end{figure}

To this end, let us recall how the measurement of an observable $A$ with eigenvalues $\pm 1$ changes the quantum state $\ket{\psi}$ being measured.  
The measurement is described by two orthogonal projectors  
\begin{align}
    P_{\pm} = \frac{1}{2} \left(I \pm A\right),
\end{align}  
where $I$ is the identity operator on the space of the $n$-qubit state $\ket{\psi}$.  
One can verify that $P_{\pm}$ are projectors, satisfying $P_{\pm}^2 = P_{\pm}$, and that they are orthogonal, meaning $P_{\pm} P_{\mp} = 0$.  
The post-measurement state $\ket{\psi^{\pm}} = \frac{1}{2}(I \pm A) \ket{\psi}$ is projected into an eigenstate of $A$ with eigenvalue $\pm 1$, depending on the measurement outcome.  
Such an observable $A$ can be measured using an ancilla system.  
To achieve this, one applies a controlled-$A$ gate between the $n$-qubit system and a single ancilla qubit initially prepared in the $\ket{+}$ state.  
Then, the ancilla is measured in the $X$ basis; see also \sfigref{fig:measurement_of_observable}{a}. 
One can verify that this circuit effectively measures the observable $A$ through the measurement of the ancilla.  
To see this, note that the ancilla state is initialized in $ \ket{+} $, such that the stabilizer group is generated by a single element: $ \mathcal{S} = \langle S_1 = X_3 \rangle $, where $ \langle \cdot \rangle $ denotes the set of operators that generate $ \mathcal{S} $.
After applying the controlled-$A$ operation, this stabilizer is updated according to the Heisenberg picture: $X_3 \mapsto X_3 A$.  
Since $S_1$ commutes with the $X$-basis measurement of the ancilla, the measurement outcome corresponds to the eigenvalue of $A$ with respect to $\ket{\psi}$.

Before continuing, let us consider a concrete example: the measurement of $A = ZZ$; see also \sfigref{fig:measurement_of_observable}{b}.  
For a general two-qubit state $\ket{\psi}= \sum_{i, j \in \{0, 1\}} c_{ij} \ket{i\, j}$, the post-measurement state is  
\begin{align}
    \ket{\psi^{\pm}} = \frac{1}{2}(I \pm ZZ) \ket{\psi} = \begin{cases}
        \frac{1}{2} \left[ c_{00} \ket{00} + c_{11} \ket{11} \right], & m = +1, \\
        \frac{1}{2} \left[ c_{01} \ket{01} + c_{10} \ket{10} \right], & m = -1, 
    \end{cases}
\end{align}  
where $m = \pm 1$ denotes the measurement outcome of the ancilla qubit. 
In this example, the measurement projects the state into one of two subspaces: the even-parity subspace ($\langle ZZ \rangle = 1$) or the odd-parity subspace ($\langle ZZ \rangle = -1$).  
This observation is noteworthy, as it is a crucial component in quantum error correction with stabilizer codes, which will be described in \cref{ssec:stabilizer_codes} --- it underpins the discretization of errors.

The example above does not capture the full picture — more generally, a measurement may not commute with all elements of the stabilizer group.  
We will see an example of this below.  
To update operators in the Heisenberg picture after measuring an observable $M \in \mathcal{P}_n$, apply the following rules~\cite{gottesman_heisenberg_1998}.  
First, identify an element $N \in \mathcal{S}$ that anti-commutes with $M$.  
Remove $N$ from the stabilizer group and replace it with $M$ in the updated stabilizer group $\mathcal{S}'$.  
Then, for each remaining $S \in \mathcal{S} \setminus \{N\}$, add $N S$ to $\mathcal{S}'$ if $S$ anti-commutes with $M$, and add $S$ unchanged otherwise.  
Apply the same procedure to the logical operators $\overline{X}_j$ and $\overline{Z}_j$ being evolved in the Heisenberg representation.  
This update ensures that the post-measurement state lies in the $+1$ (or $-1$) eigenspace of $M$, consistent with the observed measurement outcome.

Importantly, as long as a quantum circuit consists only of gates from the Clifford group, together with initialization and measurement in the Pauli basis, the circuit can be efficiently simulated on a classical computer.  
This result is known as the Gottesman-Knill theorem~\cite{gottesman_heisenberg_1998}.  
It is the foundational reason why we can simulate the behavior of large quantum error-correcting codes on classical hardware --- enabling the efficient benchmarking of codes such as those studied in \refpaper{VI}, \refpaper{VII}, \refpaper{VIII}, and \refpaper{IX}.

\section{Measurement-based Quantum Computing}
As an alternative to the gate-based or circuit-based model of quantum computation~\cite{deutsch_quantum_1997} described above, the measurement-based approach of quantum computing was proposed in the early 2000s~\cite{raussendorf_one-way_2001, raussendorf_measurement-based_2003, nielsen_quantum_2003, leung_quantum_2004}.
While different variants of this model of quantum computation exist, we will restrict the discussion here to the graph state model and note that all variants can be unified~\cite{childs_unified_2005}.
In the graph state or cluster state model, one begins with a sufficiently large resource state and performs adaptive single-qubit measurements.
The adaptivity in the measurement bases is based upon the fact that earlier measurement outcomes will influence the measurement bases at later stages in the computation.
In comparison to gate-based computation, \gls{mbqc} is often thought of as trading space for time, as it only requires shallow entangling circuits but resource states consisting of a large number of qubits.
However, this is only partially correct, as there is no requirement for the full cluster state to exist before the computation begins, and it is sufficient for the cluster state to be extended while the computation proceeds.
Furthermore, we note that one might argue that if one is concerned with fault-tolerant quantum computing, modern computation schemes employ the measurement-based approach of quantum computing in practice~\cite{litinski_game_2019}.

\subsection{Cluster States and Measurements \label{ssec:cluster_states_measurements}}
We will begin with the general definition of a graph state.
To this end, recall that a graph $G = (V, E)$ consists of a set of vertices $V$ and edges $E$ between pairs of vertices.
If one refers to a graph state, one typically considers that each vertex of $G$ represents a qubit initialized in the state $\ket{+}$ and each edge $(v_1, v_2) \in E$ corresponds to a $\CZ(v_1, v_2)$ gate applied between qubits $v_1$ and $v_2$.
The graph state is then described by the state vector $\ket{G}$ given by
\begin{align}
\label{eq:graph_state_CZ}
    \ket{G} = \prod_{(v_1, v_2) \in E} \CZ(v_1, v_2) \ket{+}^{\otimes \lvert V \rvert},
\end{align}
where $\lvert V \rvert$ is the number of vertices in the graph.
Equivalently, we can describe the graph state in terms of its stabilizers, that is, for every qubit $v \in V$, there is a stabilizer
\begin{align}
    \label{eq:graph_state_stabilizer}
    S_v = X_v \prod_{u \in N(v)} Z_u,
\end{align}
where $N(v)$ is the set of neighboring qubits of $v$.
It is easy to show that all stabilizers commute and that they are indeed stabilizers of the state $\ket{G}$.

The Heisenberg representation is convenient for describing how measurements in the Pauli basis affect the graph state.
Let us illustrate this based on a small example, the five-qubit star graph given by the graph and stabilizers
\begin{align}
    \raisebox{-0.45\totalheight}{\includegraphics{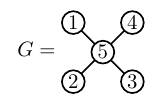}}, \hspace{2em}
    \mathcal{S} = \langle X_1 Z_5, X_2 Z_5, X_3 Z_5, X_4 Z_5, Z_1 Z_2 Z_3 Z_4 X_5 \rangle.
\end{align}
Measuring qubit 5 with outcome $m^{(5)}_P \in \{\pm1 \}$ with $P \in \{X, Y, Z \}$ produces the stabilizer groups
\begin{align}
     m^{(5)}_X: \mathcal{S} &= \langle m^{(5)}_X X_5, X_1 X_2, X_1 X_3, X_1 X_4, m^{(5)}_X Z_1 Z_2 Z_3 Z_4  X_5 \rangle , \label{eq:example_five_qubit_x} \\
     m^{(5)}_Y: \mathcal{S} &= \langle m^{(5)}_Y Y_5,  X_1 X_2, X_1 X_3, X_1 X_4, m^{(5)}_Y Z_1 Z_2 Z_3 Z_4 Y_5 \rangle,  \label{eq:example_five_qubit_y} \\
      m^{(5)}_Z: \mathcal{S} &= \langle m^{(5)}_Z Z_5,  m^{(5)}_Z X_1 Z_5, m^{(5)}_Z X_2 Z_5, m^{(5)}_Z X_3 Z_5, m^{(5)}_Z X_4 Z_5, m^{(5)}_Z Z_1 Z_2 Z_3 Z_4 Z_5 \rangle. \label{eq:example_five_qubit_z}
\end{align}
Thus, ignoring the measurement outcome, a $Z$ measurement effectively ``deletes'' the measured qubit.
This is not surprising, as after all, the measurement commutes with the entangling circuit of the graph state.
An $X$ measurement creates a stabilizer group that can be separated into generators consisting solely of Pauli $X$ and $Z$ operators, respectively.
This will be an important detail when we discuss fault-tolerant graph states in the later sections of this thesis.

\subsection{The teleportation primitive}
\begin{figure}[!tb]
    \centering
    \includegraphics{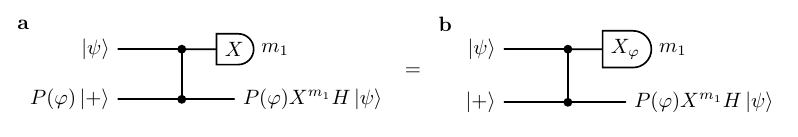}
    \caption{The one-bit teleportation primitive. The left-hand side of the equality shows how to teleport an arbitrary phase gate $P(\varphi)$ onto the input state $\ket{\psi}$ by preparing the second qubit in the state $P(\varphi)\ket{+}$.
    The right-hand side shows an equivalent circuit that only requires preparation of the state $\ket{+}$ and instead requires measurement of the first qubit in the rotated basis $P(\varphi) X P^{\dagger}(\varphi)$.
    }
    \label{fig:teleportation_cicuit}
\end{figure}

In the following, we discuss briefly an important underlying idea of \gls{mbqc}, that is, the one-bit teleportation circuit, see also \sfigref{fig:teleportation_cicuit}{a}.
This circuit consists of two input qubit states, the first one prepared in an arbitrary state $\ket{\psi}$ and the second one prepared in $\ket{+}$.
The states are entangled through the application of the $\CZ$ gate.
Finally, upon measurement of the first qubit in the Pauli $X$ basis, the state of the second 
qubit conditioned on the measurement outcome $m_1$ is $X^{m_1} H \ket{\psi}$.
Thus, up to local Clifford operations, the state of the first qubit has been teleported onto the second qubit.
Note that a measurement in the $Z$ basis would disentangle the qubits, and the second qubit would always be in the state $\ket{+}$ independent of the measurement outcome.
The circuit can be slightly generalized by considering the following modification.

Instead of preparing the second qubit in the state $\ket{+}$, prepare the state $P(\varphi) \ket{+} = (\ket{0} + e^{i \varphi} \ket{1}) / \sqrt{2}$, with $P(\varphi)$ the arbitrary phase gate.
Then, the output of the one-bit teleportation circuit is modified, and one obtains $P(\varphi)X^{m_1}H\ket{\psi}$ conditioned on the measurement outcome.
Importantly, this means that even if we are unable to perform the phase gate $P(\varphi)$, as long as we are able to prepare the state $(\ket{0} + e^{i \varphi} \ket{1}) / \sqrt{2}$, we can apply the gate to state $\ket{\psi}$ by performing the teleportation circuit and teleporting the gate $P(\varphi)$ onto it.
This is also known as gate-teleportation, see Refs~\cite{nielsen_programmable_1997, nielsen_quantum_2003} for generalizations.
Note that for $\varphi = \pi / 4$ one realizes the non-Clifford $T$ gate, see Eq.~\eqref{eq:t_gate_computational_basis}.

\begin{figure}[H]
    \centering
    \includegraphics{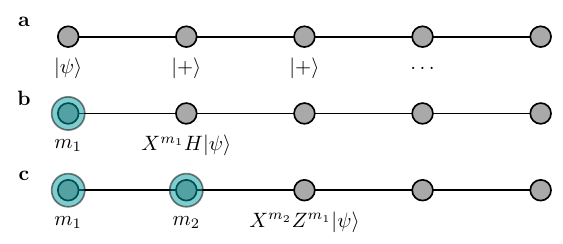}
    \caption{
    Illustration of measurement-based quantum computing. 
    A five-qubit cluster state is shown, with qubits (nodes) entangled as indicated by edges. 
    Measurement proceeds left to right, with measured qubits shown in blue.  
    \textbf{a} The left-most qubit is initialized in the state $\ket{\psi}$; others are in $\ket{+}$.  
    \textbf{b} Measuring the first qubit in the $X$ basis teleports $\ket{\psi}$ one node right, up to a Hadamard gate and a Pauli correction $X^{m_1}$.  
    \textbf{c} Measuring the next qubit in the $X$ basis further teleports the state, resulting in $X^{m_2} Z^{m_1}\ket{\psi}$ on the third qubit.
    }
    \label{fig:example_mbqc_teleportation}
\end{figure}

Lastly, let us mention that if one cannot prepare the initial state $P(\varphi)\ket{+}$, it is possible to derive an equivalent circuit that prepares the second qubit in the state $\ket{+}$ and measures the first qubit in the rotated basis $X_{\varphi} = P(\varphi) X P^{\dagger}(\varphi)$, see also \sfigref{fig:teleportation_cicuit}{b}.
Thus, one can defer the choice of which gate should be applied until a basis for the destructive measurement is chosen and the measurement is performed.
By concatenating multiple one-bit teleportation circuits, one can achieve arbitrary single-qubit unitary operations even if one is restricted to a finite set of measurement angles $\varphi$, see also \figref{fig:example_mbqc_teleportation} for a simplified example.

\subsection{Measurement-induced entanglement}
\begin{figure}[!hb]
    \centering
    \includegraphics{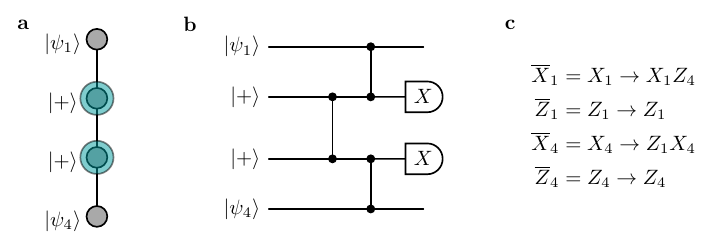}
    \caption{\textbf{a} Cluster state representation of the remote entanglement primitive. \textbf{b} The remote entanglement primitive from the gate-based view consists of a low-depth circuit and measurement of the central qubits in the $X$ basis.
    \textbf{c} Evolution of the unmeasured qubits in the Heisenberg picture.
}
    \label{fig:remote_CNOT_circuit}
\end{figure}

A key building block for understanding \gls{mbqc} is what we refer to as the \emph{measurement-induced entanglement primitive}.
This protocol enables the entanglement of two information-carrying qubits via intermediate measurements, without requiring direct interaction between them.
A prominent instance of this primitive implements the $\CZ$ gate, as illustrated in \sfigref{fig:remote_CNOT_circuit}{a}.

Using the rules of Heisenberg operator evolution, one can verify that the equivalent gate-based circuit shown in \sfigref{fig:remote_CNOT_circuit}{b} realizes the same conjugation of logical operators as a controlled-phase gate acting between $\ket{\psi_1}$ and $\ket{\psi_4}$.
This equivalence is depicted in \sfigref{fig:remote_CNOT_circuit}{c}.

Combined with the teleportation primitive, measurement-induced entanglement suffices to perform arbitrary quantum computation on a two-dimensional graph state.
We will return later to the requirements for universal and fault-tolerant quantum computation within this model.
However, it is worth emphasizing here that this entangling operation plays a foundational role in fault-tolerant quantum computing more broadly, even outside the \gls{mbqc} formalism.
In particular, this primitive underlies the realization of the logical $\CX$ through lattice surgery in the surface code~\cite{horsman_surface_2012}, and it generalizes naturally to the context of \gls{qldpc} codes through recent developments in generalized lattice surgery~\cite{ide_fault-tolerant_2024, cowtan_ssip_2024, he_extractors_2025}.

\clearpage

\section{Classical Error Correction \label{sec:classical_error_correction}}
Classical error correction addresses the problem of transmitting messages through a noisy channel.
Typically, a message consists of a sequence of bits, known as a \emph{bit string}, which can also be referred to as a \emph{word} in the context of information processing.
Each element of a bit string belongs to the finite field $\mathrm{GF}(2)$, commonly denoted as $\mathbb{F}_2$, containing the elements $0$ and $1$. 
Consequently, a bit string of length $n$ can be viewed as a vector in the vector space $\mathbb{F}_2^n$.

When Alice wants to communicate with Bob, she sends a message $\vec{u}$, a bit string, through the channel. 
Bob then receives the message, which is called the \emph{received vector} and denoted by $\vec{v}$.
In the ideal case of a noiseless channel, Bob receives exactly the message Alice sent, that is, $\vec{u} = \vec{v}$.
However, in the general case where noise is present, the message is corrupted during transmission. 
The error $\vec{e}$ that occurs during this process is defined as the difference between the sent and received messages:
\begin{align}
    \label{eq:classical_ec_error_def}
    \vec{e} = \vec{u} - \vec{v}.  %
\end{align}
It is crucial to note that the error vector $\vec{e}$ is unknown to both Alice and Bob. The goal of classical error correction is to infer the error $\vec{e}$ based only on the received vector $\vec{v}$, and to recover the original message $\vec{u}$.

\subsection{Classical Noise Channels \label{sec:classical_noise_channels}}

\begin{figure}[!b]
    \centering
    \includegraphics{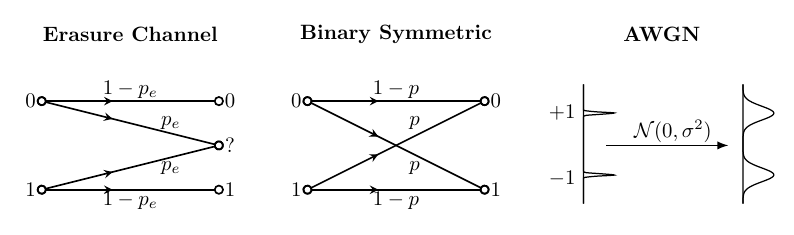}
    \caption{Pictorial representation of three different classical noise channels, from left to right, the erasure channel, the binary symmetric channel, and the additive white Gaussian noise channel.}
    \label{fig:classical_noise_channels}
\end{figure}

There is a plethora of different ways a noisy channel may act upon the bit string $\vec{u}$. 
We will focus on the three most relevant ones for this thesis, which share a common feature, namely, that they act on each bit in the bit string individually.
We also show a pictorial representation of them in \figref{fig:classical_noise_channels}.

\paragraph{The erasure channel.}
The erasure channel can be considered the simplest form of a noisy channel.
Each bit in a bit string is \emph{erased} with a certain probability $p_e$.
An erasure is typically represented by replacing the binary value with a ``?'', indicating complete uncertainty about the value of that bit.
Maybe counter-intuitively, an error $\vec{e}$ due to an erasure channel is, in many cases, easier to correct than errors stemming from other channels. %
Intuitively, the reason is that while for erased bits there is complete uncertainty about their value, we have precise information about which bits are erased and which ones are not, and thus, which ones are not to be trusted and which ones can be trusted.

\paragraph{The binary symmetric channel.}
The binary symmetric channel can be considered one of the most important models for noisy communication,
and we will mostly focus on this channel.
The action of this channel is to flip the value of each transmitted bit with probability $p < 0.5$, changing it from $0$ to $1$ or from $1$ to $0$, depending on the transmitted word $\vec{u}$.
As the name implies, the channel is symmetric with respect to $0$ and $1$ such that we can assign a probability to an error $\vec{e}$.
To this end, it is instructive to define the \emph{weight} $\weight{\vec{e}}$ of an error $\vec{e} \in \mathbb{F}_2^n$ as the number of nonzero components of $\vec{e}$, also known as the \emph{Hamming weight} of $\vec{e}$.
In the simplest case, if the channel acts identically on all bits, the probability for a given error $\vec{e}$ to occur is proportional to $p^{\weight{\vec{e}}}$, making errors of small weight more probable.
This is important as one cannot correct arbitrary errors, and our goal is to be able to correct errors that only affect a small fraction of components of $\vec{v}$.

\paragraph{The additive white Gaussian noise channel.}
The additive white Gaussian noise channel, also AWGN channel, describes a basic model that aims to model various random processes in nature and is particularly useful for satellite-based communication, for example.
It is distinct from the erasure and the binary symmetric channel, as this channel considers a continuous type of noise in comparison to the previously considered discrete channels.
The name implies specific characteristics, that is,
\begin{itemize}
    \item \textit{additive} --- the noise term acts additively to the transmitted signal. For this noise model, it is common to assume the transmitted bit $x$ to have values $\pm 1$ instead of $0/1$ such that the noisy ``bit'' $y$ is obtained as
    \begin{align}
        y = x + \delta,
    \end{align}
    where $\delta \in \mathbb{R}$ is a random variable connected to the noise,
    \item \textit{white} --- refers to the idea that the power spectral density is uniform across a frequency range, as this is a noise model relevant for wireless communication,
    \item \textit{Gaussian} --- the noise, i.e., the random variable $\delta \sim \mathcal{N}(\mu, \sigma^2)$, follows a Gaussian distribution characterized by a mean $\mu$ and variance $\sigma^2$.
\end{itemize} 
In practice, the received vector $\vec{y}$ is usually quantized, and the magnitude of the elements of $\vec{y}$ is used as \emph{soft-information} to infer the send message $\vec{x}$. 
This quantization allows one to treat this channel as a binary symmetric channel, but with a varying prior from shot to shot and for each received bit.
Thus, for the following discussion, we restrict our attention to discrete noise channels.
However, the definitions and techniques introduced here can be naturally extended to the continuous setting.

\subsection{Fundamentals of Error-Correcting Codes}
All these channels effectively specify a set of errors $E$.
We say that with respect to $E$ two words $\vec{u}_1$ and $\vec{u}_2$ are \emph{distinguishable} if and only if
\begin{align}
    \vec{u}_1 + \vec{e}_1 \neq \vec{u}_2 + \vec{e}_2, \quad \forall \vec{e}_1, \vec{e}_2 \in E.
\end{align}
Clearly, for bit strings of length $n$ not all words, that is, all elements of $\mathbb{F}_2^n$, can be distinguished if $E$ is nontrivial.
Instead, we should restrict to a subset $C$ of words from $\mathbb{F}_2^n$.
The subset $C$ then is an \emph{error-correcting code} of length $n$ and the elements of $C$ are called \emph{code words}.
If the size of $C$ is $\lvert C \rvert = 2^k$ then we say that $C$ encodes $k$ (logical) bits.
A set of errors $E$ is correctable by $C$ if all pairwise combinations of code words are distinguishable with respect to $E$.
Suppose $E(t)$ contains errors of bounded weight, that is, $\weight{\vec{e}} \leq t, \forall \vec{e} \in E$.
If $C$ corrects $E(t)$ but does not correct $E(t+1)$ then we say that $C$ is a $t$-\emph{error-correcting code}.

It is more common, however, to characterize $C$ by its \emph{distance} $d(C)$.
The distance of two code words $\vec{c}_1$ and $\vec{c}_2$ is defined as $d(\vec{c}_1, \vec{c}_2) = \weight{\vec{c}_1 - \vec{c}_2}$.
The code distance $d(C)$ is the minimum distance between any pair of code words, that is,
\begin{align}
    \label{eq:def_classical_distance}
    d(C) = \min_{\vec{c}_1, \vec{c}_2 \in C, \vec{c}_1 \neq \vec{c}_2} d(\vec{c}_1, \vec{c}_2) = \min_{\vec{c}_1, \vec{c}_2 \in C, \vec{c}_1 \neq \vec{c}_2} \weight{\vec{c}_1 - \vec{c}_2}.
\end{align}
A code $C$ is $t$-error-correcting if and only if its minimum distance $d$ satisfies $d \ge 2t + 1 $. 
This follows because if two distinct code words were indistinguishable by errors of weight at most $t$, then the error difference would have weight at most $ 2t $, contradicting the definition of the distance.
For a code $C$ of length $n$ that encodes $k$ (logical) bits with distance $d$, $C$ is typically denoted as an $[n, k, d]$ code.

\subsection{Linear Block Codes}
There are many classes of codes, but we will focus on \emph{linear (block) codes}.
These codes have certain properties that make them convenient and extendable to quantum error-correcting codes.
A linear $[n, k, d]$ code is a $k$-dimensional subspace $C$ of $\mathbb{F}_2^n$ with distance $d = \min_{\vec{c} \in C / \{ \mathbf{0} \}} \weight{\vec{c}}$.
The code $C$ can be efficiently specified by a full-rank\footnote{$H$ doesn't need to be full rank; redundant parity checks are allowed and define the same code. However, there always exists a full-rank matrix $\tilde{H}$ that specifies the same code. 
Nevertheless, including redundant checks can be relevant in certain settings, for example, in the classical theory of locally testable codes~\cite{goldreich_short_2005}, or in quantum error correction schemes where redundancy enables single-shot decoding~\cite{bombin_single-shot_2015}.} $(n-k) \times n$ \emph{parity-check matrix} $H$ over $\mathbb{F}_2$, that is,
\begin{align}
    \label{eq:def_linear_code_via_pcm}
    C = \left\lbrace \vec{c} \in \mathbb{F}_2^{n} \mid H \vec{c} = 0\right\rbrace,
\end{align}
where we assume that any vector is a column vector.
The parity-check matrix, therefore, identifies correctable errors independent of the transmitted code word, i.e., $H(\vec{c}_1 + \vec{e}) = H(\vec{c}_2 + \vec{e}) = H\vec{e}$.
The product $H\vec{e}$ is referred to as the \emph{syndrome} $\vec{s}$ and 
\begin{align}
    \label{eq:syndrome_equation}
    \vec{s} = H \vec{e},
\end{align}
is the syndrome equation.

Through the parity-check matrix, linear codes admit a natural interpretation.
To this end, consider an arbitrary element $\vec{x} \in \mathbb{F}_2^n$ that we write as $x_1 x_2\dots x_n$.
For $\vec{x}$ to be a code word as specified by Eq.~\eqref{eq:def_linear_code_via_pcm}, $\vec{x}$ needs to fulfill a set of constraints specified by the rows of the parity-check matrix $H$.
These constraints are also sometimes referred to as \emph{parity constraints} as all algebra is done over $\mathbb{F}_2$ such that a constraint is either fulfilled ($0$) or not ($1$).
As there are only $n-k$ constraints for $n$ variables, the linear system of equations $H \vec{x} = \vec{0}$ is under-constrained, leaving $k$ degrees of freedom that are connected to the $k$ encoded bits.
This redundancy is a necessary ingredient for error-correcting codes, and one can derive criteria for an $[n, k, d]$ code to exist, see, e.g., Refs.~\cite{macwilliams_theory_1978, richardson_modern_2008}.

The relation between redundancy and encoded bits becomes even more apparent if one considers an alternate definition of the code space $C$.
To this end, note that the rows of $H$ do not form a basis of the code space $C$ but instead its orthogonal complement $C^{\perp}$, that is, the subspace of vectors that are orthogonal to all vectors in $C$.
Instead, the matrix whose rows specify a basis for $C$ is called the \emph{generator matrix} $G \in \mathbb{F}_2^{k \times n}$ and it satisfies $G H^T = 0$ such that it is determined by $H$.
Given the generator matrix $G$ and the check matrix $H$, we can define the $n \times n$ invertible \emph{encoding} matrix $V$ over $\mathbb{F}_2$ as~\cite{poulin_quantum_2009-1} 
\begin{align}
    \label{eq:def_classical_encoding_matrix}
    V = \mqty(G \\ (H^{-1})^T ),
\end{align}
where, by slight misuse of notation, we denote by $H^{-1}$ the right inverse of $H$ over $\mathbb{F}_2$ such that $H H^{-1} = \mathbbm{1}_{n - k}$, with $\mathbbm{1}_{n - k}$ the identity operation on $\mathbb{F}_2^{n-k}$.
Then, the code space can be equivalently defined in terms of the encoding matrix as
\begin{align}
    \label{eq:def_linear_code_encoding_matrix}
    C = \left\lbrace \vec{c} = (\vec{b} : \vec{0}_{n-k})^T V \mid \vec{b} \in \mathbb{F}_2^k \right\rbrace,
\end{align}
where the notation $(\vec{a} : \vec{b})$ denotes the concatenation of $\vec{a}$ followed by $\vec{b}$ as column vectors.
The representation of the code space through the encoding matrix is rather uncommon, however, it is beneficial to understand the connection between error-correcting codes and quantum stabilizer codes.
For this note that for a noisy code word $\vec{v} = \vec{c} + \vec{e}$ resulting from the binary symmetric channel, we can use the inverse encoding matrix $V^{-1}$ to decompose $\vec{v}$ into the syndrome $\vec{s} \in \mathbb{F}_2^{(n-k)}$ and a logical error $\vec{\ell} \in \mathbb{F}_2^{k}$ as $\vec{e}^T V^{-1} = (\vec{\ell} : \vec{s})$.
The reason that both the logical error $\vec{\ell}$ and the syndrome $\vec{s}$ appear here is that an uncorrectable error from the set $E(t+1)$ typically will have the same syndrome as a correctable error from $E(t)$.
As the encoding matrix is invertible, opposed to the parity-check matrix $H$, the logical error $\vec{\ell}$ encodes the additional information that is not contained in $\vec{s}$ that is necessary to reconstruct $\vec{e}$.

We will delay a detailed discussion on decoding algorithms that can be used to obtain an estimate of the error $\vec{e}$ until \cref{sec:decoding}.
Here, we simply note that a key advantage of linear codes is that many of them can be constructed such that their structure allows one to efficiently infer the error from the syndrome.
An instructive example for this is the $[7, 4, 3]$ code, also known as the Hamming code, that does not require a dedicated decoding algorithm in the conventional sense.
The parity-check matrix of the Hamming code is given by
\begin{align}
    \label{eq:hamming_code_example}
    H = \mqty[
    1 & 0 & 1 & 0 & 1 & 0 & 1 \\
    0 & 1 & 1 & 0 & 0 & 1 & 1 \\
    0 & 0 & 0 & 1 & 1 & 1 & 1 \\
    ].
\end{align}
As the columns are the binary representation of the numbers from one to seven, the syndrome $\vec{s} = s_1 s_2 s_3$ directly gives the error location in binary representation (for a single error). 

\subsection{Tanner graphs \label{ssec:tanner_graphs}}
\begin{figure}[!ht]
    \centering
    \includegraphics{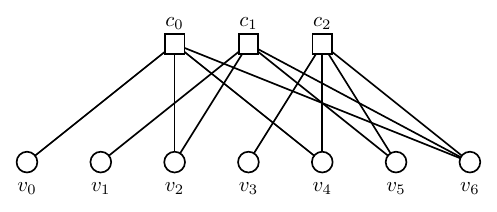}
    \caption{Tanner graph of the $[7, 4, 3]$ Hamming code defined in Eq.~\eqref{eq:hamming_code_example}. Variable nodes $v_j$ are represented as circles and check nodes $c_i$ as squares.}
    \label{fig:tanner_graph_hamming_code}
\end{figure}

The Tanner graph is a useful tool for visualizing the structure of the parity-check matrix $H$.
To this end, one can associate $H$ with the biadjacency matrix of a bipartite graph $G = (V_v \cup V_c, E)$ consisting of variable nodes $V_v$, check nodes $V_c$, and edges $E$ between $V_v$ and $V_c$.
In particular, there is an edge $(v_i, c_j) \in E$ between a variable node $v_i \in V_v$ and a check node $c_j \in V_c$ if and only if $H[j, i] = 1$, where $H[j, i]$ denotes the $j^{\mathrm{th}}$ row and $i^{\mathrm{th}}$ column of the matrix $H$.
As the variable nodes correspond to the bits of the linear codes, we call them bit nodes synonymously. 
In \figref{fig:tanner_graph_hamming_code} we show the Tanner graph of the Hamming code defined in Eq.~\eqref{eq:hamming_code_example}.
In this representation, we typically represent bit nodes as circle nodes and check nodes as squares (or boxes).
As Tanner graphs contain the structure of the parity-check matrix, they are incredibly useful for the formulation of decoding algorithms.
They can be extended to the case of quantum codes, and they play an important role in fault-tolerant measurement-based quantum computing as well.

\section{Quantum Error Correction \label{sec:quantum_error_correction}}
The role of quantum error correction is to protect quantum information from noise.  
However, this task is significantly more intricate than its classical counterpart.  
For some time, it was believed that quantum error correction might be impossible due to the continuous nature of quantum errors, reminiscent of error accumulation in classical analog computing.  
Fortunately, this pessimism was misplaced.  
As is often the case in science, the word ``impossible'' was eventually replaced by the less daunting ``difficult.''

In 1995, Peter Shor~\cite{shor_scheme_1995}, and independently, Andrew Steane~\cite{steane_error_1996, steane_multiple-particle_1996} demonstrated that it is possible to measure a carefully chosen set of observables that detect the presence of errors without disturbing the encoded quantum information.  
These measurements not only reveal whether an error has occurred, but also discretize the noise: they project the state into one of several orthogonal subspaces, effectively collapsing a continuous error into a discrete one.  
Importantly, these observables do not distinguish between different logical states, but only between error-free and erroneous subspaces.
One can then use measurement outcomes to infer which kind of error has occurred.
It turns out that many tools of classical coding theory can be applied from here.

\subsection{Noise channels \label{ssec:qec_noise}} As in the classical case, designing quantum error-correcting codes requires choosing a noise model that reflects the types of errors the code is meant to protect against.  
A physically motivated example is the single-qubit depolarizing channel, defined as  
\begin{align}
    \label{eq:def_single_qubit_noise}
    \mathcal{E}_1(\rho) = (1 - p)\rho + \frac{p}{3}(X \rho X + Y \rho Y + Z \rho Z),
\end{align}  
where $\rho$ is the density operator of the quantum system.  
This channel leaves the state unchanged with probability $1 - p$, and applies a random single-qubit Pauli error $X$, $Y$, or $Z$ with probability $p/3$.  
It captures decoherence due to independent noise on each qubit.  

However, this model is typically insufficient for describing realistic noise in quantum circuits.  
Two-qubit gates, such as the controlled-NOT ($\CX$) gate, often introduce correlated noise.  
Such effects can be modeled using the two-qubit depolarizing channel,  
\begin{align}
    \label{eq:def_two_qubit_noise}
    \mathcal{E}_2(\rho) = (1 - p)\rho + \frac{p}{15} \sum_{E \in \widetilde{\mathcal{P}}_2 / \{II\}} E \rho E^{\dagger},
\end{align}  
where $\widetilde{\mathcal{P}}_2$ denotes the effective two-qubit Pauli group $\mathcal{P}_2 / \{\pm 1, \pm i \}$.  

In circuit-level noise models, noisy single-qubit gates are typically modeled as the ideal gate followed by the single-qubit depolarizing channel~\eqref{eq:def_single_qubit_noise}, 
and noisy two-qubit gates as the ideal gate followed by the two-qubit depolarizing channel~\eqref{eq:def_two_qubit_noise}.

Although such discrete noise models may appear overly simplistic --- given that quantum systems evolve continuously --- the act of measurement discretizes this evolution in a physically meaningful way.
To understand this, consider again the Bell state~\eqref{eq:def_bell_plus} $\ket{\Phi^{+}} = (\ket{00} + \ket{11})/\sqrt{2}$.
Suppose the first qubit undergoes a small rotation generated by the Pauli $X$ operator, leading to the evolved state %
\begin{align}
\exp(i \delta X_1 t / 2) \ket{\Phi^{+}} = \cos(\delta t) I_1 \ket{\Phi^{+}} + i \sin(\delta t) X_1 \ket{\Phi^{+}}.    
\end{align} 
Now, imagine measuring the $ZZ$ parity as described in \cref{ssec:stabilizer_circuits_measurements}.
If $\delta t \ll 1$, with probability $\cos^2(\delta t)$ close to one, we obtain the outcome $+1$, and the state collapses back to the original Bell state.
With small probability $\sin^2(\delta t)$, we instead obtain the outcome $-1$, projecting the state onto $X_1 \ket{\Phi^{+}}$, which is equivalent to a discrete Pauli error on the first qubit.

This example illustrates a key point: even continuous errors become discrete upon measurement of appropriate observables that anti-commute with the (generator of the) error.
More generally, because any quantum operation can be decomposed into a superposition of Pauli operators, correcting all Pauli errors of weight less than $t$ ensures the correction of arbitrary errors affecting at most $t$ qubits.

\subsection{Stabilizer codes \label{ssec:stabilizer_codes}}
Quantum stabilizer codes~\cite{gottesman_stabilizer_1997} can be viewed as the quantum analog of classical linear block codes.  
Abstractly, a stabilizer code is defined by an Abelian subgroup $\mathcal{S}$ of the $n$-qubit Pauli group $\mathcal{P}_n$, with the condition that $-I \notin \mathcal{S}$.  
The group $\mathcal{S}$ is called the \emph{stabilizer group}, and its elements are referred to as \emph{stabilizers}.  
The code space is defined as the joint $+1$-eigenspace of all elements in $\mathcal{S}$,  
\begin{align}
    \label{eq:def_stabilizer_code}
    C = \left\{ \ket{\psi} \mid S \ket{\psi} = \ket{\psi}, \; \forall S \in \mathcal{S} \right\}.
\end{align}
The stabilizer group is typically specified by a set of independent generators $S_1, S_2, \dots, S_r \in \mathcal{P}_n$, such that $\mathcal{S} = \langle S_1, S_2, \dots, S_r \rangle$.  
Each generator defines a constraint that any valid code state must satisfy.  
If the stabilizer group has $r$ independent generators, the code encodes $k = n-r$ logical qubits into $n$ physical qubits.  
This mirrors the role of parity-check constraints in classical linear codes, a connection that will be made explicit later.

Having defined the stabilizer group and its generators, we now turn to the logical operators of the code.  
Logical operators correspond to symmetries of the code space that preserve the code space but act non-trivially on the logical qubits.  
The logical operators $\mathcal{L}$ are the set of operators that commute with $\mathcal{S}$ but are themselves not elements of $\mathcal{S}$, or more formally, the logical operators are given by %
\begin{align}
    \label{eq:logical_operators_stabilizer_code_normalizer}
    \mathcal{L} \cong \mathcal{N}(\mathcal{S}) / \mathcal{S} = \{ L  \in \mathcal{P}_{n} \mid S L = L S \; \forall S \in \mathcal{S} \} / \mathcal{S},
\end{align}
where we write $\mathcal{N}(\mathcal{S})$ for the normalizer of $\mathcal{S}$ in ${\mathcal{P}_n}$.

The distance $d$ of the code is defined as the minimum weight of a non-trivial logical operator, i.e., the smallest number of qubits on which such an operator acts non-trivially. 
We use the notation $[\![ n, k, d]\!]$ to refer to a stabilizer code with $n$ qubits, $k$ encoded logical qubits, and distance $d$.
For a code encoding $k$ logical qubits, there will be $2k$ logical operators, an encoded Pauli $\overline{Z}_j$ and $\overline{X}_j$ for each logical qubit.
As the minimum weight of $\overline{Z}_j$ and $\overline{X}_j$ are not necessarily equal, one also encounters the notation $[\![n, k, d_X, d_Z]\!]$ for stabilizer codes.

In the absence of errors, measuring the generators of the stabilizer group $\mathcal{S}$ will deterministically have outcome $+1$.
This is because all code states $\ket{\psi} \in C$ are stabilized by every $S_i \in \mathcal{S}$, i.e., $S_i \ket{\psi} = \ket{\psi}$.
However, in the presence of errors, any generator $S_i$ that anti-commutes with an error $E \in \mathcal{P}_n$ results in a measurement outcome $-1$, as $S_i E \ket{\psi} = -E g_i\ket{\psi_i} = - E\ket{\psi}$.
By measuring all generators $S_i$ of the stabilizer group, we obtain a list of measurement outcomes $\vec{\sigma}(E)$ known as the syndrome.
We typically view $\vec{\sigma}(E)$ as a bit string $\vec{s} \in \mathbb{F}_2^{n-k}$ that encodes the commutation relations of the errors with the generators, that is,
$E S_i + (-1)^{s_i} S_i E = 0$ and make the dependence of $\vec{s}$ on the error $E$ implicit.

We note that the elements of the stabilizer group, together with the logical operators, do not generate the full Pauli group $\mathcal{P}_n$.  
As a result, we cannot represent any error $E$ as a product of a logical operator $L \in \mathcal{L}$ and a stabilizer $S \in \mathcal{S}$, as we could in the classical case (see the discussion around Eq.~\eqref{eq:def_linear_code_encoding_matrix}).  
This discrepancy arises because $\mathcal{P}_n$ contains $4^n$ elements (up to phases), whereas the stabilizer group $\mathcal{S}$ contains $2^{n - k}$ elements and the logical operators $\mathcal{L}$ contain $2^{2k}$ independent elements.  
Together, $\mathcal{S}$ and $\mathcal{L}$ contribute $2^{n + k}$ elements, leaving the remaining $4^n / 2^{n + k} = 2^{n - k}$ elements to be called \emph{pure errors} or \emph{destabilizers}, denoted $\mathcal{T}$.  
These pure errors are formally defined as
\begin{align}
    \mathcal{T} \cong \mathcal{N}(\mathcal{L}) / \mathcal{S}.
\end{align}
Together with the stabilizers $\mathcal{S}$ and logical operators $\mathcal{L}$, the destabilizers $\mathcal{T}$ complete the set of operators that span the full $n$-qubit Pauli group $\mathcal{P}_n$.  
Any operator $E \in \mathcal{P}_n$ can therefore be represented as
\begin{align}
    \label{eq:def_pauli_TLS_decomposition}
    E = T L S, 
\end{align}
where $T \in \mathcal{T}$, $L \in \mathcal{L}$, and $S \in \mathcal{S}$.

While the notion of pure errors is useful for the optimal decoding of stabilizer codes, as we will discuss later, their introduction also provides insight into an alternative characterization of a stabilizer code. 
Alternatively to Eq.~\eqref{eq:def_stabilizer_code} a stabilizer code can be specified through a Clifford transformation $V$ on $n$ qubits such that
\begin{align}
    \label{eq:def_stabilizer_code_encoding_unitary}
    C = \{ \ket{\psi} = V (\ket{\varphi} \otimes \ket{0_{n-k}}) \mid \ket{\varphi} \in \mathbb{C}^{2^k} \},
\end{align}
where $V$ is known as the encoding unitary, analogous to the encoding matrix in the classical linear code definition in Eq.~\eqref{eq:def_linear_code_encoding_matrix}.  
Since the encoding matrix is an element of the Clifford group, it can be implemented using Clifford operations, such as $\CX$, $S$, and $\Ha$, only.
This encoding unitary implicitly determines the logical operators, stabilizer generators, and the generators of the pure errors.  
These generators are obtained by conjugating the single-qubit Pauli operators, as follows:
\begin{align}
    \label{eq:encoding_unitary_xz_output}
    V: \begin{cases}
    Z_j \to \overline{Z}_j = VZ_jV^{\dagger }, &\textrm{for } j = 1\dots k,\\
    X_j \to \overline{X}_j = V X_j V^{\dagger}, &\textrm{for } j = 1\dots k, \\
    Z_{j + k} \to g_j = VZ_{j+k}V^{\dagger }, &\textrm{for }j = 1 \dots n - k, \\
    X_{j+k} \to T_j = V X_{j+k} V^{\dagger}, &\textrm{for } j = 1 \dots n - k.
    \end{cases}
\end{align}

\paragraph{Binary symplectic representation.} To make the connection between classical linear codes and stabilizer codes more explicit, we now introduce the \emph{binary symplectic representation} due to Ref.~\cite{calderbank_quantum_1997}.
This representation defines a map $\operatorname{Mat}_{\mathcal{B}}: \mathcal{P}_n \to \mathbb{F}_2^{2n}$ such that a $n$-qubit Pauli operator can be written as a vector with elements in $\mathbb{F}_2$, that is, 
\begin{align}
    \label{eq:def_binary_symplectic_rep}
    \alpha \bigotimes_{i = 1}^{n}X^{x_i}Z^{z_i} \to (x_1, x_2, \dots, x_n \mid z_1, z_2, \dots, z_n).
\end{align}
We will typically ignore the phases \( \alpha \in \{\pm 1, \pm i\} \) of elements in the Pauli group. 
The reason is that in the context of quantum error correction, we typically identify Pauli operators that differ only by a phase $\alpha$, as such phases do not affect the syndrome.
Additionally, whenever the stabilizer group $\mathcal{S}$ that specifies the stabilizer code C~\eqref{eq:def_stabilizer_code} contains an element $S_i$ with phase $-1$, we can replace $S_i$ with $-S_i$ without changing properties of the code.
Hence, the phase $\alpha$ is irrelevant for error correction, but it is relevant for the simulation of Clifford circuits with measurements~\cite{gottesman_heisenberg_1998}.

Through the binary symplectic representation, we can specify the stabilizer group $\mathcal{S} = \langle S_1,\dots, S_r\rangle$ that defines a stabilizer code $C$ through Eq.~\eqref{eq:def_stabilizer_code} by an $r \times 2n$ binary parity-check matrix $H$ where the $i^{\mathrm{th}}$ row is $\bsr{S_i}$.
Multiplying two Pauli strings $P_1, P_2 \in \mathcal{P}_n$ corresponds to a simple addition within the binary symplectic representation, i.e., $\bsr{P_1} + \bsr{P_2} = \bsr{P_1 P_2}$.
$P_1$ and $P_2$ commute if and only if the symplectic inner product vanishes, i.e., $\bsr{P_1} \Lambda \bsr{P_2}^{T} = 0$, where $\Lambda$ is the symplectic matrix and given by
\begin{align}
    \Lambda = \mqty[0 & \mathbbm{1}_n \\ \mathbbm{1}_n & 0],
\end{align}
with $\mathbbm{1}_n$ the $n \times n$ identity matrix.
Hence, for $\mathcal{S}$ to be Abelian we require that $H \Lambda H^{T} = H_X H_Z^T + H_Z H_X^T = 0$ where we used that we can write $H = [H_X \mid H_Z]$.
Similar to the classical case, the rank of the parity-check matrix is related to the number of encoded logical qubits, that is, it determines the number of independent generators of $\mathcal{S}$, such that $k = n - \RANK(H)$.
It is instructive to consider the five-qubit code as an example.
The stabilizer group of the five-qubit code can be written as $\mathcal{S} = \langle XZZXI, IXZZX, XIXZZ, ZXIXZ\rangle$, which has the following binary symplectic representation
\begin{align}
    H = \left[ \begin{array}{ccccc|ccccc}
    1 & 0 & 0 & 1 & 0  & 0 & 1 & 1 & 0 & 0 \\
    0 & 1 & 0 & 0 & 1  & 0 & 0 & 1 & 1 & 0 \\
    1 & 0 & 1 & 0 & 0  & 0 & 0 & 0 & 1 & 1 \\
    0 & 1 & 0 & 1 & 0  & 1 & 0 & 0 & 0 & 1
        \end{array} \right].
\end{align}
It can be verified that the five-qubit code has parameters $[\![5, 1, 3 ]\!]$.

\paragraph{Calderbank-Shor-Steane codes.} A particularly important class of stabilizer codes are \gls{css} codes~\cite{calderbank_good_1996, steane_error_1996, steane_multiple-particle_1996} whose defining property is that their stabilizer group admits a set of generators $S_1, S_2, \dots, S_r$ such that they are either $X$-type or $Z$-type, that is, their check matrix $H$ admits a block diagonal structure, 
\begin{align}
    \label{eq:def_css_code_check_matrix}
    H = \mqty[H_X & 0 \\ 0 & H_Z],
\end{align}
where $H_X \in \mathbb{F}_2^{m_x \times n}$ and $H_Z \in \mathbb{F}_2^{m_z \times n}$.
This particular structure simplifies the commutativity condition to 
\begin{align}
    \label{eq:css_code_commutation_condition}
    H_X H_Z^T = 0
\end{align}
and the number of encoded logical qubits becomes $k = n - \RANK(H_X) - \RANK(H_Z)$.
The additional structure inherent to \gls{css} codes significantly simplifies their construction, making them a powerful tool in quantum error correction.
As a result, many breakthrough results in the field have been achieved through \gls{css} codes.
These results have relied upon what is today often referred to as the ``\gls{css}-to-homology" correspondence, a mathematical framework that employs \emph{chain complexes}, objects that are studied in homological algebra, to represent not only \gls{css} codes but also classical linear codes.
This formulation has become the modern perspective on (quantum) error correction and is expected to influence the field further.
We will give a brief introduction to this formalism after reviewing how fault-tolerance properties of \gls{css} stabilizer codes can be lifted to the measurement-based setting.

\subsection{Fault-tolerant graph states \label{ssec:ftgs}}
We have seen previously that universal quantum computation can be achieved in the measurement-based model by using a two-dimensional cluster state as a resource.
This model mimics the circuit model in that logical gates are implemented through sequences of single-qubit measurements.
However, to protect such computations against noise, additional redundancy is required.
Unlike in the circuit model, where stabilizer measurements can be used to identify and correct errors during computation, the measurement-based approach requires that fault tolerance can be built directly into the resource state.
This raises the question: how can one design such a resource state so that it supports fault-tolerant quantum computation?
One answer is given by the method of \emph{foliation}, which produces a fault-tolerant cluster state from any \gls{css} code.
This method, due to Bolt \emph{et al.}~\cite{bolt_foliated_2016}, generalizes the construction of Raussendorf \emph{et al.}~\cite{raussendorf_topological_2007} for the surface code.

Recall that a \gls{css} code is characterized by two parity-check matrices $H_Z$ and $H_X$.
As described in \cref{ssec:tanner_graphs}, with each of these matrices one can associate a bipartite graph $G = (V_b \cup V_c, E)$.
We can also associate with such a graph $G$ a graph state or cluster state by identifying the vertex sets $V_b$ and $V_c$ with code qubits and check qubits, respectively.
The edges $E$ in the graph state indicate between which qubits a $\CZ$ gate is applied, see \cref{ssec:cluster_states_measurements}.

Consider the graph state $G_Z$ associated with the $Z$ parity-check matrix $H_Z$ as described above.
It is straightforward to see that the stabilizers centered on the ancilla qubits of the graph state can be written as 
\begin{align}
    S_{c_j} = X_{c_j} \bigotimes_{i = 1}^{n} Z_{b_i}^{H[j, i]},
\end{align}
for any $c_j \in V_c$ and the notation $H[j, i]$ denotes the $i^{\mathrm{th}}$ entry in the $j^{\textrm{th}}$ row associated to the check $c_j$.
Thus, upon measurement of the ancilla qubits in the Pauli $X$ basis, the resulting graph state is an eigenstate of the stabilizer group described by $H_Z$.
Importantly, the graph state $G_Z$ is also an eigenstate of the stabilizer group generated by $H_X$.
The reason is that $H_X$ and $H_Z$ must commute and thus $X$- and $Z$-stabilizers must overlap on an even number of code qubits.
As a result, when constructing $X$-type stabilizers associated with elements $\mathcal{S}_X$ from graph state stabilizers of code qubits $S_b = X_b \otimes_{a \in N(b)} Z_a$, any ancilla qubit $a$ appears an even number of times.
Since logical $X$ operators can be seen as a special form of $X$-type stabilizer, the same argument applies to them.
Hence, measurement of the ancilla qubits in the graph state projects the remaining qubits into the logical dual basis code state of the \gls{css} code described by $H_X$ and $H_Z$.

\begin{figure}[!ht]
    \centering
    \includegraphics{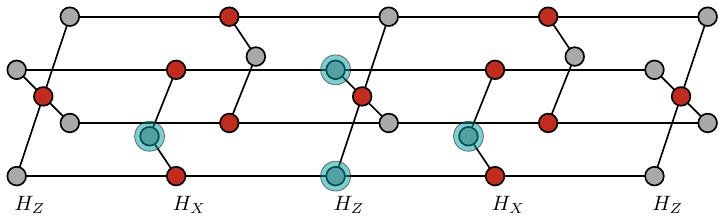}
    \caption{Foliation of the $[\![4, 1, 2]\!]$ \gls{css} code described by check matrices $H_Z = [1\ 1\ 1\ 1]$ and $H_X = \vcenter{\hbox{\scriptsize$\begin{bmatrix} 1 & 1 & 0 & 0 \\ 0 & 0 & 1 & 1 \end{bmatrix}$}}$. The fault-tolerant graph state shown contains $L = 2$ foliation layers.
    Gray and red nodes indicate primal and dual qubits, respectively.
    Highlighted in blue is a set of $X$ measurement outcomes that yield a detector of the foliated code, that is, a product of measurement outcomes that is deterministic in the absence of errors. 
}
    \label{fig:foliated_412_code}
\end{figure}

By layering the graph states $G_Z$ and $G_X$ in an alternating pattern and entangling code qubits between adjacent layers by $\CZ$ gates, one obtains the \emph{foliated code}, see \figref{fig:foliated_412_code} for an example.
The foliated code is a fault-tolerant graph state that inherits properties of the underlying \gls{css} code and can protect against $Z$-type errors on code and ancilla qubits.
Within this fault-tolerant graph state, one distinguishes between two types of qubits: \emph{primal} and \emph{dual} qubits.
Code qubits in $G_Z$ and ancilla qubits in $G_X$ are considered \emph{primal}, while code qubits in $G_X$ and ancilla qubits in $G_Z$ are considered \emph{dual}.
This assignment aligns with the alternating structure of the foliated code and reflects the duality between $X$- and $Z$-type checks.
Parity-check operators, also called \emph{detectors}, can be identified within the graph state and inferred from sets of single-qubit $X$ measurement outcomes.
The parity of individual detectors yields a syndrome $\vec{s}$ which can be used together with a decoder to infer the presence and location of $Z$ errors.
We will not reproduce here the expression for the detectors of the foliated code and instead refer to Ref.~\cite{bolt_foliated_2016} for the original derivation or to \refpaper{IX} for a derivation based on chain complexes and the hypergraph product, which we introduce in the next section.

Lastly, we mention that foliated codes do not belong to the class of stabilizer codes described in \cref{ssec:stabilizer_codes}, but instead more closely resemble subsystem \gls{css} stabilizer codes~\cite{poulin_stabilizer_2005}.
Furthermore, while we described the foliation of \gls{css} codes, the approach can be extended to ordinary stabilizer codes, see Ref.~\cite{brown_universal_2020}.

\section{Chain Complexes \label{sec:chain_complexes}}
For a more thorough introduction to chain complexes, we refer the interested reader to Refs.~\cite{hatcher_algebraic_2002, breuckmann_quantum_2021}.
For the purpose of this thesis, a \emph{chain complex} $\mathcal{C}$ of length $n$ is defined as a sequence of $n+1$ $\mathbb{F}_2$-vector spaces $C_i$ and $n$ linear maps $\partial_i^C$ called boundary operators, 
\begin{align}
    \mathcal{C} = \{ 0 \} \overset{\partial_{n+1}^{C}}{\longrightarrow} C_n \stackrel{\partial_n^{C}}{\longrightarrow} \dots \stackrel{\partial_1^{C}}{\longrightarrow} C_0 \overset{\partial_0^{C}}{\longrightarrow} \{ 0 \}, 
\end{align}
with the composition of boundary operators fulfilling 
\begin{align}
    \label{eq:chain_complex_boundary_composition}
    \partial_i^{C} \partial_{i+1}^{C} = 0.
\end{align}
Here, the spaces $C_i$ always have finite dimension $n_i$, and a basis of $C_i$ can be identified with $\mathbb{F}_2^{n_i}$.
We suppress the superscripts if no distinction is necessary, as well as make implicit the trivial boundary operators $\partial_{n+1}: \{0\} \to C_n$ and $\partial_0: C_0 \to \{0\}$ that are treated formally as zero $n_n \times 0$ and $0 \times n_0$ matrices, respectively.

It is common to refer to elements of the kernel $Z_i(\mathcal{C}) \coloneqq \ker \partial_i$ as $i$-cycles, and elements of the image $B_i(\mathcal{C}) \coloneqq \image  \partial_{i+1}$ as $i$-boundaries, and one has the embedding relation
\begin{align}
    B_i \subseteq Z_i \subseteq C_i.
\end{align}
The quotient vector space
\begin{align}
    H_i(\mathcal{C}) = Z_i / B_i = \ker \partial_i / \image \partial_{i+1},
\end{align}
is called the $i^\text{th}$ homology group of $\mathcal{C}$ and is a central object of interest for the study of stabilizer codes through the presented formalism.

Associated with $\mathcal{C}$ is also a cochain complex, with coboundary operators $\delta^{i}: C_i \rightarrow C_{i+1}$ defined as $\delta^{i} = \partial_{i+1}^T$. 
This cochain complex is given by 
\begin{align}
     \mathcal{C}^T = \{ 0 \} \stackrel{\delta^{-1}}{\rightarrow} C_0 \stackrel{\delta^{0}}{\rightarrow} \dots \stackrel{\delta^{n-1}}{\rightarrow} C_n \stackrel{\delta^{n}}{\rightarrow} \{ 0 \},
\end{align}
with cohomology groups defined as 
\begin{align}
    H^{i} \coloneqq \ker \delta^{i} / \image \delta^{i-1}.
\end{align}
Elements of $Z^{i} \coloneqq \ker \delta^{i}$ and $B^{i} \coloneqq \image  \delta^{i-1}$ are referred to as cocycles and coboundaries, respectively. 
Thus, the $i^\text{th}$ cohomology group of $\mathcal{C}$ is defined as 
\begin{align}
    H^{i}(\mathcal{C}) = Z^i / B^i =  \ker \delta^{i} / \image \delta^{i-1} = \ker \partial_{i+1}^{T} / \image \partial_{i}^T,
\end{align}
which is similarly important to the $i^\text{th}$ homology group of $\mathcal{C}$ in capturing the essential features of the underlying structure.

\subsection{The CSS-to-homomology correspondence}
Since the boundary maps of a chain complex $\mathcal{C}$ can be identified with matrices over $\mathbb{F}_2$, any length-1 chain complex can be viewed as a classical linear $[n, k, d]$ code by identifying $\partial_1 = H$, where $H$ maps an error vector from the space $C_1 \cong \mathbb{F}_2^n$ to the syndrome space $C_0 \cong \mathbb{F}_2^r$ with $r \geq n-k$. 
The codespace, denoted by $C$, coincides with the first homology group of $\mathcal{C}$. 
In other words, for a length-1 chain complex with boundary map $\partial_1 = H$, the first homology group is given by 
\begin{align}
    H_1(\mathcal{C}) = \ker H / \image \partial_2 = \ker H / \{ \vec{0} \},
\end{align}
and the smallest non-trivial weight element of $H_1(\mathcal{C})$ determines the distance $d$.

Notice that the defining property of a chain complex, namely the composition property of boundary operators in Eq.~\eqref{eq:chain_complex_boundary_composition}, is equivalent to the commutation condition~\eqref{eq:css_code_commutation_condition} for the check matrices of a CSS code.
  
Thus, any CSS code, $\mathcal{C}$, can be represented by a length-2 chain (sub)complex 
\begin{equation}
    \label{eq:CSS_chain_complex}
    \ldots \rightarrow C_{i+1} \stackrel{\partial_{i+1}}{\longrightarrow} C_i \stackrel{\partial_i}{\longrightarrow} C_{i-1} \rightarrow \ldots,
\end{equation}
where, by convention, $\partial_{i+1} = H_Z^T$ and $\partial_i = H_X$.
Identifying qubits with the space $C_i$, the number of qubits is $n = \dim C_i$.
  
The sets of $X$ and $Z$ logical operators are elements of the groups $H_i(\mathcal{C})$ and $H^i(\mathcal{C})$, respectively, with the smallest non-trivial weight element defining the distances $d_X$ and $d_Z$, respectively.
Intuitively, for a CSS code, a logical $Z$ operator from the group 
\begin{equation}
    H_i(\mathcal{C}) = \ker H_X / \image H_Z^T,
\end{equation}
is an element $\vec{\ell}$ of $C_i$ that does not produce an $X$-syndrome (i.e., $\vec{\ell} \in \ker H_X$) and is not generated by the $Z$-stabilizers (i.e., $\vec{\ell} \notin \image H_Z^T$).

\subsection{Quantum codes from classical codes}
Why should we explore this abstract formalism for representing a CSS code? To address this question, we first need to highlight two key points.
First, while we can represent any CSS code as a length-2 chain complex, the reverse is also true: any length-2 chain complex corresponds to a valid CSS code. 
This equivalence between chain complexes and CSS codes forms the foundation for understanding how classical codes can be used to design quantum codes.
Second, constructing \emph{good} \gls{qldpc} codes --- stabilizer codes with sparse check matrices and asymptotically linear parameters --- has proven to be significantly more challenging than in the classical case. 
For classical linear codes, Gallager demonstrated in 1960~\cite{gallager_low_1960, gallager_low-density_1962} that taking a random sparse parity-check matrix is sufficient to obtain a \emph{good} \gls{ldpc} code with parameters $[n, \Theta(n), \Theta(n)]$. 
However, the commutation condition for quantum codes prevents a similar straightforward construction. 

It was only in 2021, following a series of breakthrough results~\cite{breuckmann_balanced_2021, panteleev_asymptotically_2022, dinur_good_2022, leverrier_quantum_2022}, that the existence of \emph{good} \gls{qldpc} codes was established.
These advancements largely rely on the formalism described here and the development of product constructions for pairs of chain complexes, while also applying techniques due to Sipser and Spielman~\cite{sipser_expander_1996} for the explicit construction of good \gls{ldpc} codes in the classical setting.
Interested readers can refer to Refs.~\cite{breuckmann_quantum_2021, audoux_tensor_2019}, and we now consider a simple example --- the hypergraph product (or homological product)~\cite{tillich_quantum_2009, tillich_quantum_2014} which illustrates how to construct quantum codes from classical ones. 
For a generalization to higher-dimensional cases, see Ref.~\cite{zeng_higher-dimensional_2019}.

\paragraph{Hypergraph product codes.} Consider two length-1 chain complexes, $\mathcal{A}: A_1 \overset{H_A}{\longrightarrow} A_0$ and $\mathcal{B}: B_1 \overset{H_B}{\longrightarrow} B_0$, corresponding to two classical linear codes with parameters $[n_i, k_i, d_i]$ and $m_i \times n_i$ check matrices $H_i$ for $i \in \{A,  B\}$. The hypergraph product code $\mathcal{C} = \mathcal{A} \times \mathcal{B}$ is a length-2 chain complex with boundary operators given by:
\begin{align}
    \label{eq:hgp_checks}
    \partial_1 = H_X = \mqty(H_A \otimes \mathbbm{1}_{n_B} \mid  \mathbbm{1}_{m_A} \otimes H_B), \quad
    \partial_2^T = H_Z = \mqty(\mathbbm{1}_{n_A} \otimes H_B \mid H_A \otimes \mathbbm{1}_{m_B}),
\end{align}
and spaces given by
\begin{align}
    \label{eq:hgp_spaces}
    C_2 = A_1 \otimes B_1, \quad C_1 = A_1 \otimes B_0 \oplus A_0 \otimes B_1, \quad C_0 = A_0 \otimes B_0.
\end{align}
The resulting chain complex corresponds to a CSS quantum code with parameters $[\![n_A m_B + m_A n_B, k_A k_B^T + k_A^T k_B, \min(d_A, d_B, d_A^T, d_B^T)]\!]$, where $k_i^T$ and $d_i^T$ are the parameters of the transpose codes. We use the convention that $d=\infty$ if $k = 0$. 

Thus, through this product construction, we can generate quantum codes by combining classical codes in a structured way that preserves their error-correcting properties while satisfying the constraints of quantum error correction.

\begin{figure}
    \centering
    \includegraphics{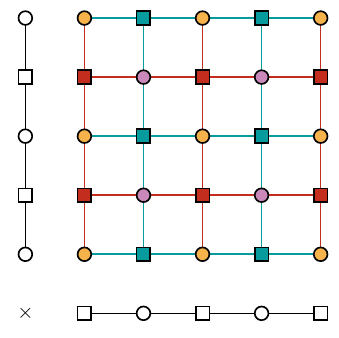}
    \caption{Hypergraph product of the $n=3$ repetition code with its transpose, yielding the surface code with rough edges on the sides and smooth edges on the top and bottom. The code has parameters $[\![13, 1, 3]\!]$. 
    Red and blue grid lines correspond to the edges in $Z$ and $X$ Tanner graphs of the check matrices of $H_Z$ and $H_X$, respectively.}
    \label{fig:hgp_rep_code_example}
\end{figure}

To exemplify this procedure and illustrate the origin of the name, let us consider a simple example, the product of two classical repetition codes.
The length-$n$ repetition code is a classical code with parameters $[n, 1, n]$, and its codewords are the all-zero and all-one vectors of length $n$, that is, $\vec{0}_n$ and $\vec{1}_n$, respectively.
The $(n - 1) \times n$ parity-check matrix of the code is given by
\begin{align}
    H_R = \left[ \begin{array}{ccccc}
         1 & 1 & &   \\
         & 1 & 1 & &   \\
         & & 1 & 1& \\
         & & & &\ddots 
    \end{array} \right],
\end{align}
where we left the zero entries blank.
Then, we define the repetition code length-1 chain complex
\begin{align}
    \mathcal{R}: R_1 \overset{H_R}{\longrightarrow} R_0,
\end{align}
where $R_1 \cong \mathbb{F}_2^{n}$ represent the $n$ bits and $R_0 \cong \mathbb{F}_2^{n-1}$ represents the $n-1$ checks.
Recall that with $H_R$ we can associate a bipartite Tanner graph.
In this case, this is the line graph, starting and ending in a variable node.
From a graphical perspective, the hypergraph product can be viewed as the Cartesian product of the Tanner graphs of $H_A$ and $H_B$. %
There will be four types of nodes: the product of a bit node with another bit node, the product of a bit and a check, the product of a check and a bit, and the product of two checks.
Visually, we represent the nodes of the different products as
\begin{center}
    \includegraphics{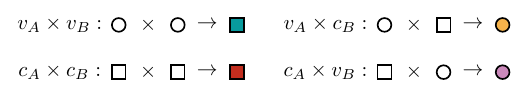}
\end{center}
where we have chosen the coloring and shapes according to the classification $V_A \times V_B \cong C_2$, $V_A \times C_B \oplus C_A \times V_B \cong C_1$, and $C_A \times C_B \cong C_0$.
Recall that by Eq.~\eqref{eq:CSS_chain_complex}, $C_2$ is the space of $Z$-syndromes, $C_1$ is the space of qubits, and $C_0$ is the space of $X$-syndromes.
There is an edge between two nodes $(a_1, a_2)$ and $(b_1, b_2)$ in $V_A \cup C_A \times V_B \cup C_B$ if either $(a_1, b_1) \in E_A$ and $a_2 = b_2$ or if $a_1 = b_1$ and $(a_2, b_2) \in E_B$.
The product of two $n=3$ repetition codes, with one of them transposed, is shown in \figref{fig:hgp_rep_code_example}, showing the Tanner graph of the surface code~\cite{fowler_surface_2012}.

\FloatBarrier

\section{Decoding of Quantum Error-Correcting Codes \label{sec:decoding}}
The redundancy in (quantum) error-correcting codes yields protection against noise. 
However, this protection is usually not autonomous in the sense that some additional work is required to restore the corrupted code words to the code space.
This task is carried out by a decoding algorithm, commonly referred to as the decoder.
We will give a formal definition of the decoder later on, and for now, describe a decoder, informally, as an algorithm that takes into account the measured syndrome information, as well as information about the noise model, to determine a correction that returns the noisy code word to the code space.
The problem that the decoder solves is known as the decoding problem.
Unfortunately, solving the decoding problem exactly turns out to be an exceptionally hard problem in general --- solving the decoding problem for stabilizer codes, which we will define more concretely later, is $\#\PTIME$-complete~\cite{iyer_hardness_2015}, while the decoding of classical linear codes is easier but still $\NP$-complete~\cite{berlekamp_inherent_1978}.
Hence, solving the decoding problem is considered (under standard hardness assumptions) intractable in general, and one can typically only find an approximate solution.  
As a result, different decoding algorithms are typically characterized by their accuracy and speed trade-offs. 

Access to highly accurate decoders is beneficial as it can lead to reduced hardware requirements, for example, lower overhead of additional qubits used for redundancy or less stringent requirements on the physical error rates present in the architecture.  
However, fault-tolerant quantum computing will require solving the decoding problem in \emph{real time} and thus the decoder must operate at least as fast as the syndrome information is generated through measurements to prevent the backlog problem~\cite{terhal_quantum_2015}.
To achieve this, decoding algorithms with linear or almost-linear complexity in the size of the parity-check matrix are desirable.  

In practice, decoders will most likely be required to be implemented on specialized hardware such as FPGAs or ASICs to achieve the necessary speed, especially for qubit architectures that generate measurements with a high rate such as quantum computers based on superconducting circuits~\cite{blais_quantum_2020} or integrated photonics~\cite{aghaee_rad_scaling_2025, bartolucci_fusion-based_2023}, likely yielding one terabit of syndrome information per second~\cite{bacon_software_2022}.  
However, fast decoders, especially all-purpose decoders, are also highly relevant for researchers to study the performance of quantum error correction protocols numerically.  
Estimating the performance of a stabilizer code under a realistic noise model typically requires millions of Monte Carlo samples to accurately estimate the probability of extremely rare events.  
Each sample requires solving the decoding problem for $10^4$ to $10^5$ or more distinct fault locations, that is, a parity-check matrix with equally many columns.

\subsection{The decoding problem \label{ssec:decoding_problem}}
We will now define the decoding problem and the condition of success.  
Quite generally, a decoder is a map $D: \{0, 1 \}^{r} \to \mathcal{P}_n$ that, given a list of measurement outcomes, returns a correction $C$ and is successful in correcting a Pauli error $E$ if $C E \in \mathcal{S}$.  
Typically, in quantum error correction, the list of measurement outcomes is related to the syndrome $\sigma(E)$ of $E$ such that one can alternatively write $D(\sigma(E)) = E / \mathcal{S}$.  
In the following, we will phrase the problem through the binary symplectic representation, which allows one to reason about the implementation of decoding algorithms much less abstractly.  
In particular, we express the Pauli error $E$ as $\vec{e} = \Lambda \bsr{E}$ where we include the symplectic matrix into the definition of $\vec{e}$ such that all products are ordinary products and not symplectic products.  
Thus, $\vec{e} = (\vec{e}_Z, \vec{e}_X)$ where $\vec{e}_X, \vec{e}_Z \in \mathbb{F}_2^{n}$ are the $Z$ and $X$ components of the error, respectively, and are differently ordered in comparison to the ordinary representation of a Pauli string.  
The syndrome equation
\begin{align}
    \vec{s} = H \vec{e},
\end{align}
then defines the syndrome $\vec{s} \in \mathbb{F}_2^{r}$ as in the classical case of linear codes, see also Eq.~\eqref{eq:syndrome_equation}, through the binary representation of the $r \times 2n$ parity-check matrix.  
As $H \vec{e}$ effectively computes the symplectic inner product of $H$ and $\vec{e}$, the $i^{\mathrm{th}}$ entry of $\vec{s}$, $s_i$, determines if $E$ commutes with the stabilizer generator $S_i$ of the code.

In addition to the syndrome, the decoder receives information about the noise model, typically in the form of error probabilities $\vec{p} \in \mathbb{R}^{2n}$ where $p_i$ is the probability that the $i^{\mathrm{th}}$ bit of $\vec{e}$ is flipped.
For an independent error model, the probability of an error $\vec{e}$ for the prior $\vec{p}$ is given by a product distribution
\begin{align}
    \label{eq:def_error_product_distribution}
    \ErrProb(\vec{e}) = \prod_{i=1}^{2n} (1 - p_i)^{1 - e_i} p_i^{e_i} = \prod_{i=1}^{2n} (1 - p_i) \prod_{i=1}^{2n} \left(\frac{p_i}{1 - p_i}  \right)^{e_{i}}.
\end{align}
The correction $\vec{c}$ returned by the decoder is valid if $H(\vec{c} + \vec{e}) = 0$, and is successful if the logical action of $\vec{c}$ and $\vec{e}$ is identical, that is, if $L \vec{c} = L \vec{e}$.
Here, $L$ is the binary representation of the logical operators, usually written as
\begin{align}
    L = \mqty(\ell_X \\ \ell_Z),
\end{align}
where $\ell_X$ and $\ell_Z$ are given by Eq.~\eqref{eq:logical_operators_stabilizer_code_normalizer} via the binary symplectic representation of $\{ \overline{X}_j \}$ and $\{ \overline{Z}_j \}$.

\paragraph{Circuit-level noise.}
In the related literature, the above definition of the decoding problem is commonly referred to as the case of \emph{perfect} syndrome measurements, with a distinction made for the case of \emph{noisy} syndrome measurements.  
Here, we avoid this distinction, as we believe it is unnecessary and potentially misleading.
Rather than distinguishing at the level of the decoding problem, we argue that the distinction should be made at the level of the quantum error-correcting code and noise model.
That is, the parity-check matrix $H$ and the logical correlation matrix $L$ should be derived from the \gls{qec} code in space-time~\cite{bacon_sparse_2017, delfosse_spacetime_2023, hillmann_single-shot_2024}, rather than from the instantaneous \gls{qec} code defined purely in space.  
In such a setting, the columns of $H \in \mathbb{F}_2^{\lvert D \rvert \times \lvert F \rvert}$ correspond to independent fault mechanisms $F$, while the rows correspond to so-called \emph{detectors} $D$, which are distinct from the stabilizers of a static \gls{qec} code\footnote{A detector is typically understood as a parity constraint on a set of measurement outcomes.}.
The entry $H[j, i]$ of the parity-check matrix --- also known as the \emph{detector check matrix} --- is $1$ if the $i^{\mathrm{th}}$ fault mechanism flips the parity of the $j^{\mathrm{th}}$ detector.  
Similarly, $L \in \mathbb{F}_2^{\lvert K \rvert \times \lvert F \rvert}$ encodes the logical correlations $K$ in space-time, with $L[j, i] = 1$ indicating that the $i^{\mathrm{th}}$ fault mechanism flips the $j^{\mathrm{th}}$ logical correlation.

Importantly, the space-time formulation typically leads to a significantly larger \gls{qec} code than its static counterpart.  
This increase in size stems primarily from the need to repeat syndrome measurements over time to maintain fault tolerance against measurement errors and the additional fault locations in the circuit.  
Each additional round of syndrome extraction introduces new fault mechanisms and new detectors into the space-time code, effectively adding an extra (discrete) temporal dimension to the structure of the code, see also \cref{ssec:ftgs} and \figref{fig:foliated_412_code}. 
Thus, a code in space-time can be understood as an ordinary stabilizer code at a single moment in time, but one that operates in a higher-dimensional space with a correspondingly increased number of checks and errors.

Therefore, in the following, we will not distinguish between the decoding problem arising from a code capacity noise model with perfect syndrome measurements and one due to the more realistic circuit-level noise model, see also \cref{sec:quantum_error_correction}.  
We will instead emphasize, if necessary, specific features that originate from the space-time formulation of the decoding problem.  
However, for most purposes of decoding, a code in space-time can be regarded as an ordinary, albeit larger, stabilizer code at a single moment in time.
Accordingly, we will use the terms \emph{error} and \emph{fault}, as well as \emph{check} and \emph{detector}, interchangeably throughout the text.

\vspace{1em}
We have now seen that the decoding problem in quantum error correction closely resembles that of classical linear binary codes.  
This analogy becomes even more evident when comparing the definitions of the code space via the encoding matrix, see Eq.~\eqref{eq:def_classical_encoding_matrix} for classical linear codes and Eq.~\eqref{eq:def_stabilizer_code_encoding_unitary} for stabilizer codes.  
For classical linear codes, decoding succeeds only if the correction matches the error exactly, that is, $\vec{e} = \vec{c}$.  
This ensures that the logical content remains unchanged, as can be seen explicitly by noting that $\vec{e} V^{-1} = (\ell : \vec{0}_{n-k}) = \vec{c} V^{-1}$ where $V$ is the encoding matrix as defined in Eq.~\eqref{eq:def_classical_encoding_matrix}.

In contrast, stabilizer codes allow for a more relaxed decoding criterion.  
As discussed above, decoding is successful as long as $L \vec{e} = L \vec{c}$, meaning that the logical prediction of the error and correction coincide.  
The vectors $\vec{e}$ and $\vec{c}$ may still differ by an element of the stabilizer group without affecting the encoded information.
This becomes particularly transparent when decomposing the Pauli operators $E$ and $C$ into their logical, destabilizer, and stabilizer parts, see Eq.~\eqref{eq:def_pauli_TLS_decomposition}.  
Specifically, we write $E = T_E L_E S_E$ and $C = T_C L_C S_C$.  
Successful decoding requires $L_C = L_E$, while $T_C = T_E$ holds by construction for a valid correction.  
Any residual difference between $S_C$ and $S_E$ corresponds to stabilizer elements, which act trivially on the code space. 
This logic is also reflected in the action of the inverse encoding unitary.  
After applying $V^\dagger$ to the noisy encoded state $E \ket{\psi}$, we find:
\begin{align}
    V^{\dagger} E \ket{\psi} &= V^{\dagger} T_{E} L_{E} S_{E} V (\ket{\varphi} \otimes \ket{0_{n-k}}) \\
    &= \ell_E \ket{\varphi} \otimes \bigotimes_{j=k+1}^{n} X_j^{s_{j - k}} Z_j \ket{0_{n-k}} \\
    &= \ell_E \ket{\varphi} \otimes \ket{s_1, \dots, s_{n-k}},
\end{align}
where $\ell_E \in \mathcal{P}_k$ is the logical operator obtained from $V^\dagger L_E V$, compare also Eq.~\eqref{eq:encoding_unitary_xz_output}. 
This expression shows that the logical operator $\ell_E$ acts on the logical state $\ket{\varphi} \in \mathbb{C}^{2^k}_2$, while the stabilizer component determines the syndrome information that is encoded into the state of $n-k$ additional qubits used for redundancy.

\subsection{Maximum-likelihood decoding \label{ssec:ml_decoding}}
As the decoding problem for stabilizer codes is not identical to that of classical linear codes, we must solve a more general task that corresponds to determining the most likely logical correction, that is, the maximization problem
\begin{align}
    \label{eq:def_mld_problem}
    \vec{\ell}_{C} = \argmax_{\vec{\ell} \in \image(L)} \Big[ \sum_{\vec{e} \in \mathbb{F}_2^{\lvert F \rvert}: L\vec{e} = \vec{\ell}, H \vec{e} = \vec{s}} \ErrProb(\vec{e}) \Big] = \argmax_{\vec{\ell} \in \image(L)} \Big[ \sum_{S \in \mathcal{S}} \ErrProb(T_s L S) \Big].
\end{align}
The obtained logical coset $\vec{\ell}_{C}$ is, by construction, the most likely logical error $L \vec{e}$, such that a decoder that evaluates Eq.~\eqref{eq:def_mld_problem} is called a \emph{maximum-likelihood} decoder.
Unfortunately, evaluating Eq.~\eqref{eq:def_mld_problem} is, in general, an extremely hard problem, belonging to the complexity class of $\# \PTIME$-complete problems~\cite{iyer_hardness_2015}.
Intuitively, the hardness of (naively) evaluating Eq.~\eqref{eq:def_mld_problem} becomes apparent by recognizing that $\image(L)$ contains $2^{\lvert K \rvert}$ elements and that additionally exponentially many elements $\vec{e} \in \mathbb{F}_2^{\lvert F \rvert}$ exist.
Note that by decomposing the error as $E = T_{\vec{s}} L S$, it is possible to reduce the number of elements in the sum to $2^{\lvert \mathcal{S} \rvert}$, reducing the prefactor but retaining the exponential scaling in the size of the code.
Some quantum error correction codes exhibit sufficient structure such that Eq.~\eqref{eq:def_mld_problem} can be efficiently evaluated~\cite{poulin_optimal_2006}, or allow approximate evaluations with high accuracy for modest system sizes~\cite{bravyi_efficient_2014, ferris_tensor_2014, chubb_general_2021}.
Unfortunately, at the time of writing, it is unknown whether there exist classes of codes for which the maximum-likelihood decoding problem can be solved (approximately) efficiently under a realistic noise model. 
In other words, it remains unclear whether quantum codes in space-time exist that are efficiently (approximately) optimally decodable.
As a result, the maximum-likelihood decoding problem is typically not solved directly, even approximately. 
Instead, the focus is often on solving the \emph{minimum-weight} decoding problem.

\subsection{Minimum-weight decoding \label{ssec:mw_decoding}}
Instead of predicting the most likely logical error, as the maximum-likelihood decoder does, a \emph{minimum-weight} decoder predicts the most probable \emph{physical error}.
The correction $\vec{c}$ returned by the minimum-weight decoder is obtained through
\begin{align}
    \label{eq:def_mwd_problem}
    \vec{c} = \argmax_{\vec{\tilde{c}} \in \mathbb{F}_2^{\lvert F \rvert}: H \vec{\tilde{c}} = \vec{s}} \Big[ \ErrProb(\vec{\tilde{c}}) \Big],
\end{align}
which still is an $\NP$-hard problem~\cite{berlekamp_inherent_1978, hsieh_np-hardness_2011, kuo_hardness_2012}.
We note that for classical linear codes, this is the optimal solution.

The difference in accuracy between a maximum-likelihood decoder and a minimum-weight decoder cannot be easily quantified and depends on various properties of the decoding problem.
However, a common observation is that the difference increases if the stabilizer group contains many low-weight elements~\cite{lidar_quantum_2013, fuentes_degeneracy_2021}, a common feature of \gls{qldpc} codes and quantum codes in space-time.
As in the case of the maximum-likelihood decoding problem, there exist codes with sufficient structure such that the minimum-weight decoding problem is efficiently solvable.

In many cases, for Eq.~\eqref{eq:def_mwd_problem} to be efficiently solvable, it is required that the parity-check matrix $H$ is sparse, or more specifically, has low column and row weights.
If additionally, $H$ contains no short cycles when viewed as the biadjacency matrix of a bipartite graph, also known as the \emph{decoding graph}, then iterative message-passing algorithms are efficient in approximately solving Eq.~\eqref{eq:def_mwd_problem}.
We will describe this algorithm and the decoding graph in more detail below.

Before concluding this section, we note a particular relevant structure in $H$ that allows the problem to be solvable efficiently.
Given that the column weight of $H$ is bounded by two, the minimum-weight perfect matching (MWPM) decoder~\cite{dennis_topological_2002} solves the minimum-weight decoding problem exactly in polynomial time~\cite{edmonds_matching_1973}.
Codes, for which the column weight condition is satisfied, are usually referred to as \emph{matchable}.
An important example of a matchable code is the two-dimensional toric code~\cite{kitaev_quantum_1997, dennis_topological_2002}, also known as the surface code when embedded into the plane.
Surface codes are currently the gold standard for building a fault-tolerant quantum computer due to a variety of desirable properties, such as the possibility to realize them in a 2D architecture with only nearest-neighbor connectivity, a high error-correction threshold, and the existence of efficient minimum-weight decoders even under realistic circuit-level noise models~\cite{fowler_minimum_2014}.
Unfortunately, while scalable to achieve arbitrary distance $d$, surface codes are limited to encoding a single logical qubit $k = 1$.
That is, when laid out as a two-dimensional grid of linear size $L$, the surface code is a quantum error-correcting code with parameters $[\![L^2, 1, L ]\!]$\footnote{To be precise, this is the \emph{rotated} version of the surface code~\cite{bombin_optimal_2007}.}.
The drawback that the surface code can only encode a single logical qubit is significant, as one requires roughly 1000 physical qubits for each additional surface code logical qubit~\cite{fowler_surface_2012} for a realistic physical error rate of $p = 10^{-3}$.

Lastly, in light of the discussion on circuit-level noise in \cref{ssec:decoding_problem}, it is important to clarify that matchability is not a property of a code alone, but rather of the combination of a code and an error model. 
In particular, under the general depolarizing noise model introduced in \cref{ssec:qec_noise}, optimal decoding of the surface code is as hard as decoding an arbitrary stabilizer code~\cite{fischer_hardness_2024}. 
This hardness arises because, although each qubit in the surface code is involved in only two $X$- and $Z$-type stabilizers, a Pauli $Y$ error flips four syndrome bits, two $X$ and two $Z$, making it incompatible with a matching-based decoder.

To faithfully represent all single-qubit Pauli errors as independent events, one constructs a parity-check matrix of the form
\begin{align}
\label{eq:depol_check_matrix}
H = \left[
\begin{array}{ccc}
H_X & 0 & H_X \\
0 & H_Z & H_Z
\end{array}
\right],
\end{align}
where the columns correspond to $X$, $Z$, and $Y$ errors, respectively. However, a common simplification is to represent $Y$ errors as the combination of an $X$ and a $Z$ error, thereby eliminating the third column in Eq.~\eqref{eq:depol_check_matrix}. This renders the decoding problem matchable, but introduces a critical inaccuracy: it treats $Y$ errors as two independent faults, misrepresenting their true physical probability $O(p)$ as $O(p^2)$.
Representing $Y$ errors as combinations of $X$ and $Z$ errors is common practice, not solely to enable matchability, but also because it decouples the decoding into smaller, independent subproblems. 
Beyond that, it also alleviates issues in other decoding approaches, as we will see shortly.

\subsection{Message-passing decoding \label{ssec:belief_propagation}}
Unfortunately, implementing a minimum-weight decoder by exhaustively comparing the syndrome with all valid codewords is infeasible, especially for codes with large $n$.
Instead, we describe here a decoding approach that dates back to Gallager~\cite{gallager_low_1960} in the context of classical error-correcting codes.
This algorithm is known under various names in different communities (see \textsc{Notes} of chapter 2 in~\cite{richardson_modern_2008}), and we will usually refer to it as \gls{bp} or message-passing.
More generally, the algorithm utilizes the generalized distributive law, see Ref.~\cite{aji_generalized_2000}.
In the following, we often (implicitly) assume that the $r \times c$ parity-check matrix $H$ is sparse and, more importantly, that both its row and column weights are small and constant.
In other words, the Tanner graph associated with $H$ has low-degree variable and check nodes.
Tanner graphs are a crucial tool for understanding the algorithm outlined below.

\begin{figure}
    \centering
    \includegraphics{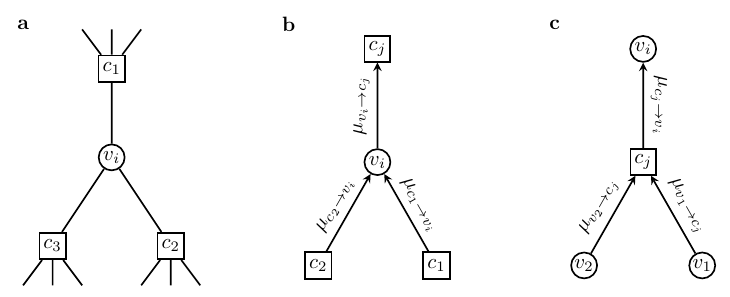}
    \caption{\textbf{a} Local subcodes or parity checks overlapping on the variable node $v_i$.
    \textbf{b} Message-passing during a single iteration from the perspective of a variable node $v_i$ sending a message to check node $c_j$.
    \textbf{c} Message-passing during a single iteration from the perspective of a check node $c_j$ sending a message to variable node $v_i$.
    }
    \label{fig:message_passing_decoding}
\end{figure}

The parity-check matrix $H$ imposes $\RANK(H)$ constraints on potential codewords $\vec{c} \in \mathbb{F}_2^{c}$.
Each of these constraints can also be interpreted as defining a local subcode involving a single check and a subset of the codeword bits, specifically, those bits connected to that given check node.  
While such a subcode has distance $d = 2$, each bit typically participates in $\colweight$ such subcodes, where $\colweight$ is the column weight of $H$, and the global code can therefore have a significantly larger distance.
This local structure is illustrated in \figref{fig:message_passing_decoding} from the perspective of a variable node $v_b$.

Within the subcode, what is the probability that the bit associated with variable node $v_i$ is in error?
More formally, we aim to evaluate the conditional probability that an error $e_i = 1$ occurred given the observed syndrome $\vec{s}$, that is,
\begin{align}
    \ErrProb[e_i =1 \mid \vec{s}].
\end{align}
Note that only a subset of the syndrome bits in $\vec{s}$ are relevant to this probability.
It turns out that the above conditional probability has a relatively simple expression when expressed as a likelihood ratio.
Let $s_c \in \{0, 1\}$ denote the observed syndrome bit for check $c$.
Then, following~\cite{gallager_low_1960}, we find
\begin{align}
    \label{eq:message_passing_conditional_property}
    \frac{\ErrProb[e_i = 0 \mid \vec{s}]}{\ErrProb[e_i = 1 \mid \vec{s}]}
    = \frac{1 - p_i}{p_i}
    \prod_{c_j \in N(v_i)} \left[
        \frac{
        1 + (-1)^{s_{c_j}} 
        \prod_{
        v_{i^{\prime}} \in N(c) \setminus \{v_i\}
        } 
        (1 - 2p_{{i^{\prime}}})
        }
        {1 - (-1)^{s_{c_j}} \prod_{v_{i^{\prime}} \in N(c) \setminus \{v_i\}} (1 - 2p_{i^{\prime}})}
    \right].
\end{align}
Here, we have excluded the variable $v_i$ from the product over $v \in N(c)$ since we are computing the conditional likelihood ratio with respect to $e_i$.
Each term in the product corresponds to the contribution of a parity check $c_j$ connected to $v_i$. 
The expression inside the product reflects the likelihood of an even or odd number of errors occurring among the neighboring variable nodes of $c_j$, excluding $v_i$.
The factor $(-1)^{s_c}$ ensures that the check contributes with the correct sign depending on whether the observed syndrome bit is $0$ or $1$.
To obtain this expression, we assume that the error on each bit is independent, and denote the corresponding vector of bitwise error probabilities by $\vec{p} = (p_1, \dots, p_c)$.
Each $p_i \in [0,1]$ is the probability that the $i^{\text{th}}$ bit is in error, i.e., $e_i = 1$.

Equation~\eqref{eq:message_passing_conditional_property} forms the basis of the message-passing decoding algorithm.
To see this, first note that the expression can be rephrased in terms of likelihood ratios $\lambda = (1 - p)/p$, such that $1 - 2p = (\lambda - 1)/(\lambda + 1)$.
Moreover, the right-hand side of Eq.~\eqref{eq:message_passing_conditional_property} can be evaluated in parallel for all bits.
However, since the equation only considers a local neighborhood in the Tanner graph, the resulting estimate is not the true marginal probability.
Fortunately, it is possible to derive an iterative algorithm based on Eq.~\eqref{eq:message_passing_conditional_property} that explores the whole graph.

This algorithm works by sending \emph{messages} between check and variable nodes according to specific update rules.
That is, for each iteration step, first, each check node $c_j$ sends the message
\begin{align}
    \label{eq:bp_check_to_bit}
    \mu_{c_j \to v_i} = \prod_{v_{i^\prime} \in N(c_j) \setminus \{ v_i \}} \big[\mu_{v_{i^\prime} \to c_j}\big]^{(1 - 2 s_{c_j})}, %
\end{align}
to all its connected variable nodes $v_i \in N(c_i)$ and, in turn, each variable node $v_i$ sends updated probabilities to its connected check nodes $c_j \in N(v_i)$ as the message
\begin{align}
    \label{eq:bp_bit_to_check}
    \mu_{v_i \to c_j} = \lambda_i \prod_{c_{j^{\prime}} \in N(v_i) \setminus \{ c_j \}} \mu_{c_{j^{\prime}} \to v_i},
\end{align}
where $\lambda_i = (1 - p_{v_i}) / p_{v_i}$ is the original likelihood of $v_i$ being in error obtained from the channel probabilities $\vec{p}$.
The messages $\mu_{v_i \to c_j}$ can be understood as \emph{a posteriori} probabilities for the $i^{\textrm{th}}$ bit to be in error given the constraints of the $j^{\mathrm{th}}$ check, that is, the $j^{\mathrm{th}}$ subcode.
The algorithm is initialized by setting the messages $\mu_{v_i \to c_j}$ as the likelihoods $\lambda_{i}$.
By inserting Eq.~\eqref{eq:bp_check_to_bit} into Eq.~\eqref{eq:bp_bit_to_check}, one almost recovers Eq.~\eqref{eq:message_passing_conditional_property} with the only modification that the product is not taken over all checks. %
This assures that there is no immediate self-interference in the calculation of the $i^{\textrm{th}}$ a posteriori probability.
That is, information sent from a variable node to a check node will never flow back through the same edge from the check node to the variable node, and the independence assumption made to derive Eq.~\eqref{eq:message_passing_conditional_property} remains valid, see also \figref{fig:message_passing_decoding}.
Thus, if this iteration process were continued until the local information has explored the whole graph, it would calculate the exact marginal probability of the $i^{\mathrm{th}}$ bit to be in error.
After the iteration steps, the marginal probability can be obtained by multiplying all messages from neighboring check nodes, that is, 
\begin{align}
    \label{eq:bp_termination}
    \mu_{v_i} =  \lambda_i \prod_{c_{j^{\prime}} \in N(v_i)} \mu_{c_{j^{\prime}} \to v_i}.
\end{align}
However, this is only true if the independence assumptions remain valid, which requires the Tanner graph not to have cycles, that is, it must be a tree. 
For a graph with cycles of length $g$, also called the girth of the graph, the independence assumption holds for $\lfloor (g-2)/4 \rfloor$ iterations~\cite{tanner_recursive_1981}.
As already pointed out by Gallager, we can ignore the lack of independence and continue the iteration process which ``\emph{is ultimately justified, of course, only by the fact that it works}''~\cite{gallager_low_1960}.
It is also common to express the algorithm in the logarithmic domain by introducing the log-likelihood ratio $L(v_i) = \log((1 - p_i) / p_i)$.
In this formulation, the product in Eq.~\eqref{eq:bp_bit_to_check} becomes a sum, which is why the algorithm is also known as the \emph{sum-product} algorithm.
In this variant of the algorithm, one can make an approximation to reduce the cost of evaluating Eq.~\eqref{eq:bp_check_to_bit}, which replaces the product with the $\min$ function, which leads to the \emph{min-sum} algorithm, see, e.g., Ref.~\cite{kschischang_factor_2001, ryan_channel_2009}.

A significant feature of the algorithm is its computational complexity.
Without parallelization, the algorithmic complexity is linear in the number of nodes in the Tanner graph, the degree of the nodes, and the number of iterations.
As noted above, the calculation of messages is completely local and can be straightforwardly parallelized. 
Additionally, for a family of (quantum) LDPC codes, typically, the check and bit degrees are constant.
Thus, a parallel implementation has at most a time complexity that is linear in the number of iterations.
As the algorithm does not have any formal convergence guarantees in the presence of cycles in the graph, a stopping criterion is required.
To this end, after each iteration step, one typically performs a hard-decoding $e_i = 0$ if $\mu_{v_i} < 0.5$ and $e_i = 1$ otherwise.
If the current estimate of the error $\vec{e}$ satisfies the syndrome equation $\vec{s} = H \vec{e}$, the algorithm is terminated.
However, for quantum error-correcting codes subject to realistic noise, it is unclear whether the average number of iterations required is bounded or scales with the size of the code, see, e.g., Ref.~\cite[App. E]{scruby_high-threshold_2024}.

Unfortunately, applying the \gls{bp} algorithm to quantum error-correcting codes yields additional issues.
One of those is related to the requirement of commutating stabilizers and is observed for many noise models.
For stabilizers to commute in a CSS code, a $X$ stabilizer $S_X$ and a $Z$ stabilizer $S_Z$ must overlap on an even number of qubits, but at least two, say $q_i$ and $q_j$.
Now, a Pauli $Y$ error on qubit $q_i$ will anti-commute with both stabilizers $S_X$ and $S_Z$ and if we represent it as an independent faults mechanism in the parity-check matrix, this leads to a cycle of length-4 $(S_Z, Y_{q_i}, S_X, Y_{q_j}, S_Z)$.
A simple solution is to not represent $Y$ errors as independent faults and represent them as a product of Pauli $X$ and $Z$ operators, and solve the decoding problems for $X$ and $Z$ faults separately.
While doing this alleviates the problem of the existence of 4-cycles, it ignores correlations in the noise model that can aid the decoding performance in general, as it misrepresents the probability of a $Y$ error as $O(p^2)$ instead of $O(p)$.
For practical considerations, one might accept this performance degradation if it is not too large for a decoder that has a linear complexity in the number of fault locations.

The second issue, however, is not as easily averted and is due to degeneracy.
To reiterate, degeneracy refers to the fact that logical observables are only well defined up to elements of the stabilizer group $\mathcal{S}$.
Equivalently, there exist combinations of faults $\{\vec{t} \mid \vec{t} \in \ker H \cap \ker L \}$ that neither flip a detector nor a logical observable and therefore can be considered stabilizers.
For a low-weight combination of faults $\vec{t}$, it is possible for two corrections $\vec{c}$ and $\vec{c} + \vec{t}$ to be both equally highly probable, such that \gls{bp} does not converge to one of them.
This situation is referred to as the \emph{split belief} problem and occurs as the \emph{a posteriori} probability distribution is not sufficiently peaked on a single configuration of faults.
Forcing the algorithm to terminate after a finite number of iterations typically yields no correction or the sum of both of the solutions, however, neither of them will return a valid correction that satisfies the syndrome, i.e., $H (\vec{c} + \vec{c} + \vec{t}) = H \vec{t} = \vec{0} \neq \vec{s} $.
It is no coincidence that we refer to such errors by the symbol $\vec{t}$ as an analogy of the issue that exists in the classical literature, where it is known as a trapping set, see, e.g., Refs.~\cite{mackay_weaknesses_2003, koetter_graph-covers_2003, raveendran_trapping_2021, di_finite-length_2002, wiberg_codes_1996}.
We will now describe a way to reduce the impact of this problem, unfortunately, at the cost of a significant increase in complexity.

\subsection{Inversion decoding \label{ssec:inversion_decoding}}
The syndrome decoding problem, introduced around Eq.~\eqref{eq:syndrome_equation}, $\vec{s} = H \vec{e}$, is concerned with finding an estimate of $\vec{e}$ given the observed syndrome $\vec{s}$.
While so far we have been mostly focused on describing the maximum-likelihood decoding problem (\cref{ssec:mw_decoding}) and the easier minimum-weight decoding problem (\cref{ssec:mw_decoding}), we have seen that due to the existence of short cycles in the decoding graph, the standard approach of decoding (quantum) \gls{ldpc} codes through \gls{bp} may fail due to non-convergence, that is, the correction $\vec{c}$ does not reproduce the syndrome $\vec{s}$.
Naively, one might expect that it is simple and efficient to always obtain a correction that agrees with the syndrome by computing
\begin{align}
    \label{eq:inversion_decoding_naively}
    \vec{c} = H^{-1} \vec{s},
\end{align}
which has cubic complexity in the size of the check matrix $H$ due to the necessity to calculate the inverse $H^{-1}$, which can be achieved by Gaussian elimination.
However, this approach does not work: a (left) inverse of $H$ does not exist, since $H$ lacks full column rank --- a necessary condition for $H$ to be the parity-check matrix of a nontrivial linear code.
Nonetheless, the conceptual idea of Eq.~\eqref{eq:inversion_decoding_naively} can be preserved by considering a more informed method.

Let $I$ be a set of column indices constructed by selecting a linearly independent subset of the columns of the check matrix $H$.
Accordingly, the size $\lvert I \rvert$ of the index set is given by $\lvert I \rvert = \RANK(H) \leq r < c$ for the $r \times c$ matrix $H$.
Then, $H_{[I]}$ is an invertible matrix formed by selecting the columns of $H$ indexed by the set $I$.
We can therefore compute a partial correction $\vec{c}^{\prime} = H_{[I]}^{-1} \vec{s}$, which specifies the correction only on the coordinates indexed by $I$.
To obtain a full correction vector $\vec{c} \in \mathbb{F}_2^{c}$, we define it component-wise as
\begin{align}
    c_j   = \begin{cases}
    c^{\prime}_j, \; \forall j\in I, \\
    0, \; \text{otherwise}.
    \end{cases}
\end{align}
Since $I$ indexes a linearly independent set of columns, each choice of $I$ defines a unique partial solution $\vec{c}^{\prime}$.
However, in general, randomly choosing the basis described by $I$ is unlikely to result in a good correction $\vec{c}$.
In other words, it is unlikely that $\vec{c}$ corresponds to the 
minimum-weight correction as defined by Eq.~\eqref{eq:def_mwd_problem}.
For instance, if $I$ does not contain the index corresponding to a symbol in error, the correction cannot reproduce the error\footnote{We note that this is not necessarily an issue for stabilizer codes as the correction must only agree up to a stabilizer with the error.}, $\vec{c} \neq \vec{e}$.

After describing the conceptual ideas involved in inversion decoding, we now turn to the \gls{osd} algorithm.
Originally introduced by Fossorier and Lin~\cite{fossorier_soft-decision_1995}, \gls{osd} serves as a post-processing step to improve upon invalid solutions produced by \gls{bp}, helping to reduce error floors in classical \gls{ldpc} codes.
To this end, \gls{osd} leverages the marginal probabilities output by the \gls{bp} decoder as soft information to guide the construction of the index set $I$, as we explain below.  
In 2019, Panteleev and Kalachev~\cite{panteleev_degenerate_2021} were the first to apply \gls{osd} in the quantum setting, introducing BP+OSD as a surprisingly effective two-stage decoder of random \gls{qldpc} codes.
Due to the generality of the (combined) decoding algorithm and its applicability even to codes in space-time, BP+OSD has become the gold standard for decoding general \gls{qldpc} codes.

We begin by describing the post-processing step, often referred to as OSD-0, which is invoked when the \gls{bp} algorithm fails to converge within a maximum number of iterations.
The OSD-0 algorithm is presented formally in \algref{alg:osd_0_algorithm}.
\begin{algorithm}[!ht]
\caption{OSD-0 Post-Processing Step}
\begin{algorithmic}[1] %
\State \textbf{Input:} Soft information \( \vec{\lambda} \) (from \gls{bp} decoder)
\State \textbf{Output:} Correction \( \vec{c} \)
\State Utilize the soft information vector \( \vec{\lambda} \) to construct an ordered list of indices \( \Pi_{\vec{\lambda}} \) (from most to least likely in error, MLE).
\State Rearrange the columns of the check matrix \( H \) according to the ordering \( \Pi_{\vec{\lambda}} \) to obtain \( H^{\prime} \).
\State Construct the set of indices \( I \) as the first \(\RANK(H) \) linearly independent columns of \( H^{\prime} \)
\State Calculate the OSD-0 solution based on the indices \( I \) by matrix inversion:
\( \vec{c}_{[I]} = H_{[I]}^{-1} \vec{s}\).
\State Obtain the correction across all bits as:
\[
    \vec{c}' = (\vec{c}_{[I]}, \vec{c}_{[J]}) = (\vec{c}_{[I]}, \vec{0}),
\]
where \( J = \bar{I} \) is the complement of \( I \).
\State Reverse the column ordering induced by \( \Pi_{\vec{\lambda}} \) to go from MLE ordering to the physical ordering, resulting in the final correction \( \vec{c} \).
\end{algorithmic}
\label{alg:osd_0_algorithm}
\end{algorithm}

There we see that the inversion decoding algorithm aims to solve the minimum-weight decoding problem~\eqref{eq:def_mwd_problem} by selecting a basis that corresponds to the bits most likely in error, that is, the MLE basis.
Indeed, if $I$ corresponds to the $\RANK(H)$ most likely bits according to the ordering $\Pi_{\vec{\lambda}}$, the OSD-0 solution is the minimum-weight solution, that is, it is the optimal solution for the decoding problem of a linear code~\cite{fossorier_reliability-based_1998}.
The attentive reader notices that the complexity of obtaining the optimal solution is polynomial, which hints that the OSD-0 solution constructed in this way will not always yield the optimal solution to the classical decoding problem.
This issue arises if the $\lvert I \rvert$ most likely error locations are not linearly independent, that is, they do not form a basis of the column space of $H$.
In that case, the set $I$ does not index the MLE basis and elements of the MLE basis are contained in $\bar{I} = J$, the complement of $I$, such that $J$ contains error locations that have larger error probabilities according to the soft information vector $\vec{\lambda}$ than some error locations contained in $I$.
To find the optimal solution, one can systematically search through all error configurations in $J$ that potentially provide a more likely estimate $\tilde{\vec{c}}$.
As this search space is exponentially large in the size of $J$, the computational cost of finding the optimal solution becomes prohibitively large for all but a few smaller codes.
Thus, in practice, only configurations with a Hamming weight up to $w$ are considered, known as \emph{order}-$w$ reprocessing.
We refer the interested reader to Refs.~\cite{panteleev_degenerate_2021, roffe_decoding_2020, fossorier_soft-decision_1995,  fossorier_reliability-based_1998} for a more complete discussion of higher-order reprocessing and briefly discuss the idea here.

In ordered statistics decoding, higher-order reprocessing refers to considering solutions for which $\vec{c}_{[J]} \neq \vec{0}$.
Given the OSD-0 solution $\vec{c}_{[I]}$, the higher-order OSD solution for a particular choice of $\vec{c}_{[J]}$ is given by 
\begin{align}
    \label{eq:osd_reprocessing_solution}
    \vec{c}' = \left(\vec{c}_{[I]} + H_{[I]}^{-1} H_{[J]} \vec{c}_{[J]}, \vec{c}_{[J]} \right),
\end{align}
which fulfills the syndrome equation for arbitrary $\vec{c}_{[J]}$.
Completing the reprocessing routine of order $w$ then involves systematically searching through the set of admissible $\vec{c}_{[J]}$ of Hamming weight $w$ and retaining the solution $\vec{c}'$ of lowest total weight.  
It is possible to choose from various search strategies that explore different admissible subspaces for $\vec{c}_{[J]}$ at finite $w$, see, e.g., Refs~\cite{fossorier_soft-decision_1995, roffe_decoding_2020}.

\glsresetall
\section{Conclusion and Outlook}
In this chapter, we have described most of the fundamentals required for fault-tolerant quantum computing.
Rather than providing a comprehensive account of all aspects of quantum computing and quantum error correction, we have focused on the concepts most crucial for understanding the contributions of this thesis. 

To this end, we have revisited classical error-correcting codes, in particular, linear block codes, to gain additional intuition about the structure of stabilizer codes.
Stabilizer codes, especially \gls{css} codes, form the backbone of most practical quantum error correction strategies today.
The correspondence between \gls{css} codes and chain complexes, the primary elements of study in homology theory, is an additional reason for the importance of \gls{css} codes.
While homology initially provided a language that explained already known properties of codes, it has since proven to be tremendously productive, yielding various \gls{qldpc} codes with good properties.
The appended work on \emph{fault complexes}, see \refpaper{IX}, aims to extend the power of this framework from (\gls{css}) codes at an instantaneous moment in time, to codes in space-time, or more generally, fault-tolerant protocols.
It intends to serve as a unifying umbrella that incorporates multiple previously distinct viewpoints, that of gate-based and measurement-based computing, while making the full suite of mathematical tools from homology theory available to analyze these systems.

Regardless of the introduction of abstract structures such as chain complexes, this chapter aimed at introducing quantum error correction not merely from a theoretical viewpoint, but as a practical tool for scalable devices.
This includes the questions of deriving low-complexity decoding algorithms that are capable of achieving the throughput and low-latency requirements that come with real-time quantum error correction.
\refpaper{VII} introduces the \emph{localized statistics decoding} algorithm, the first parallel decoding algorithm for general quantum error correction protocols that matches the performance of the current state-of-the-art belief propagation plus ordered statistics decoding algorithm whilst being substantially faster.
Central to this is a novel linear algebra routine for parallel matrix factorization, which can efficiently solve sparse linear systems that we call \emph{on-the-fly elimination}.

While low-complexity decoders will be necessary for fault-tolerant quantum computing, reducing the size of the decoding problem alleviates the issue.
A promising path to achieve this is by considering quantum error-correcting codes that can be accurately decoded over a small (constant) window of time, or equivalently, a small window of syndrome measurement cycles.
Such codes are also known as single-shot (decodable) quantum codes~\cite{bombin_single-shot_2015, campbell_theory_2019}.
Quantum radial codes, introduced in \refpaper{VIII}, are a family of \gls{qldpc} codes derived from the lifted product of classical quasi-cyclic codes. 
Numerical simulations suggest that these codes are single-shot decodable even under circuit-level noise, while additionally showing comparable error suppression to surface codes of similar
distance while using approximately five times fewer physical qubits. 
Overall, their error correction capabilities, tunable parameters, and small size make them promising candidates for implementation on near-term quantum processors.

It is worth emphasizing that this chapter has mostly omitted the discussion of a fundamental aspect of fault-tolerant quantum computing.
That is, fault-tolerant quantum computing is not only about quantum memories but also about enabling reliable computation on encoded information.
The question of fault-tolerant logical operations is largely beyond the scope of this thesis and is therefore omitted here.
However, outside of this thesis, fault-tolerant operations for \gls{qldpc} codes are an extremely active topic of research.
One illustrative approach to fault-tolerant logical operations is surface code lattice surgery~\cite{horsman_surface_2012}, where logical operations are realized by merging and splitting encoded patches through local measurements.  
Lattice surgery can be naturally interpreted within the framework of measurement-based quantum computation and connects closely to the idea of Pauli-based computation~\cite{litinski_game_2019}, where computation proceeds via sequences of multi-qubit Pauli measurements.  
Generalizing lattice surgery to \gls{qldpc} codes poses unique challenges because these codes typically encode multiple logical qubits into a single block.
As a result, addressing individual qubits becomes generally challenging.
We point the interested reader to Ref.~\cite[Sec. 3]{he_extractors_2025} for an overview of the current developments around logical operations on \gls{qldpc} codes.

Finally, the methods discussed so far inevitably lead to large resource overheads.  
While stabilizer codes offer a powerful framework for protecting quantum information, their practical implementation demands substantial redundancy in physical qubits and operations.  
This motivates the exploration of alternative approaches.  
In the next chapter, we will shift focus to bosonic quantum error correction, where the structure of continuous-variable systems may offer a path toward reducing the overheads that plague traditional qubit-based schemes.  
By leveraging the rich physics of bosonic modes, new strategies for fault-tolerant quantum computing become accessible, expanding the toolkit available for future devices.

\cleardoublepage

\chapter{Quantum Continuous Variables \label{chap:quantum_continuous_variables}}

\glsresetall
Up to this point, we have focused on finite-dimensional quantum systems.
By combining multiple two-level systems, we expanded the Hilbert space, either to enhance computational capabilities or to introduce redundancy for error correction.
However, not all physical systems are inherently two-dimensional --- or even finite-dimensional.
Many naturally occurring quantum systems are described by infinite-dimensional Hilbert spaces.
A prominent example are \gls{cv} systems, where the expectation values of observables such as position and momentum take values over a continuum.

While a mathematically rigorous treatment of continuous-variable quantum mechanics lies beyond the scope of this chapter, we will introduce the essential concepts required to understand their use in quantum information.
We will not concern ourselves with subtleties such as unphysical infinite-energy states, instead focusing on the tools and intuition needed for practical applications.
This introduction aims to provide readers, particularly those familiar with discrete-variable systems, with the prerequisites for understanding the results presented in the appended papers.

We will consider the case of a single degree of freedom, also known as a single mode, for this presentation and note that the generalization to a finite number of modes is straightforward.

\section{Fundamentals of Quantum Continuous Variables}
In continuous-variable systems of a single degree of freedom, there exists a pair of self-adjoint operators, $\hat{q}$\footnote{In the literature one will also often encounter the symbol $x$ or $X$ for this operator. Here we use $q$ to distinguish it clearly from the Pauli $X$ operator.} and $\hat{p}$, that we refer to in the following as position and momentum, respectively, that satisfy
\begin{align}
    \label{eq:def_ccr}
    \left[ \hat{q}, \hat{p}\right] = i \hbar \mathbbm{1},
\end{align}
where in the following we will work in natural units $\hbar = 1$ and make the (infinite-dimensional) identity operator $\mathbbm{1}$ implicit.
The above equation is known as the \emph{canonical commutation relation} (CCR) and the pair of operators $\hat{q}$ and $\hat{p}$ as \emph{canonical} operators, accordingly.
The naming terminology originates from classical Hamiltonian dynamics, where the commutator in Eq.~\eqref{eq:def_ccr} is replaced by the Poisson bracket.
Indeed, the above commutation relations typically arise when quantizing a simple phase space.
Typically, one aims to find a representation for operators through matrices; however, in this case, no representation with \emph{finite} dimensional matrices exists.
Informally, this can be seen by taking the trace of Eq.~\eqref{eq:def_ccr}.
The cyclic property of the trace leads to $0 = i \Tr{\mathbbm{1}}$, which is undoubtedly incorrect.
Therefore, mathematically speaking, the position and momentum operators $\hat{q}$ and $\hat{p}$ are unbounded, and hence they cannot be trace-class, finite-rank, or represented on a finite-dimensional Hilbert space $\mathcal{H}$. 
Instead, one can define representations of these operators on the Hilbert space of square-integrable functions over the real line, $\mathcal{H} = L^2(\mathbb{R})$, such that
\begin{align}
    (\hat{q} \psi)(q) = q \psi(q), \quad
    (\hat{p} \psi)(q) = -i \frac{d}{dq} \psi(q), \quad \forall \psi \in L^2(\mathbb{R}).
\end{align}
While the eigenstates of $\hat{q}$ and $\hat{p}$ do not belong to $L^2(\mathbb{R})$, we adopt the standard formalism in which we work with their  quasi-eigenstates $\ket{q}$ and $\ket{p}$, defined by
\begin{align}
    \hat{q} \ket{q} = q \ket{q}, \quad
    \hat{p} \ket{p} = p \ket{p}, \quad q, p \in \mathbb{R},
\end{align}
which, although not normalizable, are conceptually useful for analytical computations.
Note that their eigenvalues form a continuous set on the real line, constituting the basis of the terminology of \emph{quantum continuous variables}.
Position and momentum eigenstates also form a basis in the generalized sense, allowing any quantum state to be represented as a superposition of these basis elements
\begin{align}
    \ket{\psi} = \int_{-\infty}^{\infty} \psi(a) \ket{a} \, \mathrm{d}a,
\end{align}
where $\ket{a}$ denotes the eigenstate of either $\hat{q}$ or $\hat{p}$ with eigenvalue $a$, and the complex-valued function $\psi(a) \in L^2(\mathbb{R})$ is called the wavefunction in the corresponding representation.

\section{The Wigner function \label{sec:wigner_function}}
While either the position or momentum representation completely describes the quantum state, it is still useful to have a representation that includes both.
This representation is somewhat analogous to the description of a classical system within phase space.
For example, the position and momentum of a single particle along one dimension can be represented by a point in a two-dimensional phase space, one axis representing position and the other momentum.
Uncertainties in the knowledge of the position or momentum of that particle can be represented if we replace the point with a probability distribution that indicates the relative likelihood of encountering the particle with a particular combination of position and momentum.

In quantum mechanics, position and momentum are fundamentally linked by the uncertainty principle.
As a result, constructing a phase-space representation of a quantum state is less straightforward than in classical mechanics. 
Nevertheless, several useful functions have been developed to describe quantum states in phase space~\cite{husimi_formal_1940, glauber_coherent_1963, sudarshan_equivalence_1963}.
Possibly one of the most prominent of those is known as the \emph{Wigner function}, due to Wigner and Szilard~\cite{wigner_quantum_1932}, and is defined in terms of the density operator $\hat{\rho}$ by
\begin{align}
    \label{eq:def_wigner_function}
    W(q, p) = \frac{1}{\pi} \int_{- \infty}^{\infty} e^{i2py} \langle q + y \lvert \hat{\rho} \rvert q - y \rangle \dd{y}. 
\end{align}
This expression shows that the Wigner function is essentially the Fourier transform of the off-diagonal elements of \( \hat{\rho} \) in the position basis.  
Just like the representation of a quantum state in the position or momentum basis, the Wigner function contains all the information about the quantum state\footnote{Indeed, all of quantum mechanics can be recast into quantum phase space.}.
As Eq.~\eqref{eq:def_wigner_function} is invariant under complex conjugation by flipping the sign of $y$, the Wigner function is real-valued for all physical states.
This makes the Wigner function representation a valuable tool to explain the behavior of continuous-variable quantum states intuitively.
To this end, it is useful to note that the marginal distributions of the Wigner function recover the usual probability distributions, that is,
\begin{align}
    P(q) = \mel{q}{\hat{\rho}}{q} = \int_{-\infty}^{\infty} W(q, p) \, \mathrm{d}p, \quad P(p) = \mel{p}{\hat{\rho}}{p} = \int_{-\infty}^{\infty} W(q, p) \, \mathrm{d}q,
\end{align}
and consequently, the Wigner function is normalized like an ordinary probability distribution,
\begin{align}
    \int_{-\infty}^{\infty} \int_{-\infty}^{\infty} W(q, p) \, \mathrm{d}q \, \mathrm{d}p = 1.
\end{align}
However, the Wigner function is only a quasi-probability distribution, as it can take on negative values.
These negativities are typically observed in Wigner functions of superpositions of distinct states, as we will see in more detail later.

Indeed, it is possible to use the Wigner function to visually distinguish different quantum states through their Wigner functions or to gain intuition about the expectation values of operators with respect to a quantum state. 
This is due to the fact that operator expectation values can be calculated as phase-space averages of their Wigner function representation weighted by the Wigner function of the corresponding quantum state, that is,
\begin{align}
    \label{eq:wigner_phase_space_average}
    \Tr[{\hat{A} \hat{\rho}}] = \int_{-\infty}^{\infty} \int_{-\infty}^{\infty} W(q, p) W_{A}(q, p) \, \mathrm{d}q \, \mathrm{d}p,
\end{align}
where $\hat{A}$ is an arbitrary Hermitian operator and $W_{A}$ denotes its phase space representation as defined by Eq.~\eqref{eq:def_wigner_function}.
For example, let \( \hat{A} \) represent another quantum state \( \hat{\rho}_2 = \ketbra{\psi_2} \).
For two states to be orthogonal, $\left| \langle \psi_1 \mid \psi_2 \rangle \right|^2 = 0$, this orthogonality condition translates to the requirement that the overlap integral in terms of their Wigner functions vanishes
\[
\int_{-\infty}^{\infty} \int_{-\infty}^{\infty} W_1(q, p) \, W_2(q, p) \dd{q} \dd{p} = 0.
\]
This can occur in two ways. Either the Wigner functions have no overlapping support in phase space, or their product includes positive and negative contributions that exactly cancel out.
In the latter case, regions where the Wigner function takes negative values play a crucial role, allowing for cancellation due to interference effects and symmetries in the states. 

\section{Quantum Harmonic Oscillators \label{sec:quantum_harmonic_oscillator}}
A particularly relevant example where quantum continuous variables appear is during the quantization of the modes of an electromagnetic field. 
To this end, let us consider the electromagnetic field confined in a finite volume.
The vector potential of this field obeys the wave equation, and its solution is a linear combination of frequency modes with a spatial profile determined by the boundary conditions.
In terms of these modes, the Hamiltonian of the electromagnetic field reduces to that of a set of independent harmonic oscillators.
Quantization of the electromagnetic field then reduces to the quantization of these individual modes.

The Hamiltonian that describes a single quantized mode of the electromagnetic field is given by
\begin{align}
    \label{eq:quantum_harmonic_oscillator}
    \hat{H} = \frac{1}{2}\left( \hat{p}^2 + \omega^2 \hat{q}^2 \right),
\end{align}
where $\omega$ is the frequency of the mode.
The above Hamiltonian is also known as the \emph{quantum harmonic oscillator} and it can be diagonalized by introducing the bosonic annihilation $\hat{a}$ and creation  $\hat{a}^{\dagger}$ operators such that
\begin{align}
    \label{eq:def_bosonic_annihilation_creation}
    \hat{a} = \frac{1}{\sqrt{2 \omega}} \left(\omega \hat{q} + i \hat{p} \right), \quad  \hat{a}^{\dagger} = \frac{1}{\sqrt{2 \omega}} \left(\omega \hat{q} - i \hat{p} \right),
\end{align}
that fulfill the bosonic canonical commutation relations $[\hat{a}, \hat{a}^{\dagger} ] = \mathbbm{1}$.
In the following we will always work with the dimensionless version of the operators $\hat{q}$ and $\hat{p}$,  that is, we redefine $\hat{q} \to \sqrt{\omega} \hat{q}$ and $\hat{p} \to \hat{p} / \sqrt{\omega}$ such that $\hat{q} = (\hat{a} + \hat{a}^{\dagger}) / \sqrt{2}$ and $\hat{p} = i (\hat{a}^{\dagger} - \hat{a})/\sqrt{2}$.
We will also sometimes refer to $\hat{q}$ and $\hat{p}$ as field quadratures, as they yield a Hamiltonian that is a sum of squares.

Expressed in terms of the annihilation and creation operators, the quantum harmonic oscillator takes the form
\begin{align}
    \label{eq:qho_a_adag}
    \hat{H} = \omega \left(\hat{a}^{\dagger}\hat{a} + \frac12 \right),
\end{align}
but we will typically neglect the constant energy shift.
Surprisingly, even though the field quadratures represent continuous variables, the energy spectrum of the quantum harmonic oscillator is discrete, with possible energy levels $E_n = \omega(n + \frac12)$ with eigenstates $\ket{n}$ for all $n \in \mathbbm{N}$.
These states are called \emph{Fock states} and form an orthonormal basis.
Physically, we associate with $\ket{n}$ a state with $n$ bosonic excitations, which in the case of the electromagnetic field we call photons.
As the name implies, annihilation and creation operators subtract and add excitations, that is,
\begin{align}
    \label{eq:bosonic_fock_states}
    \hat{a}\ket{n} = \sqrt{n}\ket{n-1}, \quad \hat{a}^{\dagger}\ket{n}=\sqrt{n+1}\ket{n+1}, \quad \hat{a}^{\dagger}\hat{a}\ket{n} = \hat{n}\ket{n} = n \ket{n},
\end{align}
where we introduced the number operator $\hat{n}$.

\subsection{States of Quantum Harmonic Oscillators \label{ssec:states_qho}}

To gain intuition about the state space of a single bosonic mode, we now turn to specific examples of quantum states and their behavior in phase space.
We will use the Wigner function introduced in \cref{sec:wigner_function} to visualize some of those states.

\begin{figure}
    \centering
    \includegraphics{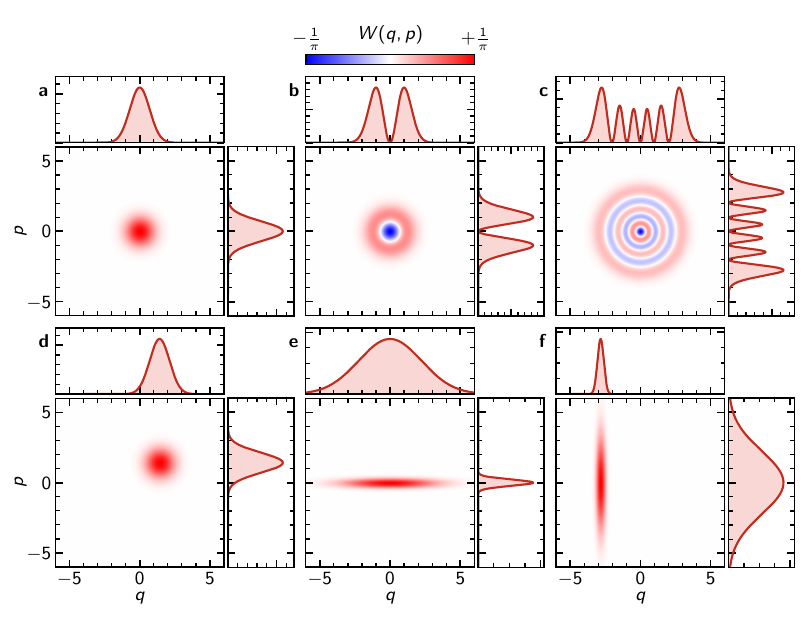}
    \caption{Wigner function $W(q, p)$ and marginals of various states of the quantum harmonic oscillator. 
    \textbf{a} The vacuum state $\ket{0}$,
    \textbf{b} Fock $\ket{1}$,
    \textbf{c} Fock $\ket{5}$,
    \textbf{d} Coherent $(\alpha = 1 + 1i)$,
    \textbf{e} Squeezed ($-10 \, \mathrm{dB}$),
    \textbf{f} Squeezed displaced ($10 \, \mathrm{dB}$, $\alpha = -2$).}
    \label{fig:states_wigner_functions}
\end{figure}

The ground state of the quantum harmonic oscillator in Eq.~\eqref{eq:qho_a_adag} is called the vacuum state, corresponding to the state $\ket{n = 0}$ in the Fock basis.
The Wigner function of the state corresponds to a rotation-symmetric Gaussian function in two dimensions that is centered at the origin of the phase space, see \sfigref{fig:states_wigner_functions}{a}.
The vacuum state saturates the Heisenberg uncertainty relations, that is, it minimizes the standard deviation in the position and momentum quadrature simultaneously.
As expected from the quadratic potential of the harmonic oscillator, Fock states with non-zero excitation number $n$ occupy a larger area in phase space and are not minimum uncertainty states.
Their Wigner function remains rotational symmetric but now displays oscillations that are also apparent in the marginals, see \sfigref{fig:states_wigner_functions}{b} and \sfigref{fig:states_wigner_functions}{c}.
These oscillations make those states highly non-classical.

A more classical-like family of states is formed by the \emph{coherent states}, which are eigenstates of the annihilation operator
\begin{align}
    \hat{a} \ket{\alpha} = \alpha \ket{\alpha}.
\end{align}
They are labeled by a complex amplitude \( \alpha \) and can be generated from the vacuum via the displacement operator
\begin{align}
    \ket{\alpha} = \hat{D}(\alpha) \ket{0}, \quad \hat{D}(\alpha) = \exp \left( \alpha \hat{a}^\dagger - \alpha^* \hat{a} \right).
\end{align}
Coherent states preserve the Gaussian shape of the vacuum in phase space but are displaced to the point \( (q = \sqrt{2}\mathrm{Re}(\alpha), p = \sqrt{2} \mathrm{Im}(\alpha)) \), see \sfigref{fig:states_wigner_functions}{d}.
They follow classical trajectories under free evolution and are often regarded as the ``most classical'' quantum states.
Importantly, they also form an overcomplete basis for the Hilbert space, satisfying the resolution of the identity
\begin{align}
    \frac{1}{\pi} \int \ketbra{\alpha} \dd^2\alpha = \mathbbm{1}.
\end{align}

A further generalization is provided by \emph{squeezed states}, which also have Gaussian Wigner functions but redistribute uncertainty between the \( \hat{q} \) and \( \hat{p} \) quadratures.
This squeezing can reduce noise below the vacuum level in one quadrature, at the expense of increased fluctuations in the conjugate variable, while still respecting the uncertainty principle.
The resulting Wigner function becomes elliptical rather than circular, see \sfigref{fig:states_wigner_functions}{e}.
Squeezed vacuum states are generated by applying a squeezing operator to the vacuum, as we will discuss in \cref{sec:cv_universal_gates}.
More generally, \emph{squeezed coherent states} combine displacement and squeezing to yield squeezed states centered away from the origin, see \sfigref{fig:states_wigner_functions}{f}, 
representing finite-energy approximations of position or momentum states dependent upon the squeezing direction.

\section{Noise Channels \label{sec:cv_noise_channels}}
If we want to use quantum continuous-variable systems for quantum information processing, we need to interact with the system in order to manipulate and control it.
However, if we are able to interact with the quantum system in a controllable way, then there must also be an environment that is interacting in an uncontrolled way with the system.
This uncontrolled interaction with an environment causes decoherence, as it does for qubit systems.
However, for continuous-variable quantum systems, due to their infinite-dimensional Hilbert space, there is a plethora of relevant decoherence mechanisms.
Under certain, typically fulfilled, assumptions, noise can be modeled through the Gorini-Kossakowski-Sudarshan-Lindblad master equation~\cite{gorini_completely_1976, lindblad_generators_1976} given by
\begin{align}
    \label{eq:def_lindlad_master_equation}
    \dv{t} \hat{\rho} = \mathcal{L}\hat{\rho} = - i \big[ \hat{H}, \hat{\rho}\big] + \sum_{j} \kappa_j \mathcal{D}[\hat{L}_j] \hat{\rho},
\end{align}
where $\mathcal{L}$ is known as the Lindbladian, and $\mathcal{D}[\hat{L}_j] \rho$ is the Lindblad dissipator acting on $\hat{\rho}$, that is $\mathcal{D}[\hat{L}_j]$ is a superoperator,  given by
\begin{align}
    \mathcal{D}[\hat{L}_j] \hat{\rho} = \hat{L}_j \hat{\rho} \hat{L}_j^{\dagger} - \frac12 \hat{L}_j^{\dagger} \hat{L}_j \hat{\rho} - \frac12 \hat{\rho} \hat{L}_j^{\dagger} \hat{L}_j.
\end{align}
The operators $\hat{L}_j$ are called jump operators which depend on the type of noise and $\kappa_j \geq 0$ are dissipation rates that characterize the strength of the noise.
Without loss of generality, we will typically assume that $\hat{H} = 0$ which can be achieved by an appropriate reference frame transformation.
In this case, we can also represent the noise channel written in the Kraus representation~\cite{ivan_operator-sum_2011, hellwig_operations_1970, sudarshan_stochastic_1961, choi_completely_1975, stinespring_positive_1955},
\begin{align}
    \label{eq:def_kraus_representation}
    \hat{\rho}(t) = \sum_k \hat{K}_k(t) \hat{\rho}(0) \hat{K}_k^{\dagger}(t), \quad \text{with } \sum_k \hat{K}_k^{\dagger}(t)\hat{K}_k(t)= \mathbbm{1},
\end{align}
which is obtained by formally integrating Eq.~\eqref{eq:def_lindlad_master_equation} up to time $t$, with $\hat{K}_k(t)$ the so-called Kraus operators.

In the following, we summarize some of the most common continuous-variable noise channels and describe intuitively their effect on quantum states.
We will usually not comment on their physical origin as this can depend on the physical system we are trying to model.
See also Refs.~\cite{scully_quantum_1997, gardiner_quantum_1991, breuer_theory_2002} for an in-depth introduction.

\paragraph{Photon Loss.}
One of the dominant noise channels in optics as well as superconducting circuits, is the loss of excitations, often referred to as photon loss in those architectures.
Losses are represented in the master equation by the Lindblad jump operator,
\begin{align}
    \hat{L}^{\mathrm{(loss)}} = \hat{a},
\end{align}
or within the Kraus representation, by the Kraus operator
\begin{align}
\label{eq:def_kraus_loss}
    \hat{K}^{\mathrm{(loss)}}_k = \frac{(1 - e^{- \kappa t})^{k / 2}}{\sqrt{k!}} e^{- \kappa \hat{n} t / 2} \hat{a}^{k}.
\end{align}
For the loss channel, the Kraus operator $\hat{K}^{\mathrm{(loss)}}_k$ describe the time evolution generated by the exact loss of $k$ photons, and $1 - \exp(-\kappa t)$ is the probability for losing a single photon.
Additionally to the factor $\hat{a}^{k}$ that maps Fock state $\ket{n}$ to Fock state $\ket{n-k}$, the other contribution is the Fock damping operator $e^{- \kappa \hat{n} t / 2}$ that leaves Fock states invariant.
Its effect is more readily observed by considering the Heisenberg evolution it generates.
Using the Baker-Campbell-Hausdorff (BCH) formula, one finds that $\hat{a} \to \hat{a} e^{- \kappa t / 2}$, contracting phase space towards the center, representing a continuous loss of energy from the system to the environment.

\paragraph{Photon Number Dephasing.}
While photon number dephasing is typically considered small in superconducting architectures, it becomes important if the system is nonlinear or is coupled to a nonlinear system, such as a two-level system.
The Lindblad jump operator is given by
\begin{align}
    \hat{L}^{\mathrm{(dephasing)}} =  \hat{a}^{\dagger} \hat{a} = \hat{n}
\end{align}
and the channel can equivalently be represented through the Kraus operators
\begin{align}
    \hat{K}^{\mathrm{(dephasing)}}_k = \frac{(\kappa_{\phi} t)^{k / 2}}{\sqrt{k!}} e^{- \kappa_{\phi} \hat{n}^2 t / 2} \hat{n}^k.
\end{align}
Notice that in this case the exponential is quadratic in $\hat{n}$ such that the BCH formula does not yield a closed-form expression as above.
While the discrete Kraus representation is sufficient to realize that the channel is energy-preserving, that is, it commutes with the harmonic oscillator Hamiltonian, additional intuition can be derived from the continuous Kraus representation of the channel given by
\begin{align}
    e^{\kappa_{\phi} \mathcal{D}[\hat{n}]t} \hat{\rho} = \int_{-\infty}^{\infty} p_{\kappa_\phi t}(\theta) e^{- i \theta \hat{n}} \hat{\rho} e^{i \theta \hat{n}} \dd{\theta}, \quad \text{with } p_{\kappa_\phi t}(\theta) = \frac{1}{\sqrt{2 \pi \kappa_{\phi} t}} \exp\Big(- \frac{{\theta^2}}{2 \kappa_{\phi} t}\Big)
\end{align}
from which dephasing can be interpreted as causing random rotations of the state with respect to an underlying Gaussian distribution of variance $\sigma^2 = \kappa_{\phi} t$.

\paragraph{Photon Gain.}
Photon gain can be considered the opposite of photon loss and is accordingly represented by the Lindblad jump operator
\begin{align}
    \hat{L}^{\mathrm{(gain)}} = \hat{a}^{\dagger},
\end{align}
and Kraus operators,
\begin{align}
        \label{eq:def_kraus_gain}
        \hat{K}^{\mathrm{(gain)}}_k = \frac{(e^{\kappa t} - 1)^{k / 2}}{ \sqrt{k!}} e^{- \kappa \hat{n} t / 2} \left(\hat{a}^{\dagger}\right)^{k}.
\end{align}
Using the BCH formula, one finds that $\hat{a} \to \hat{a} e^{\kappa t / 2}$, expanding phase space.
Typically, photon gain does not appear on its own and instead appears together with photon losses.

\paragraph{Thermal Noise.}
Thermal noise originates from the coupling of a system to a finite temperature environment and corresponds to a combination of loss and gain, requiring two Lindblad jump operators
\begin{align}
    \hat{L}_{-} = \sqrt{n_{\mathrm{th}} + 1} \hat{a}, \quad \hat{L}_{+} = \sqrt{n_{\mathrm{th}}} \hat{a}^{\dagger}, \quad \text{with } n_{\mathrm{th}} = \frac{1}{\exp(\hbar \omega / k_B T) - 1},
\end{align}
where we introduced the average thermal occupation number $n_{\mathrm{th}}$ and both jumps happen with rate $\kappa$.
The loss channel is recovered in the limit of $T \to 0$.
Since creation and annihilation operators, together with the identity, form a closed algebra, it is possible to express the Kraus operators of this channel as a product of the Kraus operators for loss and gain, see Ref.~\cite{ivan_operator-sum_2011} for details.

\paragraph{Gaussian displacements.}
The Gaussian displacement channel has Lindblad jump operators
\begin{align}
    \hat{L}_{1} = \hat{a}, \quad \hat{L}_2 = \hat{a}^{\dagger},
\end{align}
and thus is recovered from the thermal noise channel in the infinite temperature limit $T \to \infty$ and zero coupling limit $\kappa \to 0$ with $n_{\mathrm{th}} \kappa t / 2 = \sigma^2$.
Here, we introduced the variance $\sigma^2$ that appears in the continuous Kraus representation of the Gaussian displacement channel given by
\begin{align}
    \hat{\rho} \to \frac{1}{2 \pi \sigma^2} \int e^{- \lvert \alpha\rvert^2 / 2 \sigma^2} \hat{D}(\alpha) \hat{\rho} \hat{D}^{\dagger}(\alpha) \, \dd^2{\alpha},
\end{align}
giving the channel its name.
A somewhat uncommon way to write this channel is in the following form~\cite{gottesman_encoding_2001}
\begin{align}
    \dv{t} \hat{\rho} = - \frac{D}{2} \left[\hat{p}, \left[\hat{p}, \hat{\rho} \right] \right] - \frac{D}{2} \left[\hat{q}, \left[\hat{q}, \hat{\rho} \right] \right],
\end{align}
where $D = \kappa / 2 $ plays the role of a diffusion constant.
As a result, the channel can be interpreted as broadening features of the Wigner function and thus washing out rapidly oscillating parts of the phase space distribution.
We note that this channel is somewhat of a quantum analog of the additive white Gaussian noise channel discussed in \cref{sec:classical_noise_channels}.

\section{Universal Gates \label{sec:cv_universal_gates}}

A universal quantum computation generally consists of three steps: preparing an initial computational state, applying a sequence of universal operations, and finally measuring the system’s state.  
In this section, we define the notion of universality for \gls{cv} quantum computing and introduce a set of operations that enable universal quantum computation in this framework.

We have seen earlier that the concept of universality in discrete-variable (DV) quantum computing is well established: a finite set of single- and two-qubit gates can approximate any unitary operation to arbitrary accuracy.  
In the \gls{cv} setting, it was long believed that this notion does not directly carry over.  
The core difficulty lies in the fact that unitary transformations on an infinite-dimensional Hilbert space are generally characterized by infinitely many parameters and thus cannot be approximated using a finite set of elementary operations~\cite{lloyd_quantum_1999}.  
However, recent results~\cite{arzani_can_2025} have shown that any \emph{physical} single-mode unitary operation can, in fact, be approximated by a finite-degree polynomial $ P(\hat{q}, \hat{p}) $ in the quadrature operators, likely generalizable to the multimode case.
This insight opens the door to defining a meaningful notion of universality in \gls{cv} quantum computation based on polynomial Hamiltonians.

Therefore, one is interested in being able to synthesize arbitrary polynomials in the bosonic creation and annihilation operators.
To achieve this, one does not need to be able to implement any monomial $\hat{a}^{\dagger m}\hat{a}^{n}$, but instead can synthesize them from a few generating Hamiltonians $\{ \hat{H}_l \}$.
For that note, the two identities following from the Baker-Campbell-Hausdorff formula~\cite{braunstein_quantum_2005}
\begin{align}
    e^{-i \hat{A} \delta t} e^{-i \hat{B} \delta t} e^{i \hat{A} \delta t} e^{i \hat{B} \delta t}&=e^{[\hat{A}, \hat{B}] \delta t^{2}}+O\left(\delta t^{3}\right) \label{eq:Lie-group-comm}, \\
    e^{i \hat{A} \delta t / 2} e^{i \hat{B} \delta t / 2} e^{i \hat{B} \delta t / 2} e^{i \hat{A} \delta t / 2} &=e^{i(\hat{A}+\hat{B}) \delta t}+O\left(\delta t^{3}\right) \label{eq:bch-addition},
\end{align}
which are valid for two operators $\hat{A}$ and $\hat{B}$.
A necessary condition for the generating set $\{ \hat{H}_l \}$ is that it does not form a closed algebra with respect to the commutator.
One can easily convince oneself that this requires at least one of the generators $\{ \hat{H}_l \}$ to be higher than quadratic order in the bosonic operators.

In the following, we describe a set of operations that is sufficient for the above notion of universality in \gls{cv} quantum computing, see also Refs.~\cite{lloyd_quantum_1999, weedbrook_gaussian_2012, braunstein_quantum_2005}.

\paragraph{Rotation.} Arbitrary rotations in phase-space by an angle $\theta$ are obtained by the unitary
\begin{align}
    \label{eq:rotation-R}
    \hat{R}(\theta) = \mathrm{e}^{-i \theta \hat{a}^{\dagger} \hat{a}} = \mathrm{e}^{-i \theta(\hat{q}^2 + \hat{p}^2) / 2}.
\end{align}
The gate transforms the quadrature operators $\hat{q}$ and $\hat{p}$ as $\hat{q} \rightarrow \cos(\theta) \hat{q} + \sin(\theta) \hat{p}$ and $\hat{p} \rightarrow \cos(\theta) \hat{p} - \sin(\theta) \hat{q}$.
For the choice $\theta = \pi / 2$ the gate is termed the Fourier transform $\hat{F} = \hat{R}(\theta)$ since it takes the quadrature operators to its conjugate.
We remark that the Fourier transform $\hat{F}$ is the continuous-variable version of the Hadamard gate $H$.
\paragraph{Displacement.} We have already encountered the displacement operator as the operator that creates coherent states.
The displacement operator $\hat{D}(\alpha) = \exp(\alpha \hat{a}^{\dagger} - \alpha^{*} \hat{a})$ transforms the bosonic annihilation operator as $\hat{D}(\alpha) \hat{a} \hat{D}^{\dagger}(\alpha) = \hat{a} + \alpha$.
\paragraph{Squeezing.} 
The squeezing operation with real squeezing factor $s$ is given by
\begin{align}
    \label{eq:squeezing-xp}
    \hat{S}(s) = \mathrm{e}^{- i \log(s) (\hat{q} \hat{p} + \hat{p} \hat{q}) / 2}.
\end{align}
This corresponds to an operation with $\frac{10}{\log(10)}\log(s^2)\,\mathrm{dB}$ of squeezing.
The action of $\hat{S}(s)$ on the quadrature operators is given by
$\hat{q} \rightarrow s \hat{q}$ and
$\hat{p} \rightarrow s^{-1} \hat{p}$,
i.e., the operator squeezes one quadrature and stretches the conjugate one.
The operator is also commonly expressed in terms of creation and annihilation operators as
\begin{align}
    \label{eq:squeezing}
    \hat{S}(\xi) = \mathrm{e}^{\frac{\xi^{*}}{2} \hat{a}^2 - \frac{\xi}{2} \hat{a}^{\dagger 2} },
\end{align}
with complex squeezing parameter $\xi = r \mathrm{e}^{i \theta}$.
The squeezing operator implements a Bogoliubov transformation given by $\hat{a} \rightarrow \hat{a} \cosh r - e^{i \theta} \hat{a}^{\dagger} \sinh r$
from which the transformation Eq.~\eqref{eq:squeezing-xp} is obtained for $r = \lvert \log s \rvert$ and $\theta = \pi$ if $s > 1$ and $\theta = 0$ if $s < 1$.
\paragraph{Beam Splitter.} The beam splitter is usually given in terms of annihilation and creation operators of two modes $a$ and $b$ as
\begin{align}
    \label{eq:beam-splitter}
    \hat{B}(\theta) = \mathrm{e}^{\theta (\hat{a} \hat{b}^{\dagger} - \hat{a}^{\dagger} \hat{b})},
\end{align}
and we omit the representation in terms of quadrature operators.
The action of the beam splitter on the annihilation operators is given by $\hat{a} \rightarrow \sqrt{\tau} \hat{a} + \sqrt{1 - \tau} \hat{b}$ and $\hat{b} \rightarrow \sqrt{\tau} \hat{b} - \sqrt{1 - \tau} \hat{a}$
where $\tau = \cos^2 \theta$ is known as the transmissivity of the beam splitter in quantum optics.

\paragraph{Cubic Phase Gate.} The cubic phase gate is defined as~\cite{marek_deterministic_2011, yukawa_emulating_2013, arzani_polynomial_2017, miyata_implementation_2016, yanagimoto_engineering_2020, weedbrook_gaussian_2012},
\begin{align}
    \label{eq:cubic-phase-gate}
    \hat{\Gamma}(\gamma) = \exp(i \gamma \hat{q}^3),
\end{align}
and is commonly used in quantum information protocols as the necessary nonlinear gate.
The coefficient $\gamma$ is known as the cubicity of the gate. Any non-zero value for the cubicity is in principle sufficient as it can be enhanced by a squeezing transformation, i.e, 
\begin{align}
    \label{eq:squeezing-cubic-phase-gate}
    \hat{\Gamma}(\gamma^{\prime}) = \hat{S}^{\dagger}(-r) \hat{\Gamma}(\gamma) \hat{S}(-r),
\end{align}
with $\gamma^{\prime} = \gamma \mathrm{e}^{3 r}$ and $r$ the real squeezing parameter.

\section{Bosonic Quantum Error Correction}
Bosonic quantum error correction is a field in which one is interested in achieving robust quantum information processing by encoding quantum information into a subspace of the Hilbert space of one or multiple harmonic oscillators.
The basis of this idea is that instead of using many qubits to provide the redundancy required to protect the encoded information, one can directly benefit from the vastness of the harmonic oscillator Hilbert space.
While one can approach this field from an information-theoretic approach, here we are more interested in a practical approach that encodes information in a non-local way in phase space and in this way achieves protection against physical noise channels such as the ones previously described, e.g., photon loss, but also undesired nonlinearities that appear as natural consequences of the hardware platform used to realize the encoding. 

While thinking about bosonic codes, it is important to consider that those are not simply generalizations of two-dimensional qubit systems to $d$-dimensional qudit systems in an appropriate limit.
The reason that this is not the case is that while qudit Pauli operators are a natural object to study from the viewpoint of stabilizer theory, these are not naturally realized in physical realizations of bosonic systems.
What is realized, though, are products of creation and annihilation operators, $\hat{a}^{\dagger}$ and $\hat{a}$.
Even though this seems to be a major complication, it necessitates thinking more closely about the physical constraints and leads to a tighter co-design of quantum systems and quantum error correction codes.
Thinking outside the usual stabilizer theory framework leads to exploring things that have not been considered or are strictly forbidden by no-go theorems.
Particular examples are the possibility of a bias-preserving CNOT gate~\cite{guillaud_repetition_2019} or even a continuous-parameter set of transversal gates~\cite{faist_continuous_2020}, both of which are possible due to the continuous-variable nature of the system and cannot be achieved in finite-dimensional systems~\cite{eastin_restrictions_2009, guillaud_repetition_2019}.

In the following, we will focus on two prominent examples of bosonic quantum error-correcting codes encoded in a single harmonic oscillator.
Even though such an encoding possibly accesses an infinite-dimensional Hilbert space, these bosonic codes do not have an error threshold~\cite{hanggli_oscillator--oscillator_2022} that allows for arbitrarily low logical error rates --- in contrast to certain families of stabilizer codes.
As a result, they are often used as inner codes concatenated with an outer stabilizer code to achieve fault tolerance.
Furthermore, depending on the noise channel, implementing the recovery unitary to restore the noisy state to the code space can be highly complicated.
Instead, one of the two alternative approaches is typically considered: engineering the dynamics of the system such that the noisy states get autonomously restored to the code space, or employing the measurement-based paradigm of quantum error correction such that physical errors cannot spread throughout the computation.

\subsection{The cat code \label{sec:cat_code}}
\begin{figure}[!hbt]
    \centering
    \includegraphics{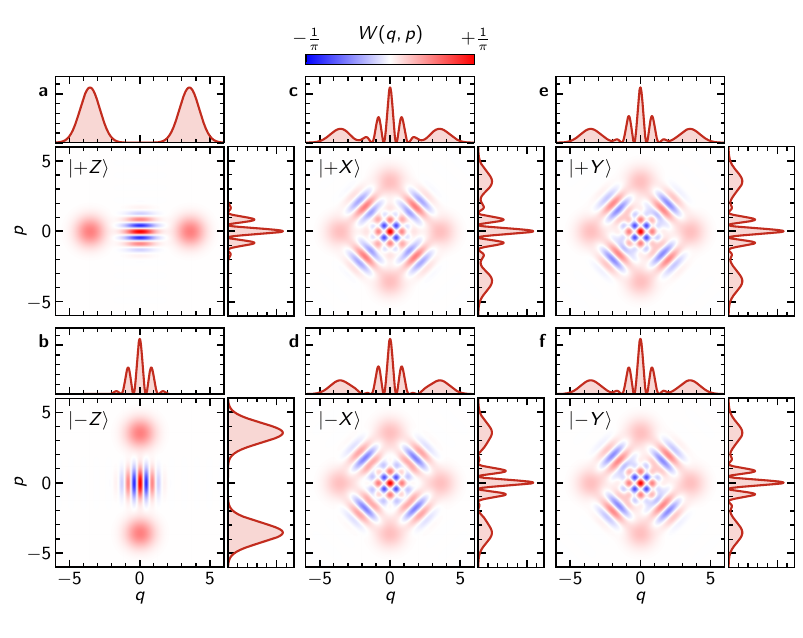}
    \caption{Wigner functions $W(q,p)$ and marginal distributions of the four-component cat code, see Eqs.~\eqref{eq:def_four-component_cat_code_zero} and~\eqref{eq:def_four-component_cat_code_one}, for the six different cardinal states given by, up to normalization, 
    \textbf{a} $\ket{+Z} = \ket{0}$, \textbf{b}  $\ket{-Z} = \ket{1}$,  
    \textbf{c} $\ket{+X} = \ket{0} + \ket{1}$, \textbf{d}  $\ket{-X} = \ket{0} - \ket{1}$,
     \textbf{e} $\ket{+Y} = \ket{0} + i \ket{1}$, \textbf{f}  $\ket{-Y} = \ket{0} - i \ket{1}$.
     The amplitude of the coherent states is $\alpha = 2.5$.
     }
    \label{fig:four_component_cat_wigner}
\end{figure}

The cat code, as we are thinking about it today, most likely originated from the work of Leghtas \emph{et al.}~\cite{leghtas_hardware-efficient_2013, leghtas_confining_2015} that proposed the bosonic cat encoding designed to protect against single photon losses.
That is, the computational basis states are defined as
\begin{align}
    \ket{0}_L &= \frac{\ket{\alpha} + \ket{- \alpha}}{\sqrt{2 (1 + e^{-2\lvert \alpha \rvert^2})}} = \frac{1}{\cosh(\alpha^2)} \sum_{n=0}^{\infty} \frac{\alpha^{2n}}{\sqrt{(2n)!}} \ket{2n},     \label{eq:def_four-component_cat_code_zero} \\
     \ket{1}_L &= \frac{\ket{i \alpha} + \ket{- i \alpha}}{\sqrt{2 (1 + e^{-2\lvert \alpha \rvert^2})}} = \frac{1}{\cosh(\alpha^2)} \sum_{n=0}^{\infty} \frac{(-1)^n \alpha^{2n}}{\sqrt{(2n)!}} \ket{2n}.     \label{eq:def_four-component_cat_code_one}
\end{align}
Notice that both computational basis states are supported on even Fock states only.
Additionally, a single photon loss would take us from the subspace of even Fock states to the subspace of odd Fock states, the error space.
While we cannot perform a measurement in the Fock basis without collapsing the state, it is possible to perform a syndrome measurement that only distinguishes even and odd Fock states.
This measurement is known as a photon number parity measurement and allows one to infer whether a photon has been lost without collapsing the states.
Intuitively, the photon number parity measurement is determined by the sign of the Wigner function at the origin of the phase space, see \Cref{fig:four_component_cat_wigner}.

It should be emphasized that it is not possible to correct the loss of two photons.
Indeed, the operator $\hat{a}^2$ acts like a Pauli $Z$ operator on the logical subspace, that is, $\hat{a}^{2} (\ket{0}_L \pm \ket{1}_L) \propto (\ket{0}_L \mp \ket{1}_L)$ as the coherent states are eigenstates to the annihilation operator.
Using our intuition that the parity measurement is determined by the sign of the Wigner function at the origin, we can also convince ourselves that the cat code protects against dephasing errors.
However, the dephasing rate should be small enough such that the computational states $\ket{0}_L$ and $\ket{1}_L$ are still distinguishable, as they are related by a $\pi / 4$ rotation in phase space.

\begin{figure}[!hbt]
    \centering
    \includegraphics[width=\textwidth]{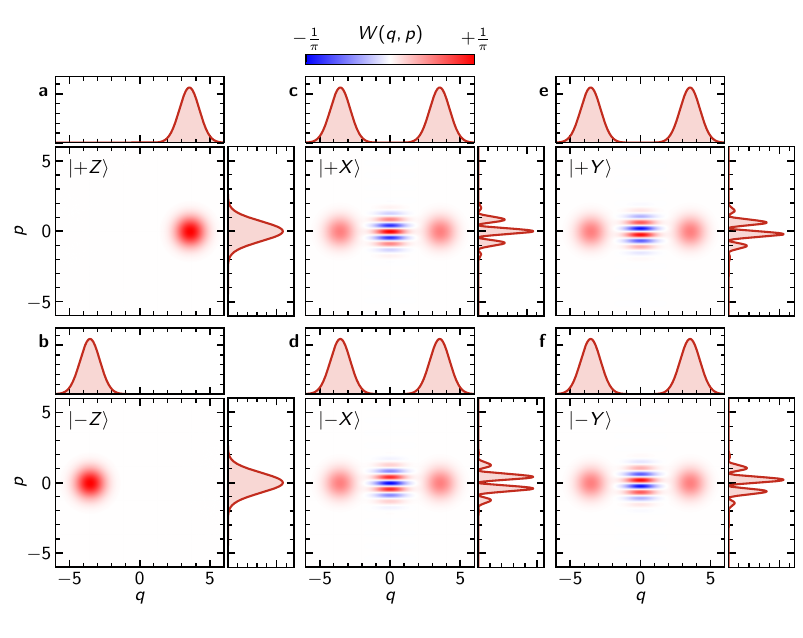}
    \caption{Wigner functions $W(q,p)$ and marginal distributions of the ordinary cat code, defined by the dual basis code states in Eq.~\eqref{eq:def_cat_code}, for the six different cardinal states given by, up to normalization, 
    \textbf{a} $\ket{+Z} = \ket{0}$, \textbf{b}  $\ket{-Z} = \ket{1}$,  
    \textbf{c} $\ket{+X} = \ket{0} + \ket{1}$, \textbf{d}  $\ket{-X} = \ket{0} - \ket{1}$,
     \textbf{e} $\ket{+Y} = \ket{0} + i \ket{1}$, \textbf{f}  $\ket{-Y} = \ket{0} - i \ket{1}$.
     The amplitude of the coherent states is $\alpha = 2.5$
     }
    \label{fig:ordinary_cat_wigner}
\end{figure}

More recently, the above bosonic quantum error correction code is more commonly referred to as the four-component cat code, while the encoding of the ``cat code'' is more likely to refer to a simpler encoding~\cite{lund_fault-tolerant_2008, jeong_efficient_2002, ralph_quantum_2003} defined via the dual basis states
\begin{align}
    \label{eq:def_cat_code}
    \ket{\pm}_{\mathrm{cat}} &=  \frac{\ket{\alpha} \pm \ket{- \alpha}}{\sqrt{2 (1 \pm e^{-2\lvert \alpha \rvert^2})}}, \\
    \ket{0}_{\mathrm{cat}} &= \frac{\ket{+}_{\mathrm{cat}} + \ket{-}_{\mathrm{cat}}}{\sqrt{2}} \approx \ket{\alpha} + O(e^{-2\lvert \alpha \rvert^2}), \\
    \ket{1}_{\mathrm{cat}} &= \frac{\ket{+}_{\mathrm{cat}} - \ket{-}_{\mathrm{cat}}}{\sqrt{2}} \approx \ket{- \alpha} + O(e^{-2\lvert \alpha \rvert^2}),
\end{align}
which is unable to correct a single photon loss, as in this case, the annihilation operator acts like a Pauli $Z$ operator on the logical subspace.
The Wigner functions of the six cardinal states of the encoding are shown in \figref{fig:ordinary_cat_wigner}.
Instead, interest in this encoding is due to Mirrahimi \emph{et al.}~\cite{mirrahimi_dynamically_2014}, Puri \emph{et al.}~\cite{puri_engineering_2017}, and related work which introduced a scheme that dynamically protects cat qubits against photon number dephasing errors, introducing stabilized cat qubits.
Here, we will sometimes refer to these ideas as the confinement of cat qubits to distinguish more clearly from stabilizer codes in the discrete-variable setting.

To describe the idea of confinement, note that the code space of a bosonic encoding is typically not stable under physical noise channels.
Thus, any logical information will eventually leak outside of the code space.
However, it is possible to stabilize the code space through appropriate confinement schemes.
To this end, one aims to engineer effective interactions whose dynamics yield a degenerate ground space isomorphic to the code space.
To achieve this, recall that the dynamics of a quantum system are either described by the Schrödinger equation, $\partial_t \ket{\psi} = - i \hat{H}(t) \ket{\psi}$, for the evolution of a closed system, or by the master equation, $\partial_t \hat{\rho} = \mathcal{L}(t) \hat{\rho}$, for an open system.
If we can find $\hat{H}$ or $\mathcal{L}$ such that the code states are eigenstates of these (super)operators with vanishing eigenvalues, then the code space is a fixed point of the evolution.
Also, the dissipative evolution generated by $\mathcal{L}$ ensures that if leakage occurs, the state will relax back to the code space.

Let us make these ideas more explicit based on the example of the cat code.
Notice that the coherent states $\ket{\pm \alpha}$ are eigenstates of the operator $\hat{F} = (\hat{a}^2 - \alpha^2)$  with eigenvalue zero such that, by linearity, $\ket{\pm}_{\mathrm{cat}}$ are eigenstates of $\hat{F}$ as well.
Then, the dissipative evolution generated by $\mathcal{D}[\hat{F}]$ and the Hamiltonian evolution generated by $\hat{H}_F = \frac{1}{2} \hat{F}^{\dagger} \hat{F}$ both continuously stabilize the code space~\cite{albert_lindbladians_2018}.
From the Knill-Laflamme conditions~\cite{knill_theory_1997}, one finds that for single photon losses, stabilized cat codes allow for an arbitrary suppression of the logical bit-flip rate $\Gamma_X$, however, at the cost of an increasing phase-flip error rate $\Gamma_Z$, that is,
\begin{align}
    \Gamma_X &\propto \abs{\mel{-\alpha}{\hat{a}}{\alpha}}^2 = \lvert \alpha \rvert^2 e^{-2\lvert \alpha\rvert^2}, \\
    \Gamma_Z &\propto \lvert \mel{+_{\mathrm{cat}}}{\hat{a}}{-_{\mathrm{cat}}} \rvert^2 = \lvert \alpha\rvert^2 \tanh(\lvert \alpha \rvert^2) \stackrel{\lvert \alpha \rvert^2 \to \infty}\sim \lvert \alpha \rvert^2.
\end{align}
We refer the interested reader to Refs.~\cite{mirrahimi_cat-qubits_2016, dubovitskii_bit-flip_2025, puri_bias-preserving_2020} for a more detailed discussion on error suppression properties of the cat qubit.

The equations above highlight a key advantage of the simpler cat qubit encoding: for sufficiently large coherent state amplitudes ($\alpha$), the bit-flip rate of the encoded qubit can be suppressed to the point of becoming practically negligible.
Consequently, the problem of correcting errors on the qubit is reduced to that of a classical bit, significantly simplifying the approach. 
This leaves only one type of error to correct, almost turning the quantum error correction challenge into a classical one and making the overall process of building a fault-tolerant quantum computer much more manageable.

We conclude this section by taking a step back and emphasizing that both the four-component cat code introduced in Eqs.~\eqref{eq:def_four-component_cat_code_zero}-\eqref{eq:def_four-component_cat_code_one} and the ordinary cat code introduced in Eq.~\eqref{eq:def_cat_code} correspond to coherent states arranged on a circle of radius $\lvert \alpha \rvert$ in phase space, with code states that are invariant under $\pi$ and $2 \pi$ rotations, respectively.
This concept was further generalized by Grimsmo, Combes, and Baragiola~\cite{grimsmo_quantum_2020}, who consider rotation-symmetric bosonic (RSB) codes composed as superpositions of arbitrary states and not just coherent states.
This framework also encompasses other well-known codes such as binomial codes~\cite{michael_new_2016}.
Using this construction, it is possible to design a code that is protected against the loss of $N-1$ photons by considering dual basis code states $(\ket{\pm})$ that possess an $N$-fold discrete rotation symmetry in phase space, that is, they are invariant under discrete rotations by an angle $2 \pi /N$.

\subsection{The Gottesman-Kitaev-Preskill code}
While cat qubits cannot be viewed as stabilizer codes in the conventional sense, the bosonic encoding proposed by Gottesman, Kitaev, and Preskill (GKP) in 2001~\cite{gottesman_encoding_2001} can be viewed as a stabilizer code.
Indeed, GKP codes can be formally derived as the $d \to \infty$ limit of $d$-dimensional qudit stabilizer codes.
Restricting to a single bosonic GKP qubit encoded in a single mode of the harmonic oscillator, the logical Pauli $X$ and $Z$ operators are given by
\begin{align}
    \label{eq:def_gkp_logicals}
    X &= \exp(\alpha {\hat {a}}^{\dagger }-\alpha ^{\ast }{\hat {a}}) = \hat{D}(\alpha), \\
    Z &= \exp(\beta {\hat {a}}^{\dagger }-\beta ^{\ast }{\hat {a}}) = \hat{D}(\beta),
\end{align}
where the complex displacements $\alpha$ and $\beta$ need to fulfill
\begin{align}
    \beta \alpha^{*} - \beta^{*}\alpha = i \pi,
\end{align}
such that $X$ and $Z$ commute up to a phase, that is, they reproduce the commutation relations of the Pauli operators, $X Z = - Z X$.
Since $X^2$ should act like the identity on the code space, one can define the stabilizer generators of the GKP encoding as $S_X = X^2 = \hat{D}(2\alpha)$ and $S_Z = Z^2 = \hat{D}(2\beta)$.
As the stabilizer generators are displacement operators with a particular amplitude, code words require a translation symmetry in phase space, which is determined by the choice of $\alpha$ and $\beta$ that can be seen as the generators of a lattice on which the states have support.
For brevity, we will restrict to the simplest and one of the most common choices, that is, the square lattice with $\alpha = \sqrt{\pi / 2}$ and $\beta = i \alpha = i \sqrt{\pi / 2}$.
For this choice, logical operators and stabilizer generators reduce to
\begin{align}
    X = e^{- i \sqrt{\pi}\hat{p}}, \quad Z = e^{i\sqrt{\pi}\hat{q}}, \quad \hat{S}_X  = e^{- i 2\sqrt{\pi}\hat{p}}, \quad \hat{S}_Z = e^{i2\sqrt{\pi}\hat{q}}.
\end{align}

Then, the ideal, infinite energy computational states are infinite trains of position eigenstates, that is,
\begin{align}
    \label{eq:def_ideal_gkp_computational_states}
    \ket{\overline{\mu}_{\mathrm{GKP}}} = \sum_{n=-\infty}^{\infty} \ket{q = (2n + \mu) \sqrt{\pi}},
\end{align}
with $\mu \in \lbrace 0, 1 \rbrace$.
Measuring the $S_X$ stabilizer is equivalent to measuring non-destructively the $q$ quadrature modulo $\sqrt{\pi}$ and reveals any displacement error along the $q$ quadrature.
If the shift is less than $\sqrt{\pi}/ 2$, applying a displacement with the same magnitude and opposite sign restores the state into the code space without applying a logical operation.
The same applies to shifts in the $p$ quadrature.

\begin{figure}[!hbt]
    \centering
    \includegraphics[width=\textwidth]{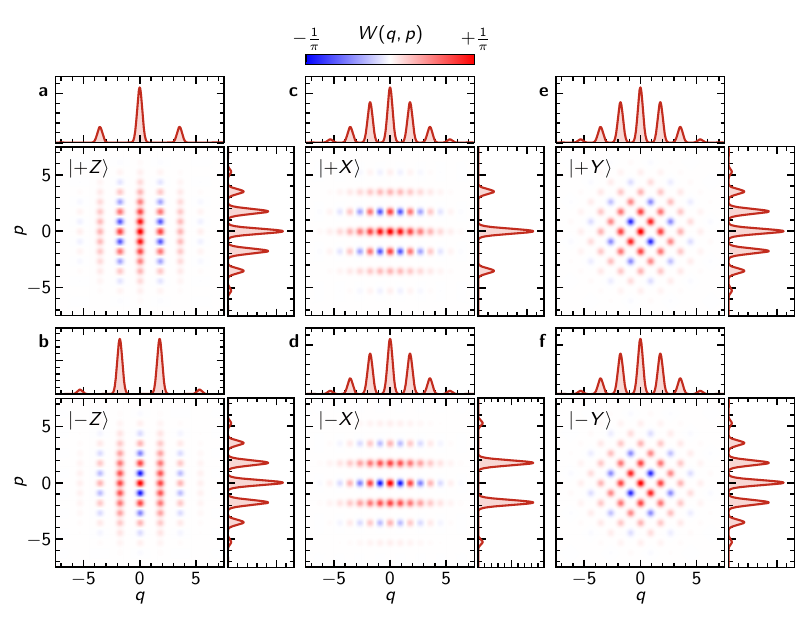}
    \caption{Wigner functions $W(q,p)$ and marginal distributions of the finite-energy Gottesman-Kitaev-Preskill code, see Eq.~\eqref{eq:logical-gkp-physical}, for the six different cardinal states given by, up to normalization, 
    \textbf{a} $\ket{+Z} = \ket{0}_{\Delta}$, \textbf{b}  $\ket{-Z} = \ket{1}_{\Delta}$,  
    \textbf{c} $\ket{+X} = \ket{0}_{\Delta} + \ket{1_{\Delta}}$, \textbf{d}  $\ket{-X} = \ket{0}_{\Delta} - \ket{1}_{\Delta}$,
     \textbf{e} $\ket{+Y} = \ket{0}_{\Delta} + i \ket{1}_{\Delta}$, \textbf{f}  $\ket{-Y} = \ket{0}_{\Delta} - i \ket{1}_{\Delta}$.
     The regularization parameter is chosen as $\Delta = 0.25$.
     }
    \label{fig:finite_energy_wigner}
\end{figure}

The ideal code states of the GKP code are unphysical and non-normalizable.
Nevertheless, it is possible to work with approximate code states for which high-energy contributions are exponentially suppressed.
The finitely squeezed code states can be expressed in various equivalent ways~\cite{matsuura_equivalence_2020}, for example, in terms of a weighted sum of squeezed coherent states as~\cite{albert_performance_2018}
\begin{align}
    \label{eq:logical-gkp-physical}
    \ket{\mu_{\Delta}} \propto \sum_{n = - \infty}^{\infty} \mathrm{e}^{- \frac{\pi}{2} \Delta^2 \left(2 n + \mu \right)^2} \hat{D}\left( \sqrt{\pi / 2} \left(2 n + \mu \right) \right) \hat{S}\left(- \ln \Delta) \right) \ket{0},
\end{align}
where $\Delta \in \left[0, 1 \right]$ and we have treated both quadratures symmetrically and point to Ref.~\cite{gottesman_encoding_2001} for the general case.
We show the six cardinal states of this encoding in \figref{fig:finite_energy_wigner}.
We mention that an alternative regularization can be achieved by applying the Fock damping operator to ideal code words, that is, $\ket{\mu_{\delta}} \propto \exp(-\delta^2 \hat{n})\ket{\mu_{\mathrm{GKP}}}$ and $ \ket{\mu_{\delta}} \approx \ket{\mu_{\Delta}}$ for $\delta \approx \Delta$ and $\delta, \Delta \ll 1$ such that it is possible to use both views for intuitive explanations. 
Both of those representations yield a pure state. 
A representation of finite energy states that yields a mixed state is known as state twirling~\cite{noh_fault-tolerant_2020} and can be viewed as sending the state through the Gaussian displacement channel with a standard deviation that corresponds to the per-peak squeezing.

A relevant trait of the ideal GKP code is that all Clifford operations --- and therefore also the syndrome extraction circuit --- can be performed by Gaussian operations, that is, unitary operations that have a generating Hamiltonian which is at most quadratic in the quadrature operators.
This makes this encoding particularly relevant for optical platforms where those operations are considered easily implementable. 
However, the Clifford operations of the finite-energy GKP code are not Gaussian anymore, indeed, they are non-unitary as they are obtained from the conjugation of the ideal operation with the non-unitary damping operator.
Applying the ideal operations to the finite energy states thus yields additional errors.
This makes the GKP encoding especially suited for measurement-based quantum computing, where the accumulation of errors is limited due to low-depth circuits followed by measurements~\cite{menicucci_universal_2006, menicucci_fault-tolerant_2014, jafarzadeh_logical_2025}.

When Gottesman, Kitaev, and Preskill proposed encoding a qubit into an oscillator 25 years ago, some considered their proposal a purely theoretical endeavor, due to the encoding being highly non-classical and therefore \emph{beyond impossible}~\cite{byron_bay_quantum_computing_workshop_gkp_2020} to realize experimentally.
Fast forward 25 years, and preparing GKP states is yet another example of why we should think twice before calling something ``impossible''  --- ``difficult'' is usually closer to the truth.
Today, GKP states have been stabilized in motional states of trapped ions~\cite{fluhmann_encoding_2019}, superconducting circuits~\cite{campagne-ibarcq_quantum_2020, sivak_real-time_2023, brock_quantum_2024}, and a universal gate set has been implemented in trapped ions as well~\cite{matsos_universal_2024}.
Even on the optical side, significant progress has been made, and low-quality states have been prepared~\cite{konno_logical_2024}.

\section{Conclusion and Outlook}
In this chapter, we reviewed the framework for describing quantum information encoded into bosonic modes.
Starting from the basic structure of states, observables, and transformations for a quantum harmonic oscillator, we established a formalism for quantum information processing with continuous variables.

Gaussian operations, for example, unitary evolution generated by a quadratic Hamiltonian, have an important role in continuous-variable quantum computing.
They play an analogous role to Clifford gates in the discrete-variable setting in the sense that they are efficiently simulatable with classical computers, assuming Gaussian input states such as squeezed coherent states.
Similar to their discrete-variable counterparts, these operations are not enough to prepare arbitrary quantum states, requiring, in this case, the addition of a non-Gaussian operation such as the cubic phase gate.

However, continuous-variable quantum systems are subject to noise, as any other quantum system.
While analog stabilizer codes exist~\cite{lloyd_analog_1998, albert_bosonic_2022}, this approach has not been considered extensively due to several complications, such as unphysical code states and a no-go theorem for protecting Gaussian states against Gaussian errors in quantum communication protocols~\cite{niset_no-go_2009}.
In practice, one therefore mostly considers discrete encodings of quantum information into quantum continuous variables, for example.
\refpaper{IV} investigates the performance of two prominent classes of bosonic codes within the measurement-based paradigm of quantum computing.
The work particularly focuses on the role of imperfect or noisy measurements when measuring the bosonic modes.
To this end, numerically exact simulations of rotation-symmetric bosonic codes and Gottesman-Kitaev-Preskill codes under realistic measurement models are performed, revealing vulnerabilities and requirements for this approach to be viable with near-term devices.

Single-mode bosonic codes alone will not suffice to achieve fault-tolerant quantum computing, and they must therefore be concatenated with discrete-variable stabilizer codes described in the previous chapter.
To bridge bosonic and discrete-variable codes, \refpaper{VI} proposes decoding strategies that explicitly exploit analog syndrome information available from bosonic qubit readout. 
These techniques apply to general concatenated architectures and reduce the need for repeated measurements, offering a promising route toward efficient fault-tolerant schemes based on those concatenated encodings.

Additionally, \refpaper{V} introduces the dissipatively stabilized squeezed cat qubit, a nonlocal encoding in phase space based on squeezed coherent states.
This approach significantly enhances the error suppression properties of the dissipative cat qubit by deforming its basis states through a squeezing transformation.
Importantly, for superconducting circuits, the same physical device that dissipatively stabilizes cat qubits can, in principle, be used to stabilize squeezed cat qubits, only requiring two additional drive tones supplied through the same drive line.
Indeed, in Ref.~\cite{rousseau_enhancing_2025}, the experimental stabilization of squeezed cat qubits and ordinary cat qubits on the same device was demonstrated, following the proposal of \refpaper{V}.

The following chapter of this thesis explores how one can utilize superconducting circuits to engineer almost arbitrary interactions and, in this way, realize quantum continuous variable systems in hardware.

\cleardoublepage

\chapter{Quantum Computing Architectures \label{chap:quantum_computing_architectures}}

\glsresetall
\section{Superconducting Circuit Architectures}
The field of \gls{cqed} emerged after the discovery of the Josephson effect~\cite{josephson_possible_1962} as a way to explore quantum effects in macroscopic systems~\cite{clarke_quantum_1988}, including the observation of quantum tunneling~\cite{devoret_measurements_1985} and the measurement of discrete energy levels~\cite{martinis_energy-level_1985}.
Later, the observation of coherent oscillations in a superconducting qubit~\cite{nakamura_coherent_1999} showcased the potential of superconducting quantum circuits as a platform for quantum information processing~\cite{divincenzo_fault-tolerant_2009}.
Today, superconducting quantum circuits provide a highly flexible and controllable platform for quantum information processing. 
Their design leverages a toolbox of fundamental circuit elements, which, when combined strategically, allow for precise control over quantum states and interactions. 
These circuits enable the realization of artificial atoms~\cite{nakamura_coherent_1999, martinis_rabi_2002, mooij_josephson_1999, pop_coherent_2014, barends_coherent_2013} with tunable energy levels, strong nonlinearities, and tailored couplings to their environment.
Achieving high-fidelity quantum operations demands both a deep understanding of the approximations underlying effective circuit models and the development of analytical and numerical techniques to systematically obtain better approximations.
Ultimately, the goal is for these refined methods to enable the engineering of quantum operations with significantly lower physical error rates.

In what follows, we begin by reviewing the Hamiltonian formulation for superconducting quantum circuits, setting the stage for a deeper understanding of these systems. 
Following this, we offer an intuitive approach to engineering effective interactions, both within the system and with its environment.
To this end, we also propose a method to systematically account for higher-order effects accurately.

\subsection{Lumped-element circuit diagrams}
\begin{figure}[!ht]
    \centering
\includegraphics{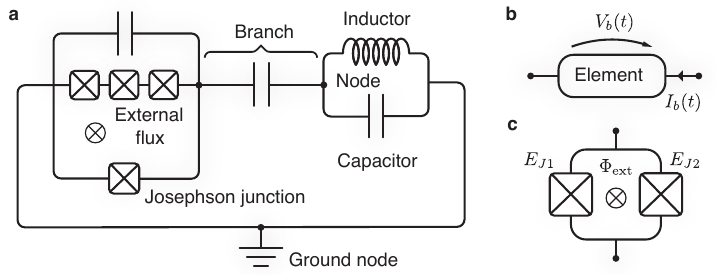}
    \caption{\textbf{a} Example of a lumped-element circuit consisting of a loop of Josephson junctions and a capacitor in parallel connected by a capacitor to another subcircuit formed by a linear inductor and a capacitor. %
    \textbf{b} An arbitrary two-terminal branch element $b$ with two nodes. The branch element is characterized by a voltage $V_b(t)$ across it and a current $I_b(t)$ through it, in the opposite direction.
    \textbf{c} A dc-SQUID, consisting of two Josephson junction with energies $E_{J1}$ and $E_{J2}$ in parallel, forming a loop. An external flux $\Phi_{\mathrm{ext}}$ is threading the loop.
    }
    \label{fig:example_lumped_element circuit}
\end{figure}

We briefly review the Hamiltonian formulation of superconducting circuits, see, for example, Refs.~\cite{blais_quantum_2020, krantz_quantum_2019, rasmussen_superconducting_2021} for more details.
In this approach, we describe an electrical circuit in terms of a graph consisting of branches that represent two-terminal lumped circuit elements.
Furthermore, the properties of the lowest frequency modes of distributed-element systems can also be described within this formalism if they are represented as lumped-element circuits.
An example of a lumped-element circuit is shown in \sfigref{fig:example_lumped_element circuit}{a} and we will discuss its different components in more detail in \cref{ssec:cqed_circuit_components}.
To derive the equations of motion of the electrical circuit, we can construct its Lagrangian in terms of the energy associated with each circuit component.

In general, each branch element $b$ of a circuit is characterized by a voltage $V_b(t)$ across it and a current $I_b(t)$ through it, see \sfigref{fig:example_lumped_element circuit}{b}, which are defined in terms of the electromagnetic fields.
The total energy stored in a branch element $b$ is obtained by integrating the power $V_b(t) I_b(t)$ over time leading to
\begin{align}
    \label{eq:branch-energy}
    \mathcal{E}_{b}(t)=\int_{-\infty}^{t} V_{b}\left(t^{\prime}\right) I_{b}\left(t^{\prime}\right) d t^{\prime}.
\end{align}

We also define the branch charge $Q_b(t)$ and the branch flux $\Phi_b(t)$ variables as
\begin{align}
    \Phi_{b}(t)=\int_{-\infty}^{t} V_{b}\left(t^{\prime}\right) \dd{t^{\prime}}, \label{eq:brach-flux} \quad
    Q_{b}(t)=\int_{-\infty}^{t} I_{b}\left(t^{\prime}\right) \dd{t^{\prime}} %
\end{align}
where we assumed that at $t^{\prime} = -\infty$ the voltage and current are both equal to zero.
The branch variables are not completely independent but related through Kirchhoff's laws,
\begin{align}
    \label{eq:def_kirchoffs_laws}
    \sum_{\text{all $b$ incident to $n$}} Q_b = q_n, \quad \quad
    \sum_{\text{all $b$ around $l$}} \Phi_b = \tilde{\Phi}_{\ell},
\end{align}
such that the number of degrees of freedom will always be less than the number of branch elements because the directed sum of the voltages must be zero.
Here, $q_n$ is the charge at node $n$ and $\tilde{\Phi}_{\ell}$ is the electromagnetic flux through the loop $\ell$.
The terminology of nodes and branches already hints at the fact that a natural language for analyzing circuits is graph theory.

Whether we want to choose a formulation in terms of branch fluxes $\Phi_b(t)$ or branch charges $Q_b(t)$ depends on the type of elements the circuit is composed of. 
In particular, this choice is determined by the element-dependent constitutive laws that relate $I_b(t)$ and $V_b(t)$.
For a capacitive element, the constitutive relation has the form
\begin{align}
    \label{eq:constitutive-relation-capacitance}
    V_b(t) = f_b(Q_b(t)),
\end{align}
where in the case of an ideal capacitor $f_b(Q_b(t)) = Q_b(t) / C$ is a linear function.
Similarly, an inductive element has a constitutive relation of the form
\begin{align}
    \label{eq:constitutive-inductance}
    I_b(t) = g_b(\Phi_b(t)),
\end{align}
with $g_b(\Phi_b(t)) = \Phi_b(t) / L$ for an ideal inductor $L$.
Typically, the relevant capacitive elements have a linear constitutive relation and we will choose to work with branch fluxes.

\subsection{Circuit components \label{ssec:cqed_circuit_components}}
We will list here the three primary components of superconducting circuits and how they contribute to the Lagrangian of the circuit, see Ref.~\cite{vool_introduction_2017} for a derivation of these energy contributions.
Additionally, voltage and current sources provide a means of (externally) manipulating the properties of the circuit in time.
Lastly, 
we treat loop configurations of Josephson junctions, such as the dc-SQUID, as primary elements.
We summarize all those elements in \cref{tab:circuit_elements}.

\paragraph{Capacitors.}
We consider capacitors as linear elements where the voltage is proportional to the charge stored on the capacitor $V(t) = q(t) / C = \dot{\Phi}(t)$, where $C$ is the capacitance.
The energy stored in a capacitor is given by 
\begin{align}
    \label{eq:def_capacitor_energy}
    E = \frac{1}{2} C \dot{\Phi}^2(t),
\end{align}
resembling a kinetic energy term within the Lagrangian.

\paragraph{Inductors.}
For a linear inductor with inductance $L$, the current is related to the magnetic flux, $I(t) = \Phi(t) / L$.
The energy stored in an inductor is given by 
\begin{align}
    \label{eq:def_capacitor_inductor} %
    E = \frac{1}{2L} {\Phi}^2(t),
\end{align}
resembling a potential energy term within the Lagrangian.

\paragraph{Josephson junctions.}
A Josephson junction consists of two superconductors separated by a thin insulating layer.
Because the layer is thin, Cooper pairs can tunnel through this barrier, leading to a current across the junction.
The current depends on the phase difference $\varphi$ between the wave functions describing the condensate of Cooper pairs on each superconductor, which is captured by the first Josephson relation~\cite{josephson_possible_1962}
\begin{align}
    I(t) = I_c \sin \varphi(t),
\end{align}
where $I_c$ is the critical current that depends on the geometry of the junction.
Additionally, the voltage between the superconductors is given by $V(t) = (\hbar /2e) \dot{\varphi}$.
This allows us to view the phase difference as a generalized flux, that is, $\Phi = \hbar \varphi / 2 e$.

Up to a constant energy shift, the energy of the Josephson junction that behaves like a nonlinear inductance takes the form 
\begin{align}
    E  = I_c \int \left( \dv{\Phi}{t'} \right) \sin\big(\frac{2 \pi}{\Phi_0} \Phi\big) \dd{t'} = - E_J \cos\big[\frac{2 \pi}{\Phi_0} \Phi(t)\big] ,
\end{align}
with $E_J = \Phi_0 I_c / 2 \pi$ the Josephson energy and $\Phi_0 = 2 \pi \hbar / 2e$ is the magnetic flux quantum.
It is common to introduce the reduced flux $\varphi(t) = 2 \pi \Phi(t) / \Phi_0$ to lighten the notation, as we will do in the following.

\paragraph{Voltage and current sources. \label{ssec:curren-voltage-sources}}
The Lagrangian formulation introduced above allows us to incorporate ideal current and voltage sources into the circuit description in terms of node fluxes.
A voltage source is modeled as a large capacitor $C_s \rightarrow \infty$ with large charge $Q_s \rightarrow \infty$ so that $Q_s / C_s = V_g$ is constant, where the constant $V_g$ is the applied voltage.
The voltage source is included in the Lagrangian by adding a kinetic energy term of the form
\begin{align}
    \label{eq:def_voltage_source}
    E = \frac12 C_g [\dot{\Phi}(t) - V_g(t)]^2.
\end{align}
Current sources are treated similarly, that is, we consider a large inductor $L_s \rightarrow \infty$ with large flux $\Phi_s \rightarrow \infty$ so that $\Phi_s / L_s = I_s$ is constant, where $I_s$ is the applied current.
Accordingly, we add a potential energy term of the form 
\begin{align}
    \label{eq:def_current_source}
    E = I_s \Phi(t),
\end{align}
to the Lagrangian.

\paragraph{dc-SQUID.}
Above, we have introduced two methods, external current and voltage sources, to change the properties of the system (in time) by classical control.
Another way to change the state of the system is by applying an external magnetic field through a loop formed by several Josephson junctions or a loop formed by a linear inductance and at least a single Josephson junction.
The presence of an external magnetic flux $\Phi_{\mathrm{ext}}(t)$ enters in the Lagrangian as a phase $\phi_{\mathrm{ext}}(t) = 2 \pi \Phi_{\mathrm{ext}}(t) / \Phi_0$ due to the flux quantization~\eqref{eq:def_kirchoffs_laws}.
The potential energy of the SQUID, shown in \sfigref{fig:example_lumped_element circuit}{c}, can be written as~\cite{riwar_circuit_2021}
\begin{align}
    \label{eq:def_general_squid}
    E = -E_{J1} \cos(\varphi + \alpha\phi_{\mathrm{ext}}) -E_{J2} \cos[\varphi + (\alpha - 1)\phi_{\mathrm{ext}}].
\end{align}
For time-independent $\phi_{\mathrm{ext}}(t)$, the parameter $\alpha \in \mathbb{R}$ is a gauge degree of freedom and can be chosen arbitrarily: any choice will give rise to the same predictions.
Recently, it has been pointed out~\cite{you_circuit_2019, riwar_circuit_2022} that additional care should be taken when considering time-dependent magnetic fields\footnote{This can be best understood by noting that the unitary transformation relating different gauges now requires additional terms in the Schrödinger equation, i.e., $H^{\prime} = \hat{U}(t) H \hat{U}(t)^{\dagger} + i \dv{t} \hat{U} \hat{U}(t)^{\dagger}  (t)$}.
We will ignore such issues here, which is typically sufficient for qualitative results and gaining an intuition.

As an example, let us continue from Eq.~\eqref{eq:def_general_squid}, choosing $\alpha = 1/2$, then one can derive that the potential energy of the dc-SQUID corresponds to the potential of a single Josephson energy with a (flux) tunable Josephson energy $E_J(\phi_{\mathrm{ext}})$, that is,
\begin{align}
    E = \underbrace{(E_{J1} + E_{J2}) \sqrt{\cos^2(\phi_{\mathrm{ext}} / 2) + \delta^2 \sin^2(\phi_{\mathrm{ext}} /2)}}_{E_J(\phi_{\mathrm{ext}})} \cos(\varphi -\varphi_0),
\end{align}
where $\varphi_0 = \arctan[\delta \tan(\phi_{\mathrm{ext}} / 2)]$ is an inconsequential flux offset and $\delta = (E_{J1} - E_{J2}) / (E_{J1} + E_{J2})$ characterizes the asymmetry of the two junctions.

\subsection{Method of nodes and the Lagrangian}
Having introduced a toolbox of components for assembling superconducting circuits, we now briefly present the methods of nodes as a practical approach to modeling most circuits containing Josephson junctions.
We will restrict this overview to static external fields, referring to~\cite{you_circuit_2019, riwar_circuit_2022} for time-dependent fields.
See Refs.~\cite{vool_introduction_2017, rasmussen_superconducting_2021-1} for a more complete presentation. 

Above, we already began utilizing the language of graph theory, and we will continue to do this here, especially to remove unnecessary degrees of freedom in a way that does not require Kirchhoff's law.
As each branch flux $\Phi_b$ corresponds to an edge $b$ in the network graph, we can equivalently represent it by two consecutive node fluxes $\phi_n$ and $\phi_{n'}$ that correspond to the vertices of $b$.
Thus, all voltages will be determined as the difference of two node fluxes, and we can arbitrarily choose one reference node called \emph{ground} to which we associate the value 0.
Additionally, we want to choose a \emph{spanning tree} $\mathcal{T}_{C}$ of the circuit $C$, that is, informally, a connected subgraph of the circuit that contains all nodes but does not contain loops.
The spanning tree naturally bipartitions the branches of the circuit into two sets, those contained within the spanning tree, and those contained in its closure (or complement) $\overline{\mathcal{T}}_C$.
Whether a branch $b$ is contained in the spanning tree or is a closure branch determines how we express it in terms of the node fluxes, that is,
\begin{align}
    \label{eq:def_spanning_tree_closure_branch_to_fluxes}
    \Phi_b = \begin{cases}
    \phi_n - \phi_{n'}, \, \forall b \in \mathcal{T}, \\
    \phi_n - \phi_{n'} + \Phi_{\mathrm{ext}}, \, \forall b \in \overline{\mathcal{T}}, \\
    \end{cases}
\end{align}
where $\Phi_{\mathrm{ext}}$ is the external flux through the loop closed by the branch.
We list the energies of the different components if they are contained in the spanning tree in \cref{tab:circuit_elements}, the energies for elements in the closure branch follow immediately from Eq.~\eqref{eq:def_spanning_tree_closure_branch_to_fluxes}.
From this, the Lagrangian of the circuit can be obtained as a signed sum of capacitive and inductive elements with the latter carrying a minus sign.
As customary, the classical Hamiltonian is obtained through the Legendre transformation of the Lagrangian, that is,
\begin{align}
    \mathcal{H}= \sum_{k} q_k \dot{\phi}_k - \mathcal{L}, \quad \text{with } q_k = \frac{\partial \mathcal{L}}{\partial \dot{\phi}_k},
\end{align}
with the generalized momentum $q_k$ also known as node charges.

\subsection{Quantization and anharmonic oscillators}
Given the classical Hamiltonian, we can obtain a quantum-mechanical description of the circuit through canonical quantization by promoting the variables to operators, that is, $\phi_k \to \hat{\phi}_k$ and $q_k \to \hat{q}_k$.
With the node flux operator $\hat{\phi}_k$ corresponding to a generalized coordinate and the node charge operator $\hat{q}_k$ corresponding to generalized momentum\footnote{We have maneuvered ourselves into a corner. The node charge operators $\hat{q}_k$ that take the role of a generalized momentum variable should not be confused with the position quadrature $\hat{q}$ introduced in \cref{chap:quantum_continuous_variables}.}, they should obey the canonical commutation relations $[\hat{\phi}_n, \hat{q}_m] = i \delta_{nm}$ where $\delta_{nm}$ is the Kronecker delta.

In the case of an ordinary LC circuit, the quantum Hamiltonian is that of the quantum harmonic oscillator.
We have seen previously, \cref{sec:quantum_harmonic_oscillator}, that the harmonic oscillator is diagonalized by introducing the bosonic annihilation $\hat{a}$ and creation $\hat{a}^{\dagger}$ operators fulfilling the canonical canonical commutation relations $[\hat{a}, \hat{a}^{\dagger}] = \mathbbm{1}$.
In this case, the flux and charge operators are expressed as
\begin{align}
    \hat{\phi} = \phi_{\mathrm{zpf}} (\hat{a} + \hat{a}^{\dagger}), \quad
    \hat{q} = -i q_{\mathrm{zpf}} (\hat{a} - \hat{a}^{\dagger}),
\end{align}
with $\phi_{\mathrm{zpf}} = \sqrt{\hbar Z_0 / 2}$ and $q_{\mathrm{zpf}} = \sqrt{\hbar / 2 Z_0}$, the zero-point fluctuations of flux and charge, respectively, and $Z_0 = \sqrt{L / C}$ is known as the impedance.
We introduce here the impedance explicitly as it plays an important role in the quantization of nonlinear circuits, as we will see in the following.

For superconducting circuits that include Josephson junctions, the Hamiltonian is inherently nonlinear, preventing direct diagonalization in terms of bosonic annihilation and creation operators. 
Nevertheless, in the weakly nonlinear regime, a common strategy is to first solve the linearized version of the circuit to obtain a basis in which the system is diagonal and then treat the junction's nonlinearity as a small perturbation. 
This approximation holds when the system remains in the low-energy regime, where flux fluctuations are confined near the minimum of the potential. 
Under this assumption, the flux across the junction can be expressed as a linear combination of the $M$ mode system, that is, 
\begin{align}
\hat{\phi} = \sum_{m = 1}^{M} \sqrt{\frac{\hbar \mathcal{Z}^{\mathrm{eff}}_m}{2}} (\hat{a}_m + \hat{a}^{\dagger}_m),
\end{align}
where $\mathcal{Z}^{\mathrm{eff}}_m$ is the effective impedance of the $m^{\mathrm{th}}$ mode of the circuit, a classical quantity, that is, in principle measurable.
This approach is known as blackbox quantization~\cite{nigg_black-box_2012}.
Note that in the junction, typically all modes of the circuit mix, allowing for the design of higher-order interactions between the modes.

For a nonlinear superconducting circuit in a parameter regime for which the black box quantization approach is valid, the linearized circuit is obtained through the Taylor expansion of the potential energy in the Hamiltonian.
For example, for a single Josephson junction, we have
\begin{align}
    E_J \cos \hat{\phi} = E_J - \frac{E_J}{2} \hat{\phi}^{2} + \frac{E_J}{24} \hat{\phi}^4 + O(\hat{\phi}^6).
\end{align}
Ignoring the constant term and incorporating the quadratic contribution into an effective linear inductance, the remaining quartic term introduces an anharmonicity to the otherwise linear system. 
Combining the Josephson junction with a capacitor results in what is commonly referred to as an anharmonic quantum oscillator\footnote{In a particular parameter regime, the resulting circuit can be viewed either as the charge qubit~\cite{nakamura_coherent_1999} or the transmon qubit~\cite{devoret_quantum_1997, koch_charge-insensitive_2007, paik_observation_2011}, see also Ref.~\cite{devoret_superconducting_2013} for a review.}.  
For a single mode, expressing the quantized Hamiltonian in terms of the bosonic operators $\hat{a}$ and $\hat{a}^{\dagger}$,
\begin{align}
    \label{eq:def_quantum_anharmonic_oscillator}
    \hat{H}_{\mathrm{AHO}} = \omega \hat{a}^{\dagger} \hat{a} + \frac{E_J}{24} \phi_{\mathrm{zpf}}^4 (\hat{a} + \hat{a}^{\dagger})^4 \approx \omega_{\mathrm{eff}} \hat{a}^{\dagger} \hat{a} - K \hat{a}^{\dagger 2} \hat{a}^2,
\end{align}
one often finds the anharmonic oscillator in the latter form.
Here, $\omega_{\mathrm{eff}} = \omega - 6 K$ is the effective frequency of the oscillator and the parameter $K$ is known as the Kerr nonlinearity.
The approximation we have performed above is known as the rotating wave approximation and will be described in more detail in the following section.

\begin{table}[!t]
    \centering
    \begin{tblr}{colspec={Q[c, m]Q[c, m]Q[c, m]Q[c, m]},
    rowspec={Q[ht=-5mm]Q[ht=-5mm]Q[ht=-5mm]Q[ht=-5mm]},
    rowsep=2pt, hline{1, 2, 6} = {-}{0.75pt},
    colsep=7pt,  %
  row{1}={font=\bfseries,rowsep=8pt}}
        Element & Symbol & Spanning tree & Black-box quantization  \\
         Capacitor &  \raisebox{-0.4\totalheight}{\includegraphics[height=10mm]{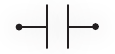}} & $\frac{C}{2}(\dot{\phi}_n - \dot{\phi}_{n'})^2$ & $q_{\mathrm{zpf}}^2 (\hat{b} - \hat{b}^{\dagger})^2$  \\ 
        Inductor & \raisebox{-0.4\totalheight}{\includegraphics[height=10mm]{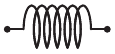}} & $\frac{1}{2L}(\phi_n - \phi_{n'})^2$ & $ \phi_{\mathrm{zpf}}^2 (\hat{b} + \hat{b}^{\dagger})^2$  \\ 
        {Josephson \\ junction} & \raisebox{-0.4\totalheight}{\includegraphics[height=10mm]{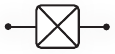}} & $-E_J \cos(\phi_n - \phi_{n'})$  &$-E_J \cos[\sum_m \phi_{m} (\hat{b}_m + \hat{b}_m^{\dagger})]$  \\
         {Current \\ source} & \raisebox{-0.4\totalheight}{\includegraphics[height=10mm]{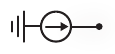}} & $I_S \phi_n$ & $I_S(t) (\hat{b} + \hat{b}^{\dagger})$ \\  
    \end{tblr}
    \caption{Circuit elements and their corresponding representations as symbols in diagrams, as Lagrangian terms if they are contained in the spanning tree,  and as terms in the Hamiltonian within the framework of black-box quantization. For the Josephson junction, $\phi_m$ denotes the zero-point fluctuations of the $m^{\mathrm{th}}$ mode mixing in the junction.}
    \label{tab:circuit_elements}
\end{table}

\section{Engineering Effective Interactions \label{sec:intro-universality-microwaves}}
The dominating energy scale in a Hamiltonian that describes a superconducting quantum circuit is typically set by the resonance frequency of the circuit $\omega_r $ which is on the order of GHz.
Any other coupling strengths, such as nonlinearities from Josephson junctions, are usually small compared to $\omega_r$ and typically on the order of tens to hundreds of kHz or MHz.
They can be seen as perturbations to the evolution of the system governed by the oscillator Hamiltonian $\hat{H}_0 = \omega_r \hat{a}^{\dagger} \hat{a}$.
Hence, any desired interaction can only be obtained effectively with respect to a specific frame, i.e., the interaction picture with respect to the free evolution $\hat{H}_0$.
Because the corresponding unitary transformation $\hat{U}(t) = \exp(i \hat{H}_0 t)$ results in a rotation of the phase space with frequency $\omega_r$, it is common to refer to this frame as the rotating frame.

In the rotating frame, the effective Hamiltonian takes the form
\begin{align}
    \hat{H}_{\mathrm{eff}}(t) = \hat{U}(t) \hat{H} \hat{U}(t)^{\dagger} + i \frac{\mathrm{d} \hat{U}(t)}{\mathrm{d} t} \hat{U}^{\dagger}(t) = \hat{U}(t) \hat{H} \hat{U}(t)^{\dagger} - \omega_r \hat{a}^{\dagger} \hat{a}.
\end{align}
The transformation $\hat{U}(t) \hat{H} \hat{U}(t)^{\dagger}$ corresponds to the replacement $\hat{a} \rightarrow \hat{a}(t) = \hat{a} \mathrm{e}^{-i \omega_r t}$ and the transformation for the creation operator $\hat{a}^{\dagger}$ is obtained by Hermitian conjugation.
Therefore, $\hat{H}_{\mathrm{eff}}(t)$ will often become time-dependent even if $\hat{H}$ is time-independent.
Because $\omega_r$ is on the order of GHz, this leads to fast oscillations which approximately average out over relevant time-scales set by a given interaction with strength $g$ on the order of MHz or less.
They can therefore be neglected if their coupling $g$ is small compared to their effective oscillatory frequency.
This is known as the rotating wave approximation (RWA)~\cite{walls_quantum_2008, schleich_quantum_2001}.

Energy preserving terms, for example, diagonal terms such as the Kerr interaction $K\hat{a}^{\dagger 2} \hat{a}^2$ are time-independent (and thus resonant) in any rotating frame.
On the other hand, energy non-conserving terms, for example, off-diagonal terms containing an unequal number of creation and annihilation operators, become time-dependent and have negligible effects.
To select these processes resonantly, it is therefore necessary to achieve an external time-dependent modulation of the relevant prefactors that counters the time-dependence of $\hat{a}^{(\dagger)}$ so that the desired interaction becomes effectively time-independent in the rotating frame.
For example, the squeezing type interaction $g_{\mathrm{sq}} (a^2 + a^{+ 2})$ requires a time modulation of the coupling strength at twice the frequency of the oscillator, that is, $g_{\mathrm{sq}} \to g_{\mathrm{sq}} \cos (2 w_r t)$.

In cQED, there are two common approaches to achieve modulation of couplings in time, which we are going to discuss now qualitatively.
The first method is based on modulating the flux through a superconducting loop, such as the SQUID, see \cref{ssec:cqed_circuit_components}.
This approach is commonly used for parametric amplification~\cite{wustmann_parametric_2013} or to engineer an exchange interaction between two modes~\cite{vrajitoarea_quantum_2020, lescanne_exponential_2020}.
Ignoring the issue of quantization in the presence of time-dependent fluxes, the potential term of the SQUID containing the external magnetic flux $\varphi_{\mathrm{ext}}(t)$ can be written as~\cite{vrajitoarea_quantum_2020}
\begin{align}
    \label{eq:flux-modulation-intro}
    \cos(\varphi - \varphi_{\mathrm{ext}}(t)) =& \cos(\varphi) \cos(\varphi_{\mathrm{ext}}(t)) - \sin(\varphi) \sin(\varphi_{\mathrm{ ext}}(t)) \nonumber \\ 
    \approx& \left(1 - \frac{1}{2!} \varphi^2 + \frac{1}{4!} \varphi^4\right) \cos(\varphi_{\mathrm{ext}}(t)) - \left( \varphi - \frac{1}{3!} \varphi^3 + \frac{1}{5!} \varphi^5 \right) \sin(\varphi_{\mathrm{ext}}(t)) \\ &+ \mathcal{O}(\varphi^6) \nonumber.
\end{align}
This results in time-dependent couplings in general. %
In principle, any arbitrary interaction can be selected by an appropriate modulation of $\varphi_{\mathrm{ext}}(t)$. 

Alternatively, effective time-dependent couplings can be obtained by applying an external current source, which leads to a displacement $\alpha(t)$ of the annihilation operator $\hat{a} \rightarrow \hat{a} + \alpha(t)$.
Here, the displacement parameter $\alpha(t)$ depends on the driving strength and the time-dependence is determined by the external modulation of the current source (see also \cref{ssec:curren-voltage-sources}).
Thus, the presence of an $n$-th order nonlinearity allows for resonantly selecting nonlinear terms of order $n-1$ or lower, which can be seen from the binomial identity,
\begin{align}
    \label{eq:displacement-drive-selction-expand}
    \left(\hat{a} + \hat{a}^{\dagger} \right)^n \rightarrow& \left(\hat{a} + \hat{a}^{\dagger} + \alpha(t) + \alpha^{*}(t) \right)^n \\ =& \sum_{k = 0}^{n} \binom{n}{k} \left(\hat{a} + \hat{a}^{\dagger} \right)^{n-k} \left( \alpha(t) + \alpha^{*}(t) \right)^k.
\end{align}

Both approaches have advantages and drawbacks depending on the concrete microwave architecture that is considered and the specific interaction that is desired.

\subsubsection{Engineering Effective Environments}
It is not always sufficient to solely engineer the interaction of closed quantum systems, as in some cases we desire to engineer the dissipative dynamics of an open quantum system. 
In particular, we aim for dynamics that do not drive the state of the system to a ``trivial'' thermal state but rather to a desired quantum state or manifold of quantum states.
For example, imagine that we wish to engineer the dissipation such that, ideally, the system of interest evolves according to the master equation,
\begin{align}
    \label{eq:engineering_diss_A}
    \dv{t}\hat{\rho} = \mathcal{D}[\hat{A}]\hat{\rho},
\end{align}
where $\mathcal{D}[\hat{A}]$ is the Lindblad dissipator with arbitrary jump operator $\hat{A}$, compare \cref{sec:cat_code} where we required $\hat{A} = \hat{a}^2 - \alpha^2$ to confine the states of the harmonic oscillator to the code space of the cat qubit.
It is possible to engineer such effective dissipation channels by starting from a larger system and adiabatically eliminating the dynamics of all but the system of interest.
To exemplify this procedure, let us consider a two-mode system $(\hat{a}$, $\hat{b}$) evolving according to the master equation
\begin{align}
    \dv{t} \hat{\rho}_{SB} = -i \left[ \hat{H}_{SB}, \hat{\rho}_{SB}\right] + \kappa_S \mathcal{D}[\hat{a}]\hat{\rho}_{SB} + \kappa_B \mathcal{D}[\hat{b}]\hat{\rho}_{SB},
\end{align}
where $\hat{H}_{SB} = g (\hat{A} \hat{b}^{\dagger} + \hat{A}^{\dagger} \hat{b})$ is the system-bath interaction with rate $g$ and $\hat{b}$ is the bosonic annihilation operator of the auxillary mode.
We emphasize that the operator $\hat{A}$ can be an arbitrary function of $\hat{a}$ and $\hat{a}^{\dagger}$.
For the system and the auxiliary mode, single-photon losses with rates $\kappa_S$ and $\kappa_B$, respectively, occur.
Consider a regime characterized by a small dimensionless parameter $\lambda$ such that $g / \kappa_B \sim \lambda$ and $\kappa_S / \kappa_B \sim \lambda^2$.
Phrased similarly, the rate of single-photon losses of the auxiliary system is the dominating energy scale of the system, and we have $\kappa_S \ll g \ll \kappa_B$.
We can therefore assume that the state of the auxiliary mode is close to the vacuum state and thus has a photon population much smaller than one.
To gain intuition, consider the case where the auxillary mode $B$ is in the vacuum state while the system $S$ is in an arbitrary state, i.e., $\hat{\rho}_{SB} = \hat{\rho}_S \otimes \ketbra{0}_B$. 
In this case, the interaction $\hat{A}\hat{b}^{\dagger}$ will generally be allowed whereas the reverse process, $\hat{A}^{\dagger} \hat{b}$, is prohibited due to $\hat{b}\ket{0} = 0$.
That is, the system $S$ is transferring an effective excitation $\hat{A}$ to the ancilla $B$, which creates a photon ($\hat{b}^{\dagger}$) of the same energy.
As the reverse process is forbidden, the system is effectively \emph{dissipating} $\hat{A}$.
As a result, the system evolves towards a state $\hat{\rho}$ that satisfies $\hat{A} \hat{\rho} = 0$.
This state must lie within the kernel of the operator $\hat{A}$.
If there is only a single state within the kernel of $\hat{A}$, the state satisfying  $\hat{A} \hat{\rho} = 0$ becomes the unique stationary state irrespective of the initial state.
If the kernel of $\hat{A}$ contains multiple states, the stationary state is not unique and will depend upon the initial state.

To make the above arguments more rigorous, let us consider an initial state of the form 
\begin{align}
    \hat{\rho}_{SB} = \hat{\rho}_{00} \ketbra{0}{0} + \lambda \left[ \hat{\rho}_{10} \ketbra{1}{0}  +  \hat{\rho}_{01} \ketbra{0}{1}\right] + \lambda^2 \hat{\rho}_{11} \ketbra{1}{1}.
\end{align}
Here we used $\hat{\rho}_{nm}$ to denote the projection of $\hat{\rho}_{SB}$ onto the Fock states $n, m$ of the ancillary mode.
We aim to derive the effective dynamics of the system alone, given by $\hat{\rho}_S = \Tr_B[\hat{\rho}_{SB}] = \hat{\rho}_{00} + \lambda^2 \hat{\rho}_{11}$ up to second-order in $\lambda$.
To this end, we assume $\lvert \lvert \hat{A} \rvert \rvert = O(1)$ in $\lambda$.
Following the derivation that can be found in Refs.~\cite{carmichael_statistical_2008, leghtas_confining_2015}, the time evolution of the different $\hat{\rho}_{mn}$ are given by
\begin{align}
    \frac{1}{\kappa_B} \dv{t}\hat{\rho}_{00} =& - i  \frac{g}{\kappa_B}(\hat{A}^{\dagger} \hat{\rho}_{10} - \hat{\rho}_{01} \hat{A}) + \frac{\kappa_S}{\kappa_B} \mathcal{D}[\hat{a}]\hat{\rho}_{00} + \hat{\rho}_{11} + \dots, \\
    \frac{1}{\kappa_B} \dv{t}\hat{\rho}_{10} =& - i  \frac{g}{\kappa_B} \hat{A} \hat{\rho}_{00} - \frac12 \hat{\rho}_{10} + \dots, \\
    \frac{1}{\kappa_B} \dv{t}\hat{\rho}_{11} =& - i  \frac{g}{\kappa_B}(\hat{A}\hat{\rho}_{01} - \hat{\rho}_{10} \hat{A}^{\dagger}) - \hat{\rho}_{11} + \dots
\end{align}
To solve this system of equations, we can make the adiabatic approximation, assuming that $\hat{\rho}_{10}$ is continuously in its steady state, that is, $\partial_t \hat{\rho}_{10} = 0$ and similarly $\partial_t \hat{\rho}_{11} = 0$.
Within this approximation, $\hat{\rho}_{10}$ and $\hat{\rho}_{11}$ are determined solely by $\hat{\rho}_{00}$ (and $\hat{A}$), and we arrive at the desired result
\begin{align}
    \label{eq:def_effective_dissipator}
    \dv{t} \hat{\rho}_S = \kappa_A \mathcal{D}[\hat{A}] \hat{\rho}_S + \kappa_{S} \mathcal{D}[\hat{a}]\hat{\rho}_S,
\end{align}
with $\kappa_A = 4 g^2  / \kappa_B$. We note that another possible approach, which is much more general than this example, is based upon the effective operator formalism for open quantum systems by Reiter and Sørensen~\cite{reiter_effective_2012}.

\section{Extracting Effective Models}
As the prospect of fault-tolerant quantum computing relies critically on the quality of quantum-gate operations, access to coherent gates with low error rates is crucial.
While the qualitative approach described above is typically a good starting point for engineering proof-of-concept effective interactions, engineering high-fidelity operations routinely requires going beyond the rotation wave approximation, and, for qubit systems, going beyond the two-level approximation.

\subsection{The Schrieffer-Wolff transformation}
The Schrieffer-Wolff transformation~\cite{schrieffer_relation_1966, luttinger_motion_1955} is an analytical tool that allows one to accurately and systematically capture effects that go beyond the rotation wave approximation and many other approximations.

The traditional perturbative approach due to Rayleigh and Schrödinger provides corrections to the energy levels and eigenstates of the Hamiltonian $\hat{H} = \hat{H}_0 + \hat{V}$ with respect to the unperturbed properties of eigenstates and eigenvalues of $\hat{H}_0$ due to a perturbation $\hat{V}$.
However, one might argue that the modern view of quantum mechanics is less concerned with states and energies, but that one rather becomes more used to thinking in terms of Hermitian operators, that is, elementary Hamiltonians, and how combining them affects properties of the system.
Fortunately, Schrieffer-Wolff perturbation theory is formulated on the level of operators and can therefore be viewed as a modern approach to perturbation theory in quantum mechanics.

In a nutshell, Schrieffer-Wolff perturbation theory is concerned with finding an anti-Hermitian generator $\hat{S} = - \hat{S}^{\dagger}$ of a unitary transformation $\hat{H}_{\mathrm{eff}} = \exp(\hat{S})\hat{H}\exp(\hat{S}^{\dagger})$ which transforms $\hat{H} = \hat{H}_0 + \hat{V}$ to a basis in which it is \emph{more} diagonal.
Here, \emph{more} diagonal refers to a transformation that suppresses the high-energy contributions of the perturbation $\hat{V}$ up to a desired order, while retaining its influence within the low-energy subspace.
To be explicit, by choosing the generator $\hat{S}$ to satisfy the operator equation $[\hat{H}_0, \hat{S}] = V$, one obtains an effective Hamiltonian
\begin{align}
    \hat{H}_{\mathrm{eff}} = \hat{H}_0 + \frac{1}{2} \comm{\hat{S}}{\hat{V}} + O(\hat{V}^3),
\end{align}
for which the perturbation $\hat{V}$ is removed to first order.
In principle, one can remove the off-diagonal part of the perturbation up to arbitrary order by an appropriate choice of $\hat{S}$.

However, historically, the difficulty of the Schrieffer-Wolff transformation was to apply it to higher orders, as no constructive methods to solve the respective operator equations existed beyond the first order~\cite{wegner_flow-equations_1994}.
\vfill

\section{Conclusion and Outlook}
In this chapter, the fundamental building blocks of superconducting circuits have been outlined as a means to engineer complex quantum systems.
The focus has been placed on how these systems can realize controllable nonlinear interactions between quantum degrees of freedom, enabling the realization of protected quantum memories, quantum gates, and tailored dissipative dynamics.

At the heart of circuit quantum electrodynamics (cQED) lies the principle that, by combining Josephson junctions with linear circuit elements, effective Hamiltonians of remarkable complexity can be ``printed'' directly onto a substrate.
However, deriving the desired dynamics from the set of available circuit components requires accurate perturbative techniques for extracting effective low-energy models that faithfully capture the essential physics.
To this end, \refpaper{II} introduces an iterative procedure for constructing the generators of the Schrieffer-Wolff transformation to arbitrary perturbative order\footnote{Around the same time, Ref.~\cite{venkatraman_static_2022} introduced a similar method.}.
This provides a systematic method to eliminate high-energy contributions while computing effective models directly at the operator level.
Crucially, it allows one to obtain an approximately linear effective Hamiltonian from an inherently nonlinear quantum system by judiciously tuning the coefficients of low-degree nonlinearities such that they cancel in the effective model.

This idea lies at the heart of \refpaper{I}, which proposes an explicit construction of a universal gate set for continuous-variable quantum computation with superconducting circuits based on the superconducting nonlinear inductive element (SNAIL)~\cite{frattini_3-wave_2017}.
The proposed architecture enables the implementation of both Gaussian and non-Gaussian gates, such as the cubic phase gate, by selectively activating interactions through parametric modulation. 
At the same time, the static, i.e., undriven, effective dynamics remain (approximately) linear.
This separation of static and driven dynamics allows for high-fidelity gate operations while suppressing spurious nonlinear effects.

The theoretical proposal of \refpaper{I} is successfully demonstrated experimentally in \refpaper{III}, showcasing the transition from theoretical design to practical realization.
The techniques developed and employed in this chapter provide a foundation for bridging theoretical constructions with experimental implementations, a central theme that runs throughout this thesis.

\cleardoublepage

\chapter{Conclusion and Perspective}

\glsresetall
This thesis has explored \gls{qldpc} codes and bosonic codes based on quantum continuous variables as
foundational elements for scalable fault-tolerant quantum computing.
Throughout, a central aim has been to bridge the gap between abstract theoretical advances and the realities of physical hardware, developing abstractions that shift experimental challenges toward more achievable targets.
By examining the interplay between code design, decoding complexity, and physical implementation constraints, this work offers insights into advancing quantum error correction strategies for near-term devices while developing conceptual foundations for scalable architectures beyond the limitations of current technology.

To this end, on the near-term level, we provide both a theoretical proposal and an experimental demonstration of a universal gate set for continuous-variable quantum computing, leveraging the hardware-native nonlinearities of the superconducting nonlinear asymmetric inductive element (SNAIL). 
This showcases the theoretically proposed versatility of the SNAIL, enabling access to both charge-driven and flux-driven interactions while remaining approximately Kerr-free.

Additionally, the dissipatively stabilized squeezed cat qubit, a noise-biased bosonic encoding, is introduced.
Compared to the ordinary cat qubit, squeezed cat qubits offer significantly improved error protection against bit-flips with very limited effect on the protection against phase-flips.
These advantages are retained under a concrete noise model derived from the proposed dissipative stabilization mechanism. 
Recently, Rousseau \emph{et al.}~\cite{rousseau_enhancing_2025} experimentally demonstrated the enhanced protection, with exceptional agreement to the theoretical proposal.

Going beyond near-term devices, this thesis proposes quantum radial codes as a high-threshold, low-overhead, and single-shot quantum memory.
While these codes cannot achieve asymptotic optimality in encoding rate and distance, the emphasis lies on realistic scenarios involving circuit-level noise and finite, practically relevant code sizes.
In this regime, their performance is competitive with bivariate bicycle codes~\cite{lin_quantum_2023, bravyi_high-threshold_2024}.
However, they offer an alternative geometric perspective that may prove advantageous for constructing fault-tolerant operations or tailored decoders.

Furthermore, as a step towards solving the problem of accurate real-time decoding of \gls{qldpc} codes, the localized statistics decoding algorithm is introduced as the first parallel decoding algorithm that matches the performance of the current state-of-the-art ordered statistics decoding algorithm for general quantum error correction protocols.
Most abstractly, the thesis introduces the concept of fault complexes, which broadens the homological framework of quantum error correction from static CSS codes to dynamic, time-evolving fault-tolerant protocols.
Fault complexes open the door to applying the full machinery of homology theory to the analysis and design of quantum error correction protocols in space-time.

However, building a fault-tolerant quantum computer remains a daunting challenge, where theoretical ideas can shape experiments by pushing the boundaries of what is necessary and what is possible, and experimental realities reshape theoretical imagination.
A timely and prominent example is a class of bivariate bicycle codes investigated by Bravyi \emph{et al.}~\cite{bravyi_high-threshold_2024}, which, with their high encoding rate and threshold, compete with the surface code, which has remained unchallenged in terms of its high error threshold for almost 20 years.
While the investigated codes require some long-range connectivity in addition to the otherwise nearest-neighbor connectivity in the plane, the significant reduction in overhead motivates the additional effort of engineering long-range couplers within a superconducting architecture. 

Studies such as the one by Bravyi \emph{et al.}~\cite{bravyi_high-threshold_2024} and the work presented on quantum radial codes in this thesis show that to build a high-threshold, low-overhead quantum computer, it is crucial to simulate potential constructions within the sheer endless design space.
The reason is that currently, the only method to accurately estimate the error threshold of a quantum error-correcting code is by direct simulation with a realistic circuit-level noise model.
However, an additional complication with these realistic noise models is that the ordering of operations during the syndrome extraction cycle becomes relevant and can significantly impact the fault distance of the code.
However, no general, efficient tools exist to determine the circuit distance of a code in the presence of noise.

A promising avenue to approach this problem is rooted within the formalism of fault complexes extended to circuit-level noise.
Intuitively, there should exist a mapping from the phenomenological noise case to the circuit-level noise case, potentially described by a chain map, that is, a homomorphism of chain complexes.
Given such a mapping, one might uncover (simple) conditions under which the fault distance is preserved, enabling more precise predictions of code performance.

Furthermore, it might be worthwhile to consider an approach to quantum error correction inspired by classical error correction.
In classical coding theory, decoder performance often dictates the direction of code development.
One might argue that such an approach has been partially followed by rephrasing the decoding problem of various topological codes such that it becomes matchable, see e.g., ~\cite{kubica_efficient_2023, brown_conservation_2022, sahay_decoder_2022}.
However, this approach has been limited and only considered minimum-weight error decoders.

The performance gap between (classical) minimum-weight error and (quantum) maximum-likelihood decoders for circuit-level noise is largely unexplored with a few exceptions, e.g., ~\cite{piveteau_tensor_2023}, due to the complexity and associated overhead of the decoder.
While certain code families~\cite{poulin_optimal_2006, iyer_hardness_2013} have efficient maximum-likelihood decoders in the static, i.e., code capacity, case, to the best of our knowledge, no efficient maximum-likelihood decoding algorithms exist for codes in space-time. 
Developing such decoders is a crucial open problem that will require further theoretical insights on the structure of the decoding problem in space-time. 
This represents an exciting area of future research, with the potential to significantly enhance the performance of quantum error correction on near-term devices and beyond.

Even though gate fidelities and coherence times for discrete-variable qubits are steadily increasing, bosonic encodings potentially offer a shortcut to lower effective physical error rates. 
A major challenge ahead is the development of scalable architectures involving multiple bosonic modes, along with the validation of corresponding theoretical error models. 
For stabilized cat qubits, the implementation of high-fidelity two-qubit gates remains a key obstacle. 
While a fully bosonic two-qubit gate between two cat qubits has not yet been demonstrated, experimental progress suggests that such a realization is within reach. 
In the meantime, two-qubit operations between hybrid cat-transmon architectures have been successfully demonstrated, enabling operation of the repetition code near the memory threshold~\cite{putterman_hardware-efficient_2025}. 
Although the current memory lifetime is not limited by the two-qubit gate fidelity, improving these operations will be critical for future scaling. 
One possible avenue is to replace stabilized cat qubits with stabilized squeezed cat qubits, which may reduce control errors related to the narrower peak structure of the squeezed cat qubit wavefunction compared to the ordinary cat qubit.
Overall, the sheer vastness of the superconducting qubit design space offers enormous potential to further enhance bosonic qubit performance --- potentially beyond what currently seems imaginable.

Fault-tolerant quantum computing stands at an extraordinary moment.
Theoretical ideas, once thought too abstract or futuristic, can now find their way into laboratory demonstrations within years, if not months.
This rapid interplay between theory and experiment creates a momentum that is both exhilarating and unprecedented.
In many ways, the current pace and scope of progress evoke the intellectual thrill that must have accompanied the discovery of quantum mechanics a century ago --- when a new formalism redefined our understanding of nature.
Likewise, the rapid development today demands that we reconsider the mathematical formalism to describe quantum error-correcting codes.
While the language of chain complexes has been tremendously fruitful for the discovery of error-correcting codes, we must ask ourselves whether thinking about codes limits our imagination.
Recent discoveries~\cite{delfosse_spacetime_2023, alam_bacon-shor_2025, shaw_lowering_2025, mcewen_relaxing_2023, bombin_unifying_2024, gidney_less_2023} point toward a broader perspective, that there must exist a more fundamental structure than codes.
This is what makes this field so exciting right now.
It’s this push beyond traditional frameworks, toward something fundamentally new, that makes this a truly exciting moment for the field.


\renewcommand{\headrulewidth}{0pt}
\fancyhead{}

\phantomsection %
\printbibliography[notcategory=fullcited,resetnumbers=true]
\topnotinctext{Other work outside the scope of this thesis:}
\papernotincluded{\protect\fullcite{pub_lu_resolving_2023}} \mybibexclude{pub_lu_resolving_2023}
\papernotincluded{\protect\fullcite{pub_alexander_generation_2024-1}} \mybibexclude{pub_alexander_generation_2024-1}
\papernotincluded{\protect\fullcite{pub_walshe_linear-optical_2025}} \mybibexclude{pub_walshe_linear-optical_2025}
\papernotincluded{\protect\fullcite{pub_aghaee_rad_scaling_2025}} \mybibexclude{pub_aghaee_rad_scaling_2025}

\paper{\citefield{pub_hillmann_universal_2020}{title}}{\fullcite{pub_hillmann_universal_2020}} \mybibexclude{pub_hillmann_universal_2020}

\paper{\citefield{pub_hillmann_designing_2022}{title}}{\fullcite{pub_hillmann_designing_2022}} \mybibexclude{pub_hillmann_designing_2022}

\paper{\citefield{pub_eriksson_universal_2024}{title}}{\fullcite{pub_eriksson_universal_2024}} \mybibexclude{pub_eriksson_universal_2024}

\paper{\citefield{pub_hillmann_performance_2022}{title}}{\fullcite{pub_hillmann_performance_2022}} \mybibexclude{pub_hillmann_performance_2022}

\paper{\citefield{pub_hillmann_quantum_2023}{title}}{\fullcite{pub_hillmann_quantum_2023}} \mybibexclude{pub_hillmann_quantum_2023}

\paper{\citefield{pub_berent_analog_2024}{title}}{\fullcite{pub_berent_analog_2024}} \mybibexclude{pub_berent_analog_2024}

\paper{\citefield{pub_hillmann_localized_2024}{title}}{\fullcite{pub_hillmann_localized_2024}} \mybibexclude{pub_hillmann_localized_2024}

\paper{\citefield{pub_scruby_high-threshold_2024}{title}}{\fullcite{pub_scruby_high-threshold_2024}} \mybibexclude{pub_scruby_high-threshold_2024}

\paper{\citefield{pub_hillmann_single-shot_2024}{title}}{\fullcite{pub_hillmann_single-shot_2024}}\mybibexclude{pub_hillmann_single-shot_2024}

\nocite{pub_hillmann_universal_2020,pub_hillmann_designing_2022,pub_eriksson_universal_2024,pub_hillmann_performance_2022, pub_hillmann_quantum_2023,pub_berent_analog_2024,pub_hillmann_localized_2024,pub_scruby_high-threshold_2024,pub_hillmann_single-shot_2024}

\end{document}